\documentclass[one column,a4paper]{article}
\usepackage{etoolbox}
\newtoggle{arxiv}
\toggletrue{arxiv}

\newtoggle{linenumbers}
\togglefalse{linenumbers}

\newtoggle{incCont} %
\togglefalse{incCont}

\newtoggle{incConj} %
\togglefalse{incConj}

\usepackage[utf8]{inputenc}
\usepackage{charter}

\usepackage{amsmath,amssymb,color,enumerate,graphicx,mathrsfs,xspace,subcaption,algorithmicx,algpseudocode,booktabs,adjustbox,cancel,threeparttable,algorithm,pdflscape,afterpage,longtable,xargs,listings}
\usepackage[normalem]{ulem}
\usepackage{longtable,lscape,tabularx}
\usepackage[usenames,dvipsnames]{xcolor}
\iftoggle{arxiv}{
\usepackage[noblocks,auth-sc]{authblk}
}{}
\usepackage[sort&compress,comma]{natbib}
 \bibpunct[, ]{[}{]}{,}{n}{}{,}%
\usepackage[colorlinks=true, allcolors=black, linkcolor=PineGreen, citecolor=blue]{hyperref}
\usepackage{orcidlink}
\usepackage{backref}
 \renewcommand*{\backref}[1]{}
 \renewcommand*{\backrefalt}[4]{%
     \ifcase #1 (Not cited.)%
     \or        (Cited on page~#2.) %
     \else      (Cited on pages~#2.) %
     \fi}

\iftoggle{linenumbers}{
    \usepackage[mathlines,displaymath]{lineno}%
    \makeatletter
    \linenumbers
    
}{
}

\iftoggle{arxiv}{
    \linespread{1}
    \allowdisplaybreaks
}{
\DoubleSpacedXII
}

\iftoggle{arxiv}{
    \usepackage{fullpage,amsthm}
    \usepackage[nameinlink]{cleveref}
    \newtheorem{definition}{Definition}
    \newtheorem{theorem}{Theorem}
    
    \newtheorem{example}{Example}
    \newtheorem{proposition}[theorem]{Proposition}
    \newtheorem{corollary}[theorem]{Corollary}
    \newtheorem{conjecture}{Conjecture}
    \newtheorem{remark}{Remark}
\newcommand{\ABSTRACT}[1]{\begin{abstract}#1\end{abstract}}
\newcommand{\KEYWORDS}[1]{}
\newcommand{\HISTORY}[1]{}
\crefname{appendix}{Appendix}{Appendices}
\newcommand{\Halmos}{}
\newcommand{\BFA}{\mathbf{A}}
\newcommand{\BFB}{\mathbf{B}}
\newcommand{\BFC}{\mathbf{C}}
\newcommand{\BFG}{\mathbf{G}}
\newcommand{\BFI}{\mathbf{I}}
\newcommand{\BFM}{\mathbf{M}}
\newcommand{\BFP}{\mathbf{P}}
\newcommand{\BFQ}{\mathbf{Q}}
\newcommand{\BFR}{\mathbf{R}}
\newcommand{\BFU}{\mathbf{U}}
\newcommand{\BFV}{\mathbf{V}}
\newcommand{\BFd}{\mathbf{d}}

\newcommand{\BFu}{\mathbf{u}}
\newcommand{\BFv}{\mathbf{v}}
\newcommand{\BFw}{\mathbf{w}}
\newcommand{\BFx}{\mathbf{x}}
\newcommand{\BFy}{\mathbf{y}}
\newcommand{\BFz}{\mathbf{z}}

\newcommand{\BFSigma}{\mathbf{\Sigma}}
\newcommand{\BFzero}{\mathbf{0}}
}{
\usepackage[nameinlink]{cleveref}
\TheoremsNumberedThrough  
\ECRepeatTheorems
\EquationsNumberedThrough 
}
\renewcommand{\bar}{\overline}
\renewcommand{\tilde}{\widetilde}
\renewcommand{\hat}{\widehat}

\crefname{assumption}{Assumption}{Assumptions}
\crefname{conjecture}{Conjecture}{Conjectures}
\crefname{lemma}{Lemma}{Lemmata}
\crefname{theorem}{Theorem}{Theorems}
\crefname{proposition}{Proposition}{Propositions}
\crefname{corollary}{Corollary}{Corollaries}
\crefname{claim}{Claim}{Claims}
\crefname{algorithm}{Algorithm}{Algorithms}
\crefname{definition}{Definition}{Definition}
\crefname{remark}{Remark}{Remarks}
\crefname{figure}{Figure}{Figures}
\crefname{section}{Section}{Sections}
\crefname{example}{Example}{Examples}
\crefname{equation}{}{}
\crefname{table}{Table}{Tables}

\newcommand{\BR}{{BR }}
\newcommand{\R}{\mathbb{R}}

\newcommand{\Z}{\mathbb{Z}}

\newcommand{\conv}{\operatorname{conv}}

\newcommand{\ceil}[1]{\left\lceil#1\right\rceil}
\newcommand{\norm}[1]{\left\Vert#1\right\Vert}

\newcommand{\Mne}{\mathbf{\zeta}}
\newcommand{\Lattice}{\mathbf{\Lambda}}
\newcommand{\Latticex}{{\Lattice_1}}
\newcommand{\Latticey}{{\Lattice_2}}

\newcommand{\covRad}[1][\Lattice]{\textcolor{black}{\mu (#1)}}

\newcommand{\Qq}[1][]{ \textcolor{black} {{\BFQ}_{#1}} }
\newcommand{\Cq}[1]{ \textcolor{black} {{\BFC}_{#1}} }
\newcommand{\dq}[1][]{ \textcolor{black} {{\BFd}_{#1}} }

\newcommand{\Rq}[1]{ \textcolor{black} {{\BFR}_{#1}} }
\newcommand{\Br}{\textcolor{black}{\mathcal{B}}}
\newcommand{\Qx}{ \Qq[1] }
\newcommand{\Qy}{ \Qq[2] }
\newcommand{\Cx}{ \Cq{1} }
\newcommand{\Cy}{ \Cq{2} }
\newcommand{\dx}{ \dq[1] }
\newcommand{\dy}{ \dq[2] }

\newcommand{\Rx}{ \Rq{1} }
\newcommand{\Ry}{ \Rq{2} }
\newcommand{\Brx}[1][\BFy]{\textcolor{black}{\Br_1\left (#1\right )}}
\newcommand{\Bry}[1][\BFx]{\textcolor{black}{\Br_2\left (#1\right )}}

\newcommandx{\proxn}{\textcolor{black}{\pi}}
\newcommandx{\prox}[2][1={\Qq},2={}]{\textcolor{black}{\proxn_{#2} \left( #1 \right) }}
\newcommandx{\xyVec}[2][1={},2={}]{\begin{pmatrix}#1\BFx#2\\#1\BFy#2\end{pmatrix}}
\newcommandx{\vecN}[1]{\begin{pmatrix}#1\end{pmatrix}}
\newcommandx{\ellip}[2][1={\eta},2={\BFu}] {\textcolor{black}{\mathcal{E}_{#1, #2}}}
\newcommand{\Flt}[1]{\textcolor{black}{ \phi _{ #1 }}}

\newcommand{\mytitle}{Best-response Algorithms for Lattice Convex-Quadratic Simultaneous Games}

\newcommand{\myname}{Sriram Sankaranarayanan}
\newcommand{\myaffil}{Operations and Decision Sciences, Indian Institute of Management Ahmedabad}
\newcommand{\myemail}{srirams@iima.ac.in}

\date{}

\begin{document}
\iftoggle{arxiv}{
    \title{\mytitle}
    \author{\myname\,\orcidlink{0000-0002-4662-3241}\thanks{\myaffil, \myemail}}
    \date{}
    \maketitle
}{
    \TITLE{\mytitle}
    \RUNAUTHOR{Sankaranarayanan}
    \RUNTITLE{\mytitle}
    \ARTICLEAUTHORS{%
    \AUTHOR{\myname}
    \AFF{\myaffil, \EMAIL{\myemail}} %
    }
}

\ABSTRACT{We evaluate the best-response (BR) algorithm for lattice convex-quadratic games, where the players have nonlinear objectives and unbounded feasible sets. We provide a sufficient condition that if certain interaction matrices (the product of the inverse of the positive definite matrix defining the convex-quadratic terms and the matrix that connects one player's problem to another's) have all their singular values less than 1, then the iterates do not diverge regardless of the initial point. We prove that if the iterates are trapped among finitely many strategies (called a trap), a relaxed version of the Nash equilibrium can be calculated by identifying a mixed-strategy Nash equilibrium of the finite game where the players' strategies are restricted to those in the trap. To establish the tightness of our sufficient condition, we also show examples where even if one singular value of one interaction matrix exceeds 1, there are infinitely many initial points from which the iterates diverge. Finally, we prove that if all the singular values of all the interaction matrices exceed 1, then the iterates diverge from every initial point except possibly a finite set of initializations. 
}

\maketitle

\section{Introduction}

Nash equilibium is a fundamental concept in game theory that predicts the outcome of strategic interactions among players. 
While computing them can be challenging, the advent of computational tools has made it possible to solve games with a large number of players and strategies.
Finding  pure and mixed-strategy Nash equilibria (PNE and MNE, respectively) is a relatively well-understood problem in two classes of games (i) finite games, where each player has finitely many strategies and (ii) convex games, where each player's feasible set is a convex set and the objective is continuous.
However, with advancements in computing technology, there has been  a surge in the study of games that deviate from these assumptions, and have structured nonconvexities in the players’ optimization problems. 
Our focus in this  paper is is to consider games, where the players' feasible sets are lattices. A lattice is a discrete additive subgroup (defined formally in \cref{sec:defnLattice}). 
For example, $\Z^{n} \subset \R^{n}$ is a lattice. 
In the games we consider, each player solves a convex-quadratic minimization problem over a lattice.
We define them formally below. 
\begin{definition}%
    A {\em Lattice Convex-Quadratic Simultaneous game} is a game of the form :
    \begin{align}
        \textbf{$\BFx$-player:}
        \min_{\BFx \in \Latticex} &: \frac{1}{2}\BFx^\top \Qx \BFx + (\Cx \BFy + \dx )^\top \BFx  \qquad  \nonumber \\
        \textbf{$\BFy$-player:}
        \min_{\BFy \in \Latticey} &: \frac{1}{2}\BFy^\top \Qy \BFy + (\Cy \BFx + \dy )^\top \BFy,   \tag{LCQS} \label{eq:intQuad}
    \end{align} 
where $\Latticex \subset \R^{n_{x}}$ and $\Latticey \subset \R^{n_{y}} $ are full-rank lattices; and $\Qx$ and $\Qy$ are symmetric positive definite matrices. 
\end{definition}
In this paper, we  explore the suitability of employing best-response (BR) algorithms for  \cref{eq:intQuad}, where in each iteration, the players play an optimal response to the strategies of the remaining players in the previous iteration. 
Typically, the iterative approach suggested by the \BR algorithm is quite easy to implement. Thus, they are of particular interest and are analyzed extensively. 
The central inquiry in the analysis of such algorithms is whether this adaptive algorithm  leads to a PNE or not.

Notably for finite games \BR algorithm converge to a PNE if the game has a {\em potential function} \citep{monderer1996potential}. Such games are called {\em potential games}.
For games that are not potential, even in the context of \emph{finite} games, the question of convergence to a PNE becomes contingent on specific conditions.
For one, a PNE might not even exist for the given game. 
	In this case, the \BR algorithm will necessary cycle among a subset of strategies. 
Alternatively, a PNE might exist, but still the \BR algorithm could become \emph{trapped} in a subset of strategies (referred to as \emph{trap} following \citet{Mimun2024}), none of which constitute a PNE.
The idea of cycling among a finite set of strategies and never escaping them was used to motivate the concept of \emph{sinking equilibria} in \citet{Goemans2005}. 
However, when the feasible set is unbounded, a third possibility arises -- the \BR algorithm may diverge, producing iterates that diverge off to infinity.
There are well-known instances of quadratic programming games with two players and one variable per player, where this happens. 
We have shown an example below, as adapted from \citet[Supplementary Material, B. Divergence of SGM]{carvalho_2020_computing}.
\begin{example} \label{ex:unbndDiverge}
    Consider the following simultaneous game. 
        \begin{align*}
\textbf{$\BFx$-player:}\min_{\BFx \in \Z} \BFx^2 - 4\BFx\BFy \qquad\qquad\qquad\qquad
\textbf{$\BFy$-player:}\min_{\BFy \in \Z} \BFy^2 - 4\BFx\BFy
        \end{align*}
    In the game given above, suppose the initial strategy for the $\BFx$-player and $\BFy$- player be $5$ and $5$ respectively. 
    Setting $\BFy=5$, the $\BFx$-player's objective becomes $\BFx^2 - 20\BFx$ whose integer minimum is $\BFx=10$. By symmetry, setting $\BFx=5$, the $\BFy$-player's optimum deviation is $\BFy=10$. 
    Thus the best response by each player gives a strategy pair $(10, 10)$. And from there on, the next best responses would be $(20, 20)$. 
    Moreover, we can observe that in this procedure, the successive iterates will be $(5 \times 2^{i+1}, 5 \times 2^{i+1})$ which diverge. 
\end{example}
The \BR algorithm diverges for this game, not due to non-existence of a Nash equilibrium as the game has a Nash equilibrium at $(0,0)$.
The divergence is not due to the game not being a potential game either as the game has an exact potential function given by $\Phi(\BFx, \BFy) = \BFx^2 - 4\BFx\BFy + \BFy^2$. 
\cref{ex:unbndDiverge} shows that the \BR algorithm can diverge even for a potential game, when the feasible sets of the player is a lattice, \emph{i.e., } a very structured but unbounded set.
This motivates our attempt in understanding the convergence properties of \BR algorithms over lattices. 

A question that follows is the motivation behind working with \emph{lattice ``convex-quadratic'' games}.
This is because, among integer-programming games with nonconvexities in them, the standard methods use a form of convexification of the feasible set. 
For example, \citet{NashStackelberg,CutnPlay} consider families of games where the objective of a player, when other players' variables are frozen, turn out to be linear\footnote{\citet{CutnPlay} calls such games {\em reciprocally bilinear.}}.  
In such cases, finding PNEs for the game where the feasible set is replaced by the convex hull of the original feasible set, helps in finding a {\em mixed-strategy Nash equilibrium} (MNE) for the original game. 
However, the central theorems in these papers fundamentally make use of linearity of the objectives, and the arguments fail when the objectives are not linear. 
Thus, we want to begin analysis of the simplest of nonlinearities, motivating convex quadratic objective functions. As far the feasible sets go, the simplest of feasible sets that is neither finite nor convex is the integer lattice -- $\Z^{n}$. 
Nevertheless, since our results hold for general lattices, we present them in maximum possible generality in this paper. 
\paragraph{Contributions. }
In the context of \cref{eq:intQuad}, , we list our contributions in this paper here below. 
\begin{enumerate}
	\item For \cref{eq:intQuad}, we provide necessary and sufficient conditions for when the \BR algorithm will not diverge, irrespective of the initial iterate. 
		This corresponds to either stopping at a PNE, or being in a \emph{trap}. 
		The sufficient condition is that all singular values of a particular set of matrices are less than $1$, a condition we call as the game having {\em positively adequate objectives} (\cref{thm:BRintConds}). 
		To hint that we have not made an unnecessarily strong assumption, we show an example (\cref{ex:unbndDivergePlus}), where \emph{all } singular values except one of them is strictly less than $1$. In this game, we show that there are an infinite number of initial points from which the \BR algorithm diverges, and infinite number of initial points from which the algorithm converges to a PNE.
We extend it with a stronger negative result that states, if all singular values of the same set of matrices are greater than $1$, then the \BR algorithm diverges from \emph{every} initial point, except from possibly finitely many points (\cref{thm:BRintDiv}). 
	We also note that these conditions for finite termination (and divergence) in \cref{thm:BRintConds,thm:BRintDiv} are dependent \emph{ only} on the structure of the objective function and \emph{not} on the structure of the feasible set, \emph{i.e., } the lattices over which the players are optimizing.
\item We show that for games with positively adequate objectives, when the algorithm gets into a \emph{trap}, and if $\Mne$ is an MNE of the restricted game where each player's strategies are restricted to those in the \emph{trap}, then $\Mne$ is a $\Delta$-MNE to the \cref{eq:intQuad} (\cref{cor:final}). 
        The paper explicitly computes the $\Delta$ that is possible. 
    \item As a form of tightness to the above result, we show the following. Let $\Delta > 0$ be given. There exist games where cycling occurs (without the said condition of positively adequate objectives holding), and given any MNE of the restricted game, at least one player has a deviation that improves their objective more than $\Delta$.
		In other words, this shows that the proof of $\Delta$-MNE as opposed to an actual MNE, or even a uniform bound on $\Delta$,  is not due to lack of tightness in analysis, but an inherent property of the \BR algorithm (\cref{thm:intFinNeg}).
	\item We perform computational experiments on two classes of problems (\cref{sec:comp}). When there are at least three players, we provide empirical evidence that our algorithm significantly outperforms the SGM algorithm \citep{carvalho_2020_computing}. 
        Moreover, while the theorems guarantee only a $\Delta$-MNE with our algorithm, empirically,  we always obtain an MNE. 
		This leads us to end the paper with a conjecture that the theorem can be strengthened to state that when we have positively adequate objectives, an MNE is always retrievable (\cref{conj:final}).	
	\item Finally, we also present  a technical result on proximity in lattice convex-quadratic optimization, a result that could be of independent interest, and on which almost all of the fundamental results in this paper are based on (\cref{thm:proxProof}). 
		In particular, using the {\em flatness theorem} and ideas based on \emph{covering radius} of a lattice, we show that the distance between the lattice minimizer and the continuous minimizer of a (strictly) convex-quadratic function  can be uniformly bounded, regardless of the linear terms in the function.
\end{enumerate}

We also provide sharper results for when the lattices are $\Z^{n}$. 
More importantly, we provide all our negative results and examples, whenever possible, with the option of choosing the matrices defining the quadratic function to be identity matrices ($\Qx = \Qy = I$) and the feasible set to be $\mathbb{Z}^n$.
This helps recognizing that the negative result can hold even in the most basic scenarios.

\subsection{Related Literature}
Our paper is related to two streams of literature: (i) Integer programming games and (ii) \BR algorithms. 
\paragraph{Integer Programming Games. }
A game, which is a strategic interaction between multiple players each with their own strategy sets and payoffs, is called an {\em integer programming game} if the feasible sets and objective functions of the players are provided in the form of an integer programming problem. 
This unique subdomain gained prominence due to the compact representation of games with discrete strategy sets. 
\citet{Carvalho2025FnT} provide a comprehensive survey on integer programming games. 

Deciding whether an integer (linear) programming game has a PNE (or an MNE) is known to be $\Sigma_2^p$-hard \citep{carvalho_2020_computing}. 
Despite the strong complexity bounds, algorithms that match the lower-bounds predicted by the complexity results have been of interest in the literature.
Typically, such algorithms assume either assume the best-response optimization problem to have linear objective functions or assume compact feasible sets or assume both. 
\citet{carvalho_2020_computing} propose the {\em SGM algorithm}, an abbreviation of {\em Sampled Generation Method} where they iteratively  generate more and more feasible points and compute an MNE for the finite subset, until such an MNE is also an MNE for the entire problem. 
\citet{cronert_equilibrium_2020} extend the algorithm into an {\em exhaustive-SGM} or eSGM algorithm, where exhaustively {\em all} MNEs of a game are enumerated, when the players' decisions are all discrete.     
These algorithms, while practical and fast, require that the feasible sets are compact. 
If not for compactness, the algorithm might never exhaust enumerating all possible strategies. 

On the other hand, algorithms that don't rely on compactness of the feasible sets, typically rely on the structure of the feasible set and use a convexification approach. 
\citet{NashStackelberg} identify an inner approximation-based algorithm to solve a class of problems they refer to as {\em NASP} (Nash game Among Stackelberg Players). 
As a counterpart to the inner-approximation algorithm, \citet{CutnPlay} proposes an outer-approximation algorithm for IPGs as well as a large class of separable games. 
Both these algorithms iteratively improve the approximation of the convex hull and their proofs of correctness fundamentally rely on the fact that the best-response optimization problem has linear objectives. %
In contrast, \citet{Harks2024}  provide a convexification-based algorithm for {\em generalized} Nash equilibrium problem, where each players' feasible set also depends on the decisions of the others. 

\paragraph{The \BR algorithm.} 
Another stream of literature that is relevant to this work is the one on \BR algorithms.
\BR algorithm have been of interest starting from the seminal paper by \citet{monderer1996potential}. 
The authors in this paper define a potential function, which when minimized over a compact set of feasible strategies, yields a PNE to the game. 
When players play the best response to the opponents in the game, they can be interpreted as descent steps in the potential function. 
\citet{Voorneveld2000} extends the concept of potential function games and define {\em best-response potential games} allowing infinite paths of improvement also. 

The \BR algorithm has been studied when we don't have a potential function too. 
\citet{Goemans2005} define a new concept of \emph{sinking equilibria} which is of particular interest when one works with the \BR algorithm. 
This corresponds to a subset of strategy profiles, which if entered, has zero probability of leaving and exploring other strategies.
To assess the possibility that the algorithm will generate iterates that are {\em trapped} among some strategies, as opposed to converging to a PNE, the algorithm's convergence property on randomly generated games have been analyzed.
 In particular, \citet{Amiet2021,Mimun2024} analyze when the \BR algorithm converge to PNEs and when they cycle among a \emph{trap} in the context random finite games. 
While the literature on finite games primarily tries to distinguish between when there is a PNE termination as opposed to a \emph{trap} termination, our focus lies on distinguishing between divergence of iterates and non-divergence of iterates (which includes both PNE and \emph{trap} terminations).

Within the scope of the problems addressed in this paper, the best-response optimization problem entails convex-quadratic minimization over a lattice, a well-explored problem called the shortest lattice vector problem under the $\ell_2$ norm.
Given that the best-response problem is $NP$-complete, we believe that \cref{eq:intQuad} do not allow for very fast algorithms in general.

\section{Definitions and Algorithm descriptions}\label{sec:defn}

\subsection{Lattices. }\label{sec:defnLattice}
In this subsection, we define lattices and provide some of the standard results that are needed for the exposition in this paper. 
We refer the reader to \citet[Chapter 7]{barvinok2002course} for a comprehensive treatment of lattices as well as formal proofs of the results on lattices. 

\begin{figure}[t]
	\centering
	\begin{subfigure}[b]{0.45\textwidth}
		\centering
		\includegraphics[width=\textwidth]{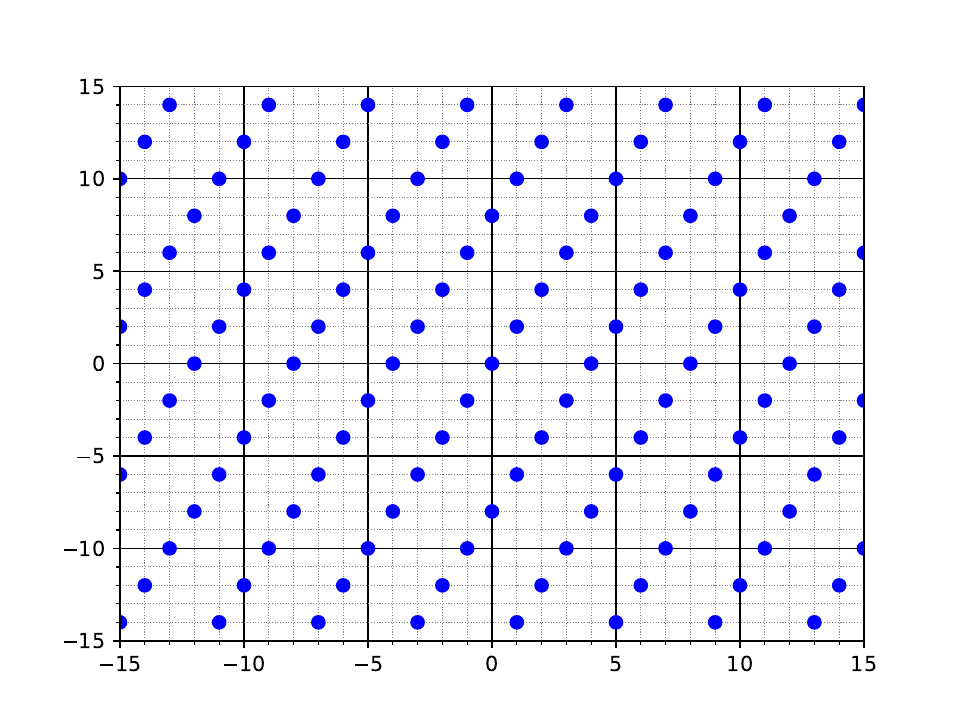}
		\caption{Plot of lattice whose generators are $\left\{ \vecN {4\\0}, \vecN{1\\2} \right\}$}\label{fig:latticePlot-1}
	\end{subfigure}
	\begin{subfigure}[b]{0.45\textwidth}
		\centering
		\includegraphics[width=\textwidth]{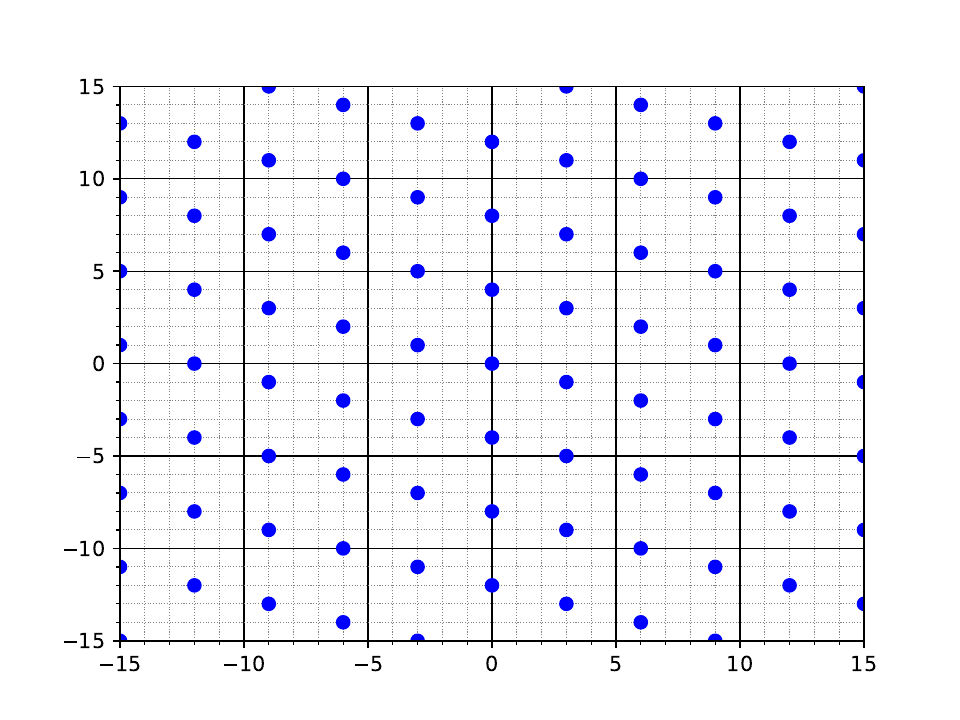}
		\caption{Plot of lattice whose generators are $\left\{ \vecN{0\\4}, \vecN{3\\3} \right\}$}\label{fig:latticePlot-2}
	\end{subfigure}
	\caption{Examples of lattices in $\R^2$}\label{fig:latticePlots}
\end{figure}

A  lattice $\Lattice$ is a set $\Lattice \subseteq \R^n$ such that (i) $0 \in \Lattice$, (ii) $\BFx, \BFy \in \Lattice $ implies $\BFx + \BFy \in \Lattice$ (iii) $\BFx \in \Lattice$ implies $- \BFx \in \Lattice$ (iv) there exists $\varepsilon > 0$ such that $\Lattice \cap B(0, \varepsilon) = \{0\}$ where $B(0, \varepsilon)$ is the ball of radius $\varepsilon$ centered at $0$.
We readily note that conditions (i)-(iii) implies that a lattice is an {\em additive subgroup} and (iv) implies that it is a discrete set. 
A standard result in the geometry of lattice is that if $\Lattice \subseteq \R^n$ is  lattice, then, there exist linearly independent vectors, $d^1, \dots, d^k \in \R^n$ such that $\Lattice = \{\sum_{i=1}^k \lambda_i d^i \mid \lambda_i \in \Z \text{ for }i=1,\dots,k\}$.
Conversely, if $d^1, \dots, d^k \in \R^n$ are linearly independent vectors, then the set $\{\sum_{i=1}^k \lambda_i d^i \mid \lambda_i \in \Z \text{ for }i=1,\dots,n\}$ is a lattice. 
An immediate implication is that \cref{eq:intQuad} allows the feasible set of, for example, the $\BFx$-player to be of the form $\{\BFx : A \BFu = \BFx ; \BFu \in \Z^{n}\}$.
The set of vectors $\{d^1, \dots, d^k\}$ is called a \emph{basis} for the lattice.
The number of linearly independent vectors $k$ is called the \emph{rank} of the lattice.
For the purposes of this paper, we consider only full-rank lattices, \emph{i.e., } lattices $\Lattice \subset \R^n$ which have a rank $n$. 
One can check that the set of all integer points, $\Z^n$ is a full-rank lattice in $\R^n$.
We show some examples of two-dimensional lattices in \cref{fig:latticePlots}.

A key concept that is used in this paper is {\em covering radius} of a lattice. Given a lattice $\Lattice \subseteq \R^{n}$, we define the {\em covering radius,} $\covRad[\Lattice] := \inf\{r \mid \cup_{\BFz\in \Lattice} B(\BFz, r) = \R^n\} $, where $B(\BFz, r)$ is a ball of radius $r$ centered at $\BFz$.
In other words, the covering radius is the smallest radius of a ball, which when centered at each point in the lattice covers the entire space.
Equivalently, it can be shown that $\covRad[\Lattice] = \sup \{\Vert \BFx - \BFz \Vert : \BFx \in \R^{n}; \BFz \in \Lattice\}$.
Next, given a lattice, $\Lattice\subseteq \R^{n}$ with full rank, the {\em dual lattice} $\Lattice^{*} := \{\BFx \in \R^{n} | \BFx^{\top} \BFz \in \Z \text{ for all }\BFz \in \Lattice \}$. 
The dual lattice is also lattice. 
One can check that the dual of the dual lattice is the original lattice. \emph{i.e., }$\Lattice^{**} = \Lattice$.
One can also confirm that $(\Z^n)^* = \Z^n$. \emph{i.e., } the integer lattice is self-dual.
With these definitions,  we present the {\em flatness theorem}. 
\begin{theorem}[Flatness Theorem]\label{thm:flat}
	Given a lattice $\Lattice \subset \R^n$ with full rank, and a convex set $C \subseteq \R^{n}$ such that $C \cap \Lattice = \emptyset$, there exists a direction $\BFz \in \Lattice^{*}$ such that 
	$
	\max \left\{ \BFz^\top \BFx \mid \BFx \in C \right\} - 
	\min \left\{ \BFz^\top \BFx \mid \BFx \in C \right\}   \le  \Flt{n}
	$, where $\Flt{n}$ is called the \emph{flatness constant}, and does not depend upon the lattice $\Lattice$ or the convex set $C$.
\end{theorem}
The theorem states that if a convex set $C \subseteq \R^n$ has no lattice points in its interior, then there exists a direction  $\BFz \in \Lattice^* \setminus \{0\}$ along which the convex set $C$ is \emph{flat}. 
We refer the readers to \citet[Pg. 317, Theorem 8.3]{barvinok2002course} for a proof of the flatness theorem where $\Flt{n} \le n^{5/2}$ is derived. 
More recently, \citet{reisRothvoss2024} show that $\Flt{n}\sim \mathcal{O}(n (\log 2n)^3)$. 
Nevertheless,  stronger ($\mathcal{O}(n)$)  bounds are conjectured \citep{celaya2022,rudelson2000}.
We state all our results in terms of $\Flt{n}$, so if stronger bounds are found for $\Flt{n}$, they are directly applicable here.

\subsection{The game and equilibria. }

For the purposes of simplicity, we assume that there are only two players as defined in \cref{eq:intQuad}. 
The results in this paper can be generalized to multiple (finitely many) players as discussed in \cref{sec:multiPlayer}\iftoggle{arxiv}{}{ of the electronic companion}.
Also, where applicable, we provide sharper results for the special case when $\Latticex = \Z^{n_x}$ and $\Latticey = \Z^{n_y}$. 
In this paper, we refer to each feasible point for a player as a \emph{strategy}, and a probability distribution over any finite subset of strategies as a \emph{mixed-strategy}. With that, we define $\Delta$-Mixed-strategy Nash Equilibrium (MNE). 
\begin{definition}[$(\Delta_x, \Delta_y)$-Nash equilibrium]
    Given \cref{eq:intQuad}, a mixed-strategy pair $\Mne = (\Mne^\BFx, \Mne^\BFy)$ is a $(\Delta_x, \Delta_y)$-mixed-strategy Nash equilibrium  ($(\Delta_x, \Delta_y)$-MNE) if \begin{subequations}
        \begin{align*}
            \mathbb{E}_{(\BFx,\BFy) \sim \Mne} \left[ \frac{1}{2} \BFx^\top \Qx \BFx + (\Cx \BFy + \dx)^\top \BFx \right]
            &\leq
            \mathbb{E}_{\BFy \sim \Mne^\BFy} \left[ \frac{1}{2} \BFx^\top \Qx \BFx + (\Cx \BFy + \dx)^\top \BFx \right] + \Delta_x
            & \forall \BFx\in \Latticex
            \\
            \mathbb{E}_{(\BFx,\BFy) \sim \Mne} \left[ \frac{1}{2} \BFy^\top \Qy \BFy + (\Cy \BFx + \dx)^\top \BFy \right]
            &\leq
            \mathbb{E}_{\BFx \sim \Mne^\BFx} \left[ \frac{1}{2} \BFy^\top \Qy \BFy + (\Cy \BFx + \dy)^\top \BFy \right] + \Delta_y
            & \forall \BFy\in \Latticey
        \end{align*}
    \end{subequations}
    Further, the finite subset of 
    $\Latticex$ and $\Latticey$
    to which $\Mne^\BFx$ and $\Mne^\BFy$ assign non-zero probability are called the $\BFx$-player's and $\BFy$-player's {\em supports} of the $(\Delta_x, \Delta_y)$-MNE respectively. 
    If $\Delta_x = \Delta_y = 0$ holds, then we call it just an MNE .
    \end{definition}
We note the subtle difference between $\varepsilon$-MNE that is popular in literature as opposed to the $\Delta$-MNE we define above. 
Typically, algorithms claim to find an $\varepsilon$-MNE, if such  solution is possible for \emph{any} given $\varepsilon > 0$, {\em i.e., } any allowable positive error. 
In contrast, our paper only guarantees solutions for a fixed value of the error term that we denote by $\Delta$. 
Besides this subtle difference, the definitions are interchangable.

\subsection{The \BR algorithm. } \label{sub:BRalgo}
    Given a game \cref{eq:intQuad}, we define the best response of the $\BFx$-player, given $\BFy$ as $\Brx \in \arg\min_x \{\frac{1}{2}\BFx^\top \Qx \BFx + \left( \Cx \BFy+ \dx \right)^\top \BFx \mid \BFx \in \Latticex  \} $ and the best response of the $\BFy$-player given $\BFx$ as $\Bry \in \arg\min_y \{\frac{1}{2}\BFy^\top \Qy \BFy + \left( \Cy \BFx+ \dy \right)^\top \BFy \mid \BFy \in \Latticey  \} $.
We note that $\Brx$ and $\Bry$ need not be singleton sets. If they are not singleton, for the purposes of the \BR algorithm, any arbitrary element from these sets can be chosen. 
We also note that the best-response optimization problems are $NP$-complete, as they are equivalent to the closest lattice-vector problem under the $\ell_{2}$ norm. 
The notion of best response immediately presents the idea of a \emph{\BR algorithm} which is formally presented in \cref{alg:BR}. 
\begin{algorithm}[h]
    \caption{The Best-Response Algorithm} \label{alg:BR}
    \begin{algorithmic}[1]
        \Require{\cref{eq:intQuad} instance $(\Qx, \Qy, \Cx, \Cy, \dx, \dy)$ and $(\hat \BFx^0, \hat \BFy^0) \in \Latticex \times \Latticey$}
        \Ensure{Two finite sets  $S_x$ and $S_y$ such that for all $\BFx\in S_{x}$, $\Bry \in S_{y}$; for all $\BFy \in S_{y}$, $\Brx \in S_{x}$}
        \State $i \gets 1$. 
        \Loop
        \State $\hat \BFx^i \gets \Brx[\hat \BFy^{i-1}] $,  $\hat \BFy^i \gets \Bry[\hat \BFx^{i-1}] $ 
        \label{alg:BR:BR}
        \If {$\hat \BFx^i = \hat \BFx^k$ and $\hat \BFy^i = \hat \BFy^k$ for some $k < i$} 
        \label{alg:BR:if}
        \State \Return 
        $S_x = \{\hat \BFx^k, \hat \BFx^{k+1}, \dots, \hat \BFx^i\}$, $S_y = \{\hat \BFy^k, \hat \BFy^{k+1}, \dots, \hat \BFy^i\}$
        \EndIf
        \State $i\gets i+1$
        \EndLoop
    \end{algorithmic}
\end{algorithm}
\cref{alg:BR} is possibly  one of the simplest  algorithms that could be considered for finding Nash equilibria for simultaneous games. 
We begin with input ``guesses'' for each player's strategy  -- $\hat \BFx^0$ and $\hat \BFy^0$.
In \cref{alg:BR:BR} we solve lattice convex-quadratic programs to identify the best response to the previous strategy of the other player. 
We repeat this until a previously-observed iterate is observed again.

Observe that it is not possible for the iterates to  converge to a point but not attain the limit, as the feasible set is discrete.
Thus, the only three outcomes are 
(i) a strategy profile is reached so that each player's strategy is the best response to the other player's strategy;
(ii) the algorithm cycling among finitely many strategies  -- {\em i.e.,} the algorithm is in a {\em trap} -- and hence terminating following \cref{alg:BR:if}; or 
(iii) diverging (non-terminating). 
Case 1 termination is the simplest, where the output can be immediately called a PNE. 
If the iterates diverge (case 3), then we have nothing to retrieve. 
In this paper, we show that, if we are in a trap (case 2), then we can retrieve a $\Delta$-MNE. 
Thus, we would want to distinguish the cases when the algorithm will terminate in Case 1 or 2, and cases when it will generate diverging iterates as in Case 3. 

\subsection{Adequate objectives. }
Before we state the main results, we define positively and negatively adequate objectives below. 
A game having (positively or negatively) adequate objectives is a property of the players' objective functions and are agnostic to the feasible sets. 
In this paper, given \cref{eq:intQuad}, we define two new matrices $\Rx := \Qx^{-1}\Cx$ and $\Ry := \Qy^{-1}\Cy$ and call them the {\em interaction matrices}. 
With that, we define positively and negatively adequate objectives. 

\begin{definition}
    \cref{eq:intQuad} is said to have \emph{positively adequate objectives} 
    if every singular value of the 
    interaction matrices  $\Rx$ and $\Ry$
    is strictly less than $1. $
\end{definition}

\begin{definition}
    \cref{eq:intQuad} is said to have \emph{negatively adequate objectives} 
    if every singular value of the 
    interaction matrices  $\Rx$ and $\Ry$
    is strictly greater than $1. $
\end{definition}
When we have positively adequate objectives, we assume that the largest singular value is $1 - \rho$ for some $\rho > 0$ and when we have negatively adequate objectives, we assume that the smallest singular value is $1 + \rho$ for some $\rho > 0$.

\paragraph{Implication of positively and negatively adequate objectives. }  
To understand the implication of adequate objectives, we note the following standard results from matrix analysis. 
First, if $\BFA$ and $\BFB$ are two matrices, then the largest singular value of $\BFA\BFB$ is at most the product of the largest singular value of $\BFA$ and the largest singular value of $\BFB$. 
Analogously, the smallest singular value of $\BFA\BFB$ is at least the product of the smallest singular value of $\BFA$ and the smallest singular value of $\BFB$.
Second, if $\BFA$ is an invertible matrix and $\sigma$ is a singular value of $\BFA$, then  $\frac{1}{\sigma}$ is a singular value of $\BFA^{-1}$.
Now, let us try to interpret what does it mean to say that the largest singular value of $\Rx$ must be less than $1$. 
This means, the largest singular value of $\Qx^{-1}\Cx$ must be less than $1$.
This condition is satisfied if the product of the largest singular value of $\Qx^{-1}$ and that of $\Cx$ is less than one. 
However, the largest singular value of $\Qx^{-1}$ is the reciprocal of the smallest singular value of $\Qx$.
Thus, essentially, if the smallest singular value of $\Qx$ is greater than the largest singular value of $\Cx$, then we have positively adequate objectives.
Now, for any matrix $\BFM$, and vector $\BFx$, 
$\norm{\BFM\BFx}_2 \le \sigma_1 \norm{\BFx}_2$ and $\norm{\BFM\BFx}_2 \ge \sigma_n \norm{\BFx}_2$ where $\sigma_1$ and $\sigma_n$ are the largest and smallest singular values of $\BFM$ respectively.
So, saying that the smallest singular value of $\Qx$ is greater than the largest singular value of $\Cx$ means, for any vector $\BFx$, $\norm{\Qx\BFx}_2 > \norm{\Cx^\top\BFx}_2$. In other words, the impact of the matrix $\Qx$ on any vector is greater than that of $\Cx$. 
But, if we observe the definition of the game in \cref{eq:intQuad}, this means to say that the matrix defining the quadratic term $\Qx$ has more impact on a vector than the matrix $\Cx$ defining the interaction. 
This is the physical interpretation of positively adequate objectives, where the terms determining the effects of one's own strategy are more impactful than the terms determining the effects of the other player's strategy.

When we consider negatively adequate objectives, the same arguments can be made in the opposite direction.
It is equivalent to saying that the smallest singular value of $\Cx$ is greater than the largest singular value of $\Qx$.
In other words, the terms determining the effects of the other player's strategy are more impactful than the terms determining the effects of one's own strategy.

We also note that in the above paragraphs, we have exclusively discussed in terms of the $\BFx$-player.
The same has to hold for the $\BFy$-player for the game to have positively or negatively adequate objectives.
However, \cref{thm:diagSing} allows us to make the above arguments individually for each player. 

\iftoggle{incCont}{\input{contQuad}}{}

\section{Main Results.}
We state the organization of results in this paper first. 
First we prove a technical result we call \emph{proximity theorem} (\cref{thm:proxProof}) that provides a bound for the maximum distance between a continuous minimizer and a lattice minimizer of a convex-quadratic function.
This bound that is insensitive to the linear terms of the convex-quadratic function, but depends on both the matrix defining the quadratic term as well as the lattice under question. 
Using this result and the assumption of positively adequate objectives, we show that \cref{alg:BR} terminates finitely (\cref{thm:BRintConds}). 
Then, we show a negative result, that if the game has negatively adequate objectives, there always exists infinitely many points from where \cref{alg:BR} generates divergent iterates (\cref{thm:BRintDiv}).
Together, \cref{thm:BRintDiv,thm:BRintConds} provide necessary and sufficient conditions for finite termination of \cref{alg:BR}. These results also indicate the existence of a sinking equilibrium (\cref{thm:sinkEq}) in the sense of \citet{Goemans2005}. 

\begin{figure}[t]
    \centering
    \includegraphics[width=0.95\textwidth]{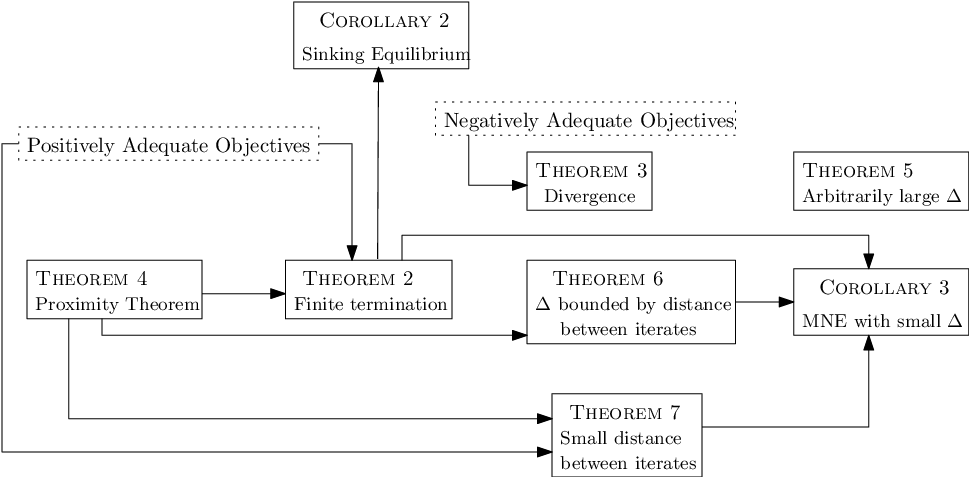}
    \caption{Implications among Theorems}\label{fig:thms}
\end{figure}

Following the results on finite termination, we consider the part where we solve a finite game restricted to the iterates about which \cref{alg:BR} cycled to obtain an MNE. 
First, we show that in general, such an MNE to the restricted finite game (over the strategies in a trap) could be arbitrarily bad for the original instance of \cref{eq:intQuad} (\cref{thm:intFinNeg}). 
    Then, we show that, the maximum profitable deviation any player in \cref{eq:intQuad} could obtain, given an MNE for the finitely restricted game, is bounded by the maximum distance between any pair of iterates generated (\cref{thm:intFinImpl}). 
    Next, we show that under our assumption of positively adequate objectives, the maximum distance between any two iterates about which cycling happens is indeed bounded (\cref{thm:closeIter}). 
    Tying these results together, we have \cref{cor:final}, which says that under the assumption of positively adequate objectives, \cref{alg:BR} can be used to obtain $\Delta$-MNE. 
    Finally, from the computational experience, we state a conjecture that in case of positively adequate objectives, the value $\Delta$ can be indeed chosen as zero. 
    For enhanced readability, we show the implications between the theorems in the manuscript in \cref{fig:thms}. 
\subsection{Necessary and sufficient conditions for finite termination. }

\begin{theorem} \label{thm:BRintConds}
    If the game \cref{eq:intQuad} has positively adequate objectives,  then \cref{alg:BR}  terminates finitely, irrespective of the initial points $\hat \BFx^0, \hat \BFy^0$. 
\end{theorem}

\cref{thm:BRintConds} is worded so that it is clear that the finite termination is irrespective of the initial points, if we have positively adequate objectives. \emph{i.e., } all the singular values of the interaction matrices are strictly less than 1.
In the below example, we establish the tightness of this result by showing that even if \emph{one} of the singular values of the interaction matrices is greater than $1$, the algorithm could diverge for some initial points.

\begin{example} \label{ex:unbndDivergePlus}
	Consider the following simultaneous game: 
        \begin{align*}
			\textbf{$\BFx$-player:}\min_{\BFx \in \Z^{n_{x}}} \sum_{i=1}^{{n_{x}}}\BFx_i^2  - 16\BFx_1\BFy_1 
\qquad\qquad\qquad
\textbf{$\BFy$-player:}\min_{\BFy \in \Z^{n_{y}} }  \sum_{i=1}^{{n_{y}}}\BFy_i^2  - \frac{1}{2}\BFx_1\BFy_1
        \end{align*}
		One can check that the singular values of $\Rx$ are $\{16,0, \dots 0\} $ and the singular values of $\Ry$ are $\{\frac{1}{2},0, \dots 0\}$, indicating that among the $n_x + n_y$ singular values, only one is greater than $1$.
		However, we can see that if we start from $(\BFx^0, \BFy^0) = \left( \vecN{4\\0\\ \vdots\\0}, \vecN{4\\0\\ \vdots\\0} \right)$, then the successive iterates are given as 
		{\small
		\begin{align*}
		\BFx^i = \vecN{
			32 \times 2^{\frac{i-1}{2}}\\0\\ \vdots \\ 0
		}
		\, 
		\BFy^i = \vecN{
			 2^{\frac{i-1}{2}}\\0\\ \vdots \\ 0
		} \, \text{if $i$ is odd; }
		\BFx^i = \vecN{
			4 \times 2^{\frac{i}{2}}\\0 \\\vdots \\ 0
		}
		\, 
		\BFy^i = \vecN{
			4 \times 2^{\frac{i}{2}}\\0\\ \vdots \\ 0
		} \, \text{if $i$ is even. }
		\end{align*}}
		Clearly, the above iterates diverge. One can easily check that if the first components of the initial iterate --- $\BFx_1^0$ and $\BFy_1^0$ ---  are both positive integer multiples of $4$, then the algorithm diverges.
		One can also observe that if the initial iterates have the first component $0$, then irrespective of what the other components are, the algorithm converges to the origin (which is a PNE). 
		\iftoggle{arxiv}{}{\hfill\Halmos}
\end{example}
In the above example, we have infinitely many initial points from which the algorithm diverges, and we have infinitely many initial points from which the algorithm converges to a PNE.
In general, the exact characteristics of convergence depend upon the precise value of the interaction matrices. 
However, a stronger negative result can be stated formally under the following stronger assumption. 
\begin{theorem} \label{thm:BRintDiv}
	If the game \cref{eq:intQuad} has negatively adequate objectives, then \cref{alg:BR} generates divergent iterates for all but finitely many feasible initial points $(\hat \BFx^0, \hat \BFy^0) \in \Latticex\times\Latticey$. 
\end{theorem}
In other words, if \emph{every} singular value of the interaction matrices is strictly greater than 1, then the algorithm diverges for all but finitely many initial points.

\begin{remark}
	The conditions of positively (and negatively) adequate objectives for finite termination (and divergence) in \cref{thm:BRintConds,thm:BRintDiv} are dependent \emph{ only} on the structure of the objective function and \emph{not} on the structure of the feasible set, \emph{i.e., } the lattices over which the players are optimizing.
	This means, the structure of the lattice does not affect the finite termination or divergence of the algorithm, but it is fundamentally driven by the interaction matrices, $\Rx$ and $\Ry$.
\end{remark}

Before proving \cref{thm:BRintDiv,thm:BRintConds}, we require an intermediate result on proximity of lattice minimizers of convex-quadratic functions to the corresponding continuous minimizers.
In that respect, we first define proximity, and then we prove that the bound is finite for any given positive definite matrix $\Qq$ and lattice $\Lattice$. 

\subsection{Proximity in integer convex-quadratic programs. }
First, we define proximity, which bounds the maximum distance between the continuous minimum and the integer minimum of a convex-quadratic function. 
The most famous of such results are the results in the context of linear programs provided in \citet{cook1986}, and recently improved by \citet{Paat2020,celaya2022}. 
Some proximity results for convex-quadratic programs \citep{granot1990}
and 
a subfamily of general convex programs \citep{moriguchi2011} are available in the literature extending the results for integer linear programs.
We provide a version of proximity result for convex-quadratic programs in this paper. 
While the results are not in full generality in line with the proximity literature in the context of integer linear programs, the version we provide here is sufficient to prove the fundamental results of the paper. 
\begin{definition}\label{def:prox}
    Let $\Qq$ be a given positive definite matrix. Let $\Lattice$ be a lattice with full rank. 
    The proximity bound of $\Qq$ on lattice $\Lattice$ with respect to the $\ell_p$ vector norm is denoted by $\prox[\Qq \mid \Lattice][p]$ and is the optimal objective value of the problem 
    \begin{subequations}
    \begin{align}
        \max _ {\dq \in \R^n  }\max_{ \substack{  \BFu \in \R^n \\ \BFv\in\Lattice }
        } &:  \norm{\BFu-\BFv}_p & \text{s.t.} \label{eq:prox:obj} \\
        \BFu &\in \arg \min \left \lbrace \frac{1}{2}\BFx^\top \Qq \BFx + \dq^\top \BFx  : \BFx \in \R^n \right \rbrace  \label{eq:prox:constrC} \\
        \BFv &\in \arg \min \left \lbrace \frac{1}{2}\BFx^\top \Qq \BFx + \dq^\top \BFx  : \BFx \in \Lattice \right \rbrace  \label{eq:prox:constr}
    \end{align}  \label{eq:prox}
    \end{subequations}
\end{definition}
\vspace{-5mm}
In this paper, we use $\ell_2$ norm predominantly, so when referring to $\prox[\Qq\mid \Lattice][2]$ we write $\prox[\Qq\mid \Lattice][ ]$. We also drop the mention of the lattice $ \Lattice$ and write only $\prox[\Qq]$ when there is no ambiguity about the lattice in question. 

In \cref{def:prox}, we consider a quadratic function whose quadratic terms (defined by $\Qq$) are fixed, but the linear terms (defined by $\dq$) are allowed to vary. 
We want to find the linear term $\dq$ such that  the distance between the continuous minimizer $\BFu$ and the lattice minimizer $\BFv$  is maximized. 
Moreover, the inner $\max$ ensures, should there be multiple integer minimizers, a minimizer which is farthest from the (unique) continuous minimizer could be chosen. 

Apriori, it is not clear if $\prox[\Qq\mid \Lattice]$ is finite for a given $\Qq$ and $\Lattice$. 
It is not clear if for a given $\Qq$ there is a sequence of $\dq^1, \dq^2,\dots$ and corresponding continuous minimizers $\BFu^1,\BFu^2,\dots$ and  lattice minimizers, $\BFv^1, \BFv^2, \dots$ of the quadratic such that $\norm{\BFu^i - \BFv^i} \to \infty$. 
However, we show that this is not the case in the following result. 
\begin{theorem}[Proximity Theorem] \label{thm:proxProof}
	Let $\Lattice$ be a lattice with full rank generated by the columns of matrix $\BFG$.  
	Moreover, let $L$ be such that $L\BFG$ has integer entries. 
    Given a positive definite matrix  $\Qq$ of dimension $n\times n$, 
    \begin{enumerate}[(i)]
		\item $
			\prox[\Qq \mid \Lattice] \le \min \left\{\frac{L \det (\BFG)\Flt{n}}{4}, \covRad\right\}\sqrt{\frac{\lambda_1}{\lambda_n}}  
			$; 
		\item The maximum difference between the optimal objective values when optimizing $\frac{1}{2} \BFx^\top \Qq \BFx + \dq^\top \BFx$  over $\R^n$ versus optimizing over $\Lattice$ is $
 \lambda_1 \min \left\{\frac{\covRad^2}{2}, \frac{L^2 \det(\BFG)^2\Flt{n}^2 }{32}\right\}
 $, 
    \end{enumerate} 
	where $\lambda_1$ and $\lambda_n$ are the largest and the smallest singular values of $\Qq$, $\covRad$ is the covering radius of the lattice $\Lattice$,  and $\Flt{n}$ is a constant dependent only on the dimension $n$. 
\end{theorem}
First we note that since $\Qq$ is a real symmetric matrix, the absolute value of its eigenvalues are its singular values. Moreover, since $\Qq$ is positive definite, all its eigenvalues are positive real numbers. So, its eigenvalues are its singular values. Thus $\lambda_1$ and $\lambda_n$ in \cref{thm:proxProof} can also be interpreted as the largest and the smallest eigenvalues of $\Qq$.  

To prove \cref{thm:proxProof}, we use the \emph{flatness theorem}. 
 
\iftoggle{arxiv}{
    \begin{proof}[Proof of \cref{thm:proxProof}.]}{
    \proof{Proof of \cref{thm:proxProof}.}}
Consider a strictly convex-quadratic function $\frac{1}{2} \BFx^\top Q \BFx + \dq^\top \BFx$. For the choice $\BFu = -\Qq^{-1}\dq$ (equivalently $\dq = - \Qq \BFu$), we can rewrite the function as $\frac{1}{2} \BFx^\top Q \BFx - \BFu^\top \Qq \BFx$. 
Neither the continuous minimizer nor the integer minimizer is sensitive to addition of  constants to the function. 
Thus the continuous and the integer minimizers of the above function are same as that of 
$\frac{1}{2} \BFx^\top Q \BFx - \BFu^\top \Qq \BFx + \frac{1}{2} \BFu^\top \BFu $. 
But this expression equals $\frac{1}{2}(\BFx-\BFu)^\top \Qq (\BFx-\BFu)$. 
Thus, it is sufficient to consider quadratic functions in this form. 
Define $f_u(\BFx):=\frac{1}{2} (\BFx-\BFu)^\top \Qq (\BFx-\BFu)$.
A useful property when considering this form is that $\BFu$ is the continuous minimizer here, and we don't need to explicitly write $\dq$ as a part of the function. 
The corresponding value of $\dq$ will be $- \Qq \BFu$. 
With this substitution, we can write $$\prox[\Qq\mid\Lattice] = \max _ {\BFu \in \R^n, \BFv \in \Lattice} \{ \norm{\BFu-\BFv} : \BFv \in \arg \min_{\BFx \in \Lattice} \{f_u(\BFx)\} \}.$$ 

Next, consider the family of ellipsoids parameterised by $\eta$ and $\BFu$, where $\ellip := \left\{ \BFx \in \R^n : f_u(\BFx) \le \eta \right\}$. 
\paragraph{Flat direction of the ellipsoid.	}
Now, let us identify the direction along which $\ellip$ is flattest, using our knowledge about the ellipsoid. 
Since $\Qq$ is symmetric, $\Qq$ has $n$ eigenvalues with corresponding eigenvectors that form an orthonormal basis of $\R^n$. We order the eigenvalues $\lambda_1 \ge \lambda_2 \ge \dots \ge \lambda_n$ and notate their corresponding orthonormal eigenvectors as $\BFw^1, \BFw^2, \dots, \BFw^n$. 
Now, consider the function $f_0(\BFx) = \frac{1}{2}\BFx^\top \Qq \BFx$. 
Using Rayleigh's theorem \citep[Pg. 234, Theorem 4.2.2, choose $S=\R^{n}$]{horn2012matrix}, one can show that $
\max \{f_0(\BFx): \norm{\BFx}_2 \le 1\} = \frac{1}{2} \lambda_1
$, and the maximizer is $\BFw^1$. 
Now, consider the ellipsoid $\ellip[\eta][0]$. Consider the point in this ellipsoid along the direction $\BFw^i$ and $-\BFw^i$ for some $i\in\{1,\dots,n\}$. 
We can observe that $\frac{1}{2}{\BFw^i}^\top \Qq \BFw^i = \frac{1}{2}{\BFw^i}^\top (\lambda_{i}{ \BFw^{i}}) = \frac{1}{2} \lambda_i $. 
Along the same line, 
$
\frac{1}{2}\sqrt{\frac{2\eta}{\lambda_i}}{\BFw^i}^\top \Qq\sqrt{\frac{2\eta}{\lambda_i}} \BFw^i  = 
\frac{1}{2}\times \left (\sqrt{\frac{2\eta}{\lambda_i}} \right) ^{2} \times \lambda_{i}
=\eta
$. 
Thus the scalings of $\BFw^i$, namely $\pm \tilde \BFw^i := \pm \sqrt{\frac{2\eta}{\lambda_i}}\BFw^i$ lie on the boundary of the ellipsoid.
So, the Euclidean distance between the two extreme points on the ellipsoid along direction $\BFw^i$ is 
$ 2 \sqrt{\frac{2\eta}{\lambda_i}}$. 
Clearly, the distance is the smallest when the denominator $\lambda_i$ is the largest, i.e., when $i=1$. 
In other words, the ellipsoid translated to the origin is flattest along the direction $\BFw^1$, and along that direction, the width is $ 2 \sqrt{\frac{2\eta}{\lambda_1}}$. 
Since the geometry of width is invariant to translations, these results apply to the original ellipsoid $\ellip$.
\paragraph{Covering radius-based argument. }
Consider a ball of radius $\sqrt{\frac{2\eta}{\lambda_i}}$ centred at $\BFu$, denoted by $B(\BFu, \sqrt{\frac{2\eta}{\lambda_i}})$. 
The ball is clearly contained in the ellipsoid $\ellip$.
Now, due to the containment $B(\BFu, \sqrt{\frac{2\eta}{\lambda_i}}) \subseteq \ellip$, %
if we choose $\eta$ so that $\ellip \cap \Lattice = \emptyset$, then the ball $B$ also satisfies $B \cap \Lattice = \emptyset$.
However, the covering radius $\covRad$ is the largest distance any point in $\R^n$ can be from the lattice $\Lattice$. 
So, if the ball $B$ has a radius larger than $\covRad$, then the ball will contain a point in $\Lattice$. 
So, the largest possible radius for the ball $B$ is $\covRad$. 
However, we have already established that the radius of the ball is $ \sqrt{\frac{2\eta}{\lambda_1}}$. 
This means, $\sqrt{\frac{2\eta}{\lambda_1}} \le \covRad$, or equivalently, $\eta \le \frac{\lambda_1\covRad^2}{2}$.
However, if $\eta$ is given, the farthest point on $\ellip[\eta][0]$ from the origin is in the direction of $\BFw^n$ with a distance  of $\sqrt{\frac{2\eta}{\lambda_n}}$. 
However, the bound on $\eta$ helps us bound the above by saying that the distance from the origin to the farthest point in the ellipsoid is at most 
$
\sqrt{\frac{2}{\lambda_n} \frac{\lambda_1\covRad^2}{2} }
= 
\covRad\sqrt{\frac{\lambda_1}{\lambda_n}} 
$
which is finite. 
While the above is a bound to the farthest point on the ellipsoid, the farthest integer point could only be possibly closer. 
So, we have $\prox[\Qq \mid \Lattice] \le \covRad\sqrt{\frac{\lambda_1}{\lambda_n}} 
$ proving (i).

\paragraph{Flatness theorem-based argument. }
We again start from the fact that the flattest direction of the ellipsoid is along $\BFw^1$, and along that direction, the width is $ 2 \sqrt{\frac{2\eta}{\lambda_1}}$. 
More interestingly, for any $\eta < f_u(\BFv)$ where $\BFv$ is the lattice minimizer of $f_u(\BFx)$, $\ellip \cap \Lattice = \emptyset$.  
Now, if $\ellip\cap\Lattice= \emptyset$, then $\ellip$ is a $\Lattice$-free convex set. 
Since $\ellip$ is a $\Lattice$-free convex set, the flatness theorem (\cref{thm:flat}) is applicable. 
In our context, this means, there exists a direction in the dual lattice, $\BFz \in \Lattice^{*} \setminus \{0\}$ along which $\ellip$ is flat, {\em i.e.,} 
$
	\max \left\{ \BFz^\top \BFx \mid \BFx \in \ellip \right\} - 
	\min \left\{ \BFz^\top \BFx \mid \BFx \in \ellip \right\}   \le  \Flt{n}
	$. 
Note that the generators of $\Lattice^{*}$ are the columns of $\BFG^{-\top}$. If $L$ is the LCM of the denominators of the entries in $\BFG$, then $L\BFG$ is an integer matrix. Moreover, this means $L\det(\BFG)\BFG^{-\top}$ is an integer matrix. If $L\det(\BFG)\BFG^{-\top}$, then $L\det(\BFG)\BFG^{-\top} \BFx \in \Z^{n}$ if $\BFx \in \Z^{n}$.
But $\BFz \in \Lattice^{*}$ means there exists an $\BFx \in \Z^{n}$ such that $\BFz = \BFG^{-\top}\BFx$. But from earlier arguments, 
we know that for any $\BFx \in \Z^{n}$, 
$L\det(\BFG)\BFG^{-\top} \BFx = L\det(\BFG) \BFz \in \Z^{n}$. 
Combining this with the result of the flatness theorem, we have 
$
	\max \left\{ L\det(\BFG)\BFz^\top \BFx \mid \BFx \in \ellip \right\} - 
	\min \left\{ L\det(\BFG)\BFz^\top \BFx \mid \BFx \in \ellip \right\}   \le L\det(\BFG) \Flt{n}
	$, where $L\det(\BFG) \BFz$ is an integer vector. 
Since there exists an (integer) direction along which the width is at most $ L\det(\BFG) \Flt{n}$, we can state that the width along the flattest direction, $\BFw^{1}$ is at most $ L\det(\BFG) \Flt{n} $. But we know that the width along this direction is $ 2 \sqrt{\frac{2\eta}{\lambda_1}}$. 
Thus, we have $ 2 \sqrt{\frac{2\eta}{\lambda_1}} \le L\det(\BFG) \Flt{n}$, implying that $\eta \le \frac{L^2 \det(\BFG)^2\lambda_1\Flt{n}^2 }{32}$.
However, like before,  if $\eta$ is given, the farthest point on $\ellip[\eta][0]$ from the origin is in the direction of $\BFw^n$ with a distance  of $\sqrt{\frac{2\eta}{\lambda_n}}$. 
So, 
$
\sqrt{\frac{2}{\lambda_n}\frac{L^2 \det(\BFG)^2\lambda_1\Flt{n}^2 }{32} }
=
\frac{L \det(\BFG) \Flt{n}}{4}\sqrt{\frac{\lambda_i}{\lambda_n}}.
$
So, we have $\prox[\Qq \mid \Lattice] \le\frac{L \det(\BFG) \Flt{n}}{4}\sqrt{\frac{\lambda_i}{\lambda_n}}.
$ 

Combining the two arguments, we have $
\prox[\Qq \mid \Lattice] \le \min \left\{\frac{L\det(\BFG)\Flt{n}}{4}, \covRad\right\}\sqrt{\frac{\lambda_1}{\lambda_n}}  
$, proving (i). 
Observing the value of the $\eta$ in each of the two arguments, we observe, 
$
\eta \le \min \left\{\frac{\lambda_1\covRad^2}{2}, \frac{L^2 \det(\BFG)^2\lambda_1\Flt{n}^2 }{32}\right\}
$ proving part (ii). 
\hfill
\Halmos
    \iftoggle{arxiv}{
        \end{proof}
    }{\endproof}

	\begin{corollary} \label{thm:proxProofZn}
    Given a positive definite matrix  $\Qq$ of dimension $n\times n$, 
    \begin{enumerate}[(i)]
		\item $
			\prox[\Qq \mid \Z^n] \le \min \left \{ \frac{1}{2},	\frac{\Flt{n}}{4}		 \right \} \sqrt{\frac{ n \lambda_1}{\lambda_n}}
			$; 
		\item The maximum difference between the optimal objective values when optimizing $\frac{1}{2} \BFx^\top \Qq \BFx + \dq^\top \BFx$  over $\R^n$ versus optimizing over $\Z^n$ is $
			\min \left \{ \frac{n}{8}, \frac{\Flt{n}^2}{32}			\right\}\lambda_1
 $, 
    \end{enumerate} 
	where $\lambda_1$ and $\lambda_n$ are the largest and the smallest singular values of $\Qq$. 
\end{corollary}
 
\iftoggle{arxiv}{
    \begin{proof}[Proof of \cref{thm:proxProofZn}.]}{
    \proof{Proof of \cref{thm:proxProofZn}.}}
	Follows by observing that the covering radius of the $\Z^n$ lattice is $\sqrt{n}/2$ and $L = \det(G) = 1$ for $\Z^{n}$.
\hfill
\Halmos
    \iftoggle{arxiv}{
        \end{proof}
    }{\endproof}

\begin{remark}
    We remark the contrast in the proximity we have in \cref{thm:proxProof} with the proximity results in \citet{granot1990}.
    \cref{thm:proxProof} provides  a bound that is insensitive to the linear terms in the quadratic function, while the bounds in \citet{granot1990} depends upon the linear terms in the objective function, but handles more general settings where there are linear constraints as well. 
    We also  note that the availability of analogous proximity results  for other families of  sets (beyond ellipsoids) will naturally extend the main results of this paper to analogous families of games.
\end{remark}

\subsection{Finite termination of \BR algorithm. }
Now we are in a position to prove \cref{thm:BRintConds}. 
\iftoggle{arxiv}{
    \begin{proof}[Proof of \cref{thm:BRintConds}.]}{
    \proof{Proof of \cref{thm:BRintConds}.}}
Given the iterates $\BFx^i$ and $\BFy^i$ in iteration $i$, the continuous minimizers of the best-response problems,  $\bar \BFx^{i+1}$ and $\bar \BFy^{i+1}$  are given by 
    \begin{align*}
        \bar \BFx^{i+1} &= -\Qx^{-1}  \left( \Cx \BFy^i + \dx  \right ) \\
        \bar \BFy^{i+1} &= -\Qy^{-1}  \left( \Cy \BFx^i + \dy  \right ) 
    \end{align*} 
The lattice optimum is at most a distance $\prox[\Qx\mid\Latticex]$ and $\prox[\Qy\mid\Latticey]$ distance away from $\bar \BFx^{i+1}$ and $\bar \BFy ^{i+1}$ for each of the players, respectively. 
If $\BFz_x(\BFx)$ and $\BFz_y(\BFy)$ denote the difference between the lattice and the continuous minimizers of each of the players' best-response problem, each iteration of \cref{alg:BR} can be modeled as application of the function $F$, where, 
    \begin{subequations}
        \begin{align}
            F \begin{pmatrix} \BFx \\ \BFy
            \end{pmatrix}
            &=  \left( 
            \begin{array}{c}
                -\Qx^{-1}  \left( \Cx \BFy + \dx  \right ) \\
                -\Qy^{-1}  \left( \Cy \BFx + \dy  \right ) 
            \end{array} \right)
            + \begin{pmatrix} \BFz_x(\BFx) \\ \BFz_y(\BFy) \end{pmatrix}
            \\
            &= 
            \begin{pmatrix}
             \BFzero & -\Rx \\
             -\Ry & \BFzero
            \end{pmatrix} 
            \xyVec
            - \begin{pmatrix} \Qx^{-1}\dx \\ \Qy^{-1}\dy \end{pmatrix}
            + \begin{pmatrix} \BFz_x(\BFx) \\ \BFz_y(\BFy) \end{pmatrix}. 
            \label{eq:t4}
        \end{align}
    Now, 
    \begin{align}
        \norm{F \xyVec}
        &= 
        \norm{
            \begin{pmatrix}
             \BFzero & -\Rx \\
             -\Ry & \BFzero
            \end{pmatrix} 
            \xyVec
            - \begin{pmatrix} \Qx^{-1}\dx \\ \Qy^{-1}\dy \end{pmatrix}
            + \begin{pmatrix} \BFz_x(\BFx) \\ \BFz_y(\BFy) \end{pmatrix}
        } \\
        &\leq
        \norm{
            \begin{pmatrix}
             \BFzero & -\Rx \\
             -\Ry & \BFzero
            \end{pmatrix} 
            \xyVec
        }
        + \norm{\begin{pmatrix} \Qx^{-1}\dx \\ \Qy^{-1}\dy \end{pmatrix}}
        + \norm{\begin{pmatrix} \BFz_x(\BFx) \\ \BFz_y(\BFy) \end{pmatrix} }
         \\
        &\leq
        \norm{
            \begin{pmatrix}
             \BFzero & -\Rx \\
             -\Ry & \BFzero
            \end{pmatrix} 
            \xyVec
        }
        + \norm{\begin{pmatrix} \Qx^{-1}\dx \\ \Qy^{-1}\dy \end{pmatrix}}
        + \norm{\begin{pmatrix} \prox[\Qx\mid\Latticex] \\ \prox[\Qy\mid\Latticey] \end{pmatrix} }
         \\
        &=
        \norm{
            - \begin{pmatrix} \Rx & \BFzero \\ \BFzero & \Ry \end{pmatrix}
	\begin{pmatrix} \BFy \\ \BFx \end{pmatrix}
        }
        + \norm{\begin{pmatrix} \Qx^{-1}\dx \\ \Qy^{-1}\dy \end{pmatrix}}
        + \norm{\begin{pmatrix} \prox[\Qx\mid\Latticex] \\ \prox[\Qy\mid\Latticey] \end{pmatrix} } \\
        &=
         \norm{
            \begin{pmatrix} \Rx & \BFzero \\ \BFzero & \Ry \end{pmatrix}
	\begin{pmatrix} \BFy \\ \BFx \end{pmatrix}
        }
        + \norm{\begin{pmatrix} \Qx^{-1}\dx \\ \Qy^{-1}\dy \end{pmatrix}}
        + \norm{\begin{pmatrix} \prox[\Qx\mid\Latticex] \\ \prox[\Qy\mid\Latticey] \end{pmatrix} } \\
        &\leq
        \left( 1-\rho \right)\norm{ \xyVec }
        + \norm{\begin{pmatrix} \Qx^{-1}\dx \\ \Qy^{-1}\dy \end{pmatrix}}
        + \norm{\begin{pmatrix} \prox[\Qx\mid\Latticex] \\ \prox[\Qy\mid\Latticey] \end{pmatrix} }
    \end{align} \label{eq:t1}

        Here, the first inequality follows from the triangle inequality of norms. 
    The second inequality follows from \cref{thm:proxProof} that 
    $\norm{\BFz_x}_2 \le \prox[\Qx\mid\Latticex]$ and $\norm{\BFz_y}_2 \le \prox[\Qy\mid\Latticey]$. 
    The equality in the next line is due to the fact that the expressions within $\norm{\cdot}$ in the first term are essentially the same and the next equality is due to positivity of norms. 
The inequality in the last line follows from the fact that (i) each singular value of $\begin{pmatrix} \Rx & \BFzero \\ \BFzero & \Ry \end{pmatrix}$ is at most the largest singular value of $\Rx$ and $\Ry$ (due to \cref{thm:diagSing} in the electronic companion), which is  $1-\rho$ for some $\rho > 0$ and (ii) \cref{thm:singularBnd} applied to $\BFM = \vecN{\Rx &\BFzero \\ \BFzero & \Ry}$. 
    \end{subequations}
    \begin{subequations}
    Now, suppose, 
    \begin{align}
    \norm{\xyVec} 
    &>
    \frac{
         \norm{\begin{pmatrix} \Qx^{-1}\dx \\ \Qy^{-1}\dy \end{pmatrix}} +
            \norm{\begin{pmatrix} \prox[\Qx\mid\Latticex] \\ \prox[\Qy\mid\Latticey] \end{pmatrix} }
    }
    {\rho} \\
    \implies \rho \norm{\xyVec} &> 
        \norm{\begin{pmatrix} \Qx^{-1}\dx \\ \Qy^{-1}\dy \end{pmatrix}} +
            \norm{\begin{pmatrix} \prox[\Qx\mid\Latticex] \\ \prox[\Qy\mid\Latticey] \end{pmatrix} } \\
    \implies \norm{\xyVec} &> 
            (1-\rho) \norm{\xyVec} +
        \norm{\begin{pmatrix} \Qx^{-1}\dx \\ \Qy^{-1}\dy \end{pmatrix}} +
            \norm{\begin{pmatrix} \prox[\Qx\mid\Latticex] \\ \prox[\Qy\mid\Latticey] \end{pmatrix} } \\
            &\geq
            \norm {F \xyVec} 
        \end{align} \label{eq:t2}
    \end{subequations}
    Here, we assume the first inequality. The second inequality is obtained by multiplying $\rho > 0$ on both sides. The third inequality is obtained by adding $(1-\rho) \xyVec$ on both sides, and the last inequality follows from \cref{eq:t1}. 
    In \cref{eq:t2},  we are saying that  $\norm {F\xyVec} < \norm{\xyVec}$ for all $\xyVec$ such that $\norm{\xyVec} > L= 
    \frac{
        \norm{\begin{pmatrix} \Qx^{-1}\dx \\ \Qy^{-1}\dy \end{pmatrix} } + 
            \norm{\begin{pmatrix} \prox[\Qx\mid\Latticex] \\ \prox[\Qy\mid\Latticey] \end{pmatrix} }
    }
    {\rho } 
$. 
    Now, we make a claim that the above statement implies that  \cref{alg:BR} has finite termination. 
    This is because, in any bounded region, in particular, in the region defined by $B = \left\{ \xyVec: \norm{\xyVec} \le L \right\}$, there are finitely many feasible $\xyVec$. 
    Now, whenever an iterate has a norm exceeding $L$, it decreases monotonically over subsequent iterations, till the norm is less than $L$ at least once, thus visiting a vector in $B$. 
    After that, it could possibly increase again. However, given that these monotonic decreases always end in some $\xyVec\in B$, after sufficiently many times, either all vectors in $B$ will be visited eventually and then returning back to $B$ will cause a second time visit of a vector,  or even before all vectors are visited,  some vector will be visited for the second time. 
    In either case, this will trigger the termination condition in \cref{alg:BR:if} of \cref{alg:BR}, leading to finite termination. 
    \hfill \Halmos
    \iftoggle{arxiv}{
        \end{proof}
    }{\endproof}
    
Given the proof, we observe that the iterates, once they get into the bounded set defined by $ B' := \left\{ \xyVec: \norm{\xyVec} \le \left ( \norm{\vecN{\Qx^{-1}\dx \\ \Qy^{-1}\dy }} + \norm{\vecN{\prox[\Qx\mid\Latticex] \\ \prox[\Qy\mid\Latticey] } } \right )  \left(1 + \frac{1}{\rho} \right) \right\} $, then the iterates will never escape this set. 
However, there are finitely many lattice points in any bounded set. 
Thus, the following corollary appears naturally. 
\begin{corollary} \label{thm:sinkEq}
	Suppose \cref{eq:intQuad} has positively adequate objectives. 
	Then, the game has a sinking equilibrium with respect to the \BR algorithm (in terms of  \citet{Goemans2005}), contained in the set $ B' := \left\{ \xyVec: \norm{\xyVec} \le \left ( \norm{\vecN{\Qx^{-1}\dx \\ \Qy^{-1}\dy }} + \norm{\vecN{\prox[\Qx\mid\Latticex] \\ \prox[\Qy\mid\Latticey] } } \right )  \left(1 + \frac{1}{\rho} \right) \right\} $. 
\end{corollary}

\vspace{2mm}
Now, we prove \cref{thm:BRintDiv} that if we have negatively adequate objectives, then there exist initial points from where the iterates from \cref{alg:BR} diverge. 
\iftoggle{arxiv}{
    \begin{proof}[Proof of \cref{thm:BRintDiv}.]}{
    \proof{Proof of \cref{thm:BRintDiv}.}}
            Let $\proxn_{\norm{\cdot}} := {\norm{ \begin{pmatrix}\prox[\Qx\mid\Latticex]\\ \prox[\Qy\mid\Latticey]\end{pmatrix} }}$. 
Let each singular value of $\Rx$ and $\Ry$ be greater than $(1+\rho)$ for some $\rho > 0$. 
Suppose the initial iterate is a vector $\xyVec$ such that $\norm{\begin{pmatrix} \hat \BFx^0 \\ \hat \BFy^0 \end{pmatrix}} > \frac{\proxn_{\norm{\cdot}}}{\rho} $. 
We now show that the subsequent iterates have monotonically increasing norms, which indicates diverging and is sufficient to prove the theorem. 

Like in the proof of \cref{thm:BRintConds}, the iterates generated by \cref{alg:BR} equals recursive application of the function $F$ given by 
        \begin{align*}
            F \begin{pmatrix} \BFx \\ \BFy
            \end{pmatrix}
            &= 
            \begin{pmatrix}
             \BFzero & -\Rx \\
             -\Ry & \BFzero
            \end{pmatrix} 
            \xyVec
         - {\begin{pmatrix} \Qx^{-1}\dx \\ \Qy^{-1}\dy \end{pmatrix} } 
            + \begin{pmatrix} \BFz_x(\BFx) \\ \BFz_y(\BFy) \end{pmatrix} \\
            F \begin{pmatrix} \BFx \\ \BFy
            \end{pmatrix}
            - \begin{pmatrix} \BFz_x(\BFx) \\ \BFz_y(\BFy) \end{pmatrix}
         + {\begin{pmatrix} \Qx^{-1}\dx \\ \Qy^{-1}\dy \end{pmatrix} } 
            &= 
            \begin{pmatrix}
             \BFzero & -\Rx \\
             -\Ry & \BFzero
            \end{pmatrix} 
            \xyVec
        \end{align*}
        where $\norm{\BFz_x}_2 \le \prox[\Qx\mid\Latticex]$ and $\norm{\BFz_y}_2 \le \prox[\Qy\mid\Latticey]$. Now, like before, 
    \begin{align*}
        \norm{F \xyVec
            - \begin{pmatrix} \BFz_x(\BFx) \\ \BFz_y(\BFy) \end{pmatrix}
         +{\begin{pmatrix} \Qx^{-1}\dx \\ \Qy^{-1}\dy \end{pmatrix} } 
        }
        &= 
        \norm{
            \begin{pmatrix}
             \BFzero & -\Rx \\
             -\Ry & \BFzero
            \end{pmatrix} 
            \xyVec
        } 
    \end{align*} 
Given that the smallest singular value of the matrix in RHS is at least $1+\rho$, we can say using \cref{thm:singularBnd},
    \begin{align*}
        \norm{F \xyVec
            - \begin{pmatrix} \BFz_x(\BFx) \\ \BFz_y(\BFy) \end{pmatrix}
         + {\begin{pmatrix} \Qx^{-1}\dx \\ \Qy^{-1}\dy \end{pmatrix} } 
        }
        &\geq (1+\rho) \norm{\xyVec} \\
        \implies
        \norm{F \xyVec}
         +\norm{ \begin{pmatrix} \BFz_x(\BFx) \\ \BFz_y(\BFy) \end{pmatrix}}
         + \norm{\begin{pmatrix} \Qx^{-1}\dx \\ \Qy^{-1}\dy \end{pmatrix} } 
        &\geq (1+\rho) \norm{\xyVec} \\
        \implies
        \norm{F \xyVec}
        + \norm{\begin{pmatrix} \prox[\Qx\mid\Latticex] \\ \prox[\Qy\mid\Latticey] \end{pmatrix} }
         + \norm{\begin{pmatrix} \Qx^{-1}\dx \\ \Qy^{-1}\dy \end{pmatrix} } 
        &\geq (1+\rho) \norm{\xyVec} %
		\end{align*}
		But this means, we have, 
		\begin{align*}
        \norm{F \xyVec} - \norm{\xyVec}
        &\geq  \rho \norm{\xyVec} 
        -\norm{\begin{pmatrix} \prox[\Qx\mid\Latticex] \\ \prox[\Qy\mid\Latticey] \end{pmatrix} }
         - \norm{\begin{pmatrix} \Qx^{-1}\dx \\ \Qy^{-1}\dy \end{pmatrix} }.
    \end{align*}
    Now, if we have $\norm{\xyVec} > 
    \frac{
        \norm{\begin{pmatrix} \prox[\Qx\mid\Latticex] \\ \prox[\Qy\mid\Latticey] \end{pmatrix} } 
         + \norm{\begin{pmatrix} \Qx^{-1}\dx \\ \Qy^{-1}\dy \end{pmatrix} } 
    }{\rho} $, the RHS in the above expression is positive. This means, in the following iteration, the norm of the iterates increase. It increases and increases without a bound indicating that the algorithm diverges. 
	Moreover, there are only finitely many points in $\Latticex\times\Latticey$, with norm at most
	$\frac{\proxn_{\norm{\cdot}}}{\rho} $, implying that the algorithm will diverge for all but finitely many feasible initial points. 
\hfill
\Halmos
    \iftoggle{arxiv}{
        \end{proof}
    }{\endproof}

	\subsection{Retrieval of an approximate equilibrium. }
It is clear that if \cref{alg:BR} terminates outputting sets such that $S_{x} = S_{y} = 1$, then the pair of strategies in the output constitute a PNE for the instance of \cref{eq:intQuad}.
However, if we have an output with a non-singleton set, it is not clear, what guarantees we could have. 
In some cases, as shown below, it could happen that an MNE could be easily found, by solving the finite game with strategies of each player restricted to $S_{x}$ and $S_{y}$. 
\begin{example}[Cycling] \label{ex:BRcycle}
    Consider the problem given as follows. 
        \begin{align}
\textbf{$\BFx$-player:}:\min_{\BFx \in \Z} \BFx^2  - 0.2 \BFy\BFx - 0.9\BFx\qquad \qquad 
\textbf{$\BFy$-player:}:\min_{\BFy \in \Z} \BFy^2  + 0.2 \BFx\BFy - 1.1\BFy \label{eq:BRcycle}
        \end{align} 
    Let us start the \BR algorithm with initial iterates $(0, 0)$. The best response for $\BFx$ is  $0$ and for $\BFy$ is $1$. 
    Now, we start the next iteration from $(0, 1)$. Now the best response for $\BFx$ is $1$. There is no change in $\BFy$'s strategy. 
    Now, we start the next iteration from $(1, 1)$. There is no change in $\BFx$'s strategy. Now the best response for $\BFy$ is  $0$. 
    Now, we start the next iteration from $(1, 0)$. Now the best response for $\BFx$ is $0$. There is no change in $\BFy$'s strategy. 
    We are now back to the strategy $(0,0)$ and the same cycle of period 4 will keep repeating.
For the problem in that example, cycling occured with  $S_x = S_y = \{0, 1\}$. 
We can find an MNE for the bimatrix game where each player's strategy is restricted to $S_x$ and $S_y$ respectively. 
    The cost matrices (payoff matrices are negative of these matrices) for the $\BFx$-player (row player) and $\BFy$-player (column player) are
    \medskip
        \begin{center}
        \begin{tabular}{|c|c|c|}
            \hline
            $\BFx\backslash \BFy$& 0 & 1\\ 
            \hline
            0 & (0, 0) & (0, -0.1) \\
            1 & (0.1, 0) & (-0.1, 0.1) \\
            \hline
        \end{tabular}
        \end{center}
    \medskip
    Here, if both players mix both their strategies with probability $0.5$, then it is an MNE for the bimatrix game. 
    This also turns out to be an MNE for the original game in \cref{eq:BRcycle}.
\end{example}

The above observation raises the following question. 
If $S_x$ and $S_y$ are the iterates returned by \cref{alg:BR} for \cref{eq:intQuad}, will there be an MNE for the \cref{eq:intQuad} whose supports are subsets of $S_x$ and $S_y$ respectively?
In other words, suppose one is in a {\em sinking equilibrium}, is the MNE of the finite game restricted to the strategies in a sinking equilibrium, in general, an approximate Nash equilibrium for the original game?
We state that this is not the case in general below. 
\begin{theorem}\label{thm:intFinNeg}
    Suppose \cref{alg:BR} terminates finitely and returns iterates $S_x$ and $S_y$ for \cref{eq:intQuad} with $|S_x| > 1$ and $|S_y|>1$. Let $\Qx = \Qy =  \BFI$,  and $\Latticex = \Latticey = \Z^{2}$.
    Given any $\Delta > 0$, there exists $\Cx$ such that an MNE of the finite game restricted to $\conv(S_x)\cap \Z^{2}$ and $\conv(S_y)\cap \Z^{2}$ is not a $\Delta$-MNE for \cref{eq:intQuad}. %
\end{theorem}
In other words, even if (i) $\Qx, \Qy$ are identity matrices, (ii) the lattices for both players are $\Z^{2}$, and (iii) we allow feasible strategies in the convex hull of $S_x$ and $S_y$, the MNE to the restricted finite game could be arbitrarily bad for \cref{eq:intQuad}.  

\iftoggle{arxiv}{
    \begin{proof}[Proof of \cref{thm:intFinNeg}.]}{
    \proof{Proof of \cref{thm:intFinNeg}.}}
Consider this game, where $M$ is a large positive even integer. 
\begin{subequations}
    \begin{align}
        \textbf{$\BFx$-player }  &: \min _ {\BFx\in \Z^2}\frac{1}{2}x_1^2+ \frac{1}{2}x_2^2 - y_1x_1 - y_2x_2 \\
        \textbf{$\BFy$-player }  &: \min _ {\BFy\in \Z^2}\frac{1}{2}y_1^2+ \frac{1}{2}y_2^2 + \frac{1}{M}x_1y_1 - (M-1)x_2y_1 - \frac{1}{M}x_1y_2- y_1
    \end{align} \label{eq:counterGame}
\end{subequations}
This completes the description of the game. 

If we start \cref{alg:BR} from $\hat \BFx^0 = (0, 1)^\top$ and $\hat \BFy^0 = (0, 1)^\top$,  the best response of $\BFx$ is $(0, 1)^\top$ while the best response for $\BFy$ is $(M, 0)^\top$. 
Given these points, the best response for $\BFx$ is $(M, 0)^\top$ and that of $\BFy$ is $(M, 0)^\top$. 
Given these points, the best response for $\BFx$ is $(M, 0)^\top$ and that of $\BFy$ is $(0, 1)^\top$.
Given these points, the best response for $\BFx$ is $(0, 0)^\top$ and that of $\BFy$ is $(0, 1)^\top$.
Thus, the following iterates will cycle between $(0, 1)^\top $ and $(M, 0)^\top$. 
    Thus,  $S_x = S_y  = \left\{ (0,1)^\top, (M, 0)^\top \right\} = \conv(S_x)\cap \Z^{2} = \conv(S_y)\cap \Z^{2}$. 

    The cost matrices (negative of payoff matrices) for both the players,  given the strategies in $S_x$ and $S_y$ are given below. 
    \medskip 
        \begin{center}
        \begin{tabular}{|c|c|c|}
            \hline
            $\BFx\backslash \BFy$& $(0,1)$& $(M,0)$\\ 
            \hline
            $\left (0, 1\right )$& $\left (-0.5, 0.5\right )$ &$ \left (0.5, -\frac{M^2}{2}\right ) $\\
            $\left (M, 0\right )$& $\left (\frac{M^2}{2}, -0.5\right )$ &$ \left (-\frac{M^2}{2}, \frac{M^2}{2}\right ) $\\
            \hline
        \end{tabular}
        \end{center}
    \medskip
    One can confirm that the above bimatrix game has no PNE, but has an MNE where the strategies are both given a probability of $0.5$ by both players. 
    The cost of both the players at this MNE is $0$. 
    However, for the game in \cref{eq:counterGame},  $\left( \frac{M}{2}, 0 \right)$ is a feasible profitable deviation for the $\BFx$-player.
    Feasibility follows from the fact that $M$ was chosen as an \emph{even} positive integer. 
    The cost $\BFx$-player incurs by playing this strategy is $\frac{M^2}{8} + 0 - \frac{M}{2}\frac{M}{2} = - \frac{M^2}{4} \ll 0$. By choosing $M$ to be arbitrarily large, we can obtain arbitrarily large profitable deviations from the MNE of the restricted game. 
\hfill
\Halmos
    \iftoggle{arxiv}{
        \end{proof}
    }{\endproof}

    We note that the family of examples described in the proof of \cref{thm:intFinNeg} are not problems that have positively adequate objectives. 
    But we also observe that the output set  generated by the game in \cref{eq:counterGame}, $S_x$ and $S_y$ have points far away from each other. 
    We show that this large distance between points within $S_x$ and $S_y$ lead to arbitrarily large values of $\Delta$. 
    The below result shows that $\Delta_x$ and $\Delta_y$ are bounded by $L_x$ and $L_y$, the maximum distance between any two iterates in $S_x$ and $S_y$ respectively. 

\begin{theorem} \label{thm:intFinImpl}
    Suppose \cref{alg:BR} terminates finitely and returns iterates $S_x$ and $S_y$ for \cref{eq:intQuad}. 
    Let $L_x= \max \left\{ \norm{\BFx^i-\BFx^j} \mid \BFx^i, \BFx^j \in S_x \right\}$ be the maximum of the norm between any two points in $S_x$.
Analgously, let $L_y$ be the maximum of the norm between any two points in $S_y$.
Then, any MNE of the finite game restricted to the strategies $S_x$ and $S_y$ is an $\left( \Delta_x, \Delta_y \right)$-MNE to \cref{eq:intQuad}, where
$\Delta_x =\lambda_1^\BFx(\prox[\Qx\mid \Latticex] + L_x)^2  $,
$\Delta_y = \lambda_1^\BFy(\prox[\Qy\mid \Latticey] + L_y)^2 $, 
$\lambda_1^\BFx$ is the largest eigenvalue of $\Qx$, and  $\lambda^y_1$ is the largest  eigenvalue of $\Qy$.
\end{theorem}
\iftoggle{arxiv}{
    \begin{proof}[Proof of \cref{thm:intFinImpl}.]}{
    \proof{Proof.}}
    Since $S_x$ and $S_y$ are sets of iterates over which \cref{alg:BR} cycles, we know that for each $\bar \BFy\in S_y$, there exists $\BFx \in S_x$ such that $\BFx\in \Brx[\bar \BFy]$.
We notate such a best response  as $\BFx^*(\bar \BFy)$ for simplicity. 
Given some $\BFy\in \R^{{n_{y}}}$, we notate the continuous minimizer of the $\BFx$-player's objective (which is $-\Qx^{-1} \left( \Cx \BFy + \dx \right)$) as $\tilde \BFx(\BFy)$.  

From \cref{thm:proxProof}, we know that for any $\BFy$,  $\norm{\BFx^*(\BFy) - \tilde{\BFx} (\BFy)} \le \prox[\Qx\mid\Latticex]$. 
Choose $\BFy' = \sum_i p_i^\BFy \bar \BFy^i$ where $\bar \BFy^i \in S_y$ 
and $p_i^\BFy$ denote the probabilities with which $\bar \BFy^i$ is played in the MNE of the restricted finite game. 
Being probabilities, $p_i^\BFy \ge 0$ with $\sum _i p_i^\BFy = 1$. 
\emph{i.e., } $\BFy'$ is convex combination of points in $S_y$. 
Now,  
$
        \tilde{\BFx}(\BFy') 
        =  \tilde \BFx\left( \sum_i p_i^\BFy \bar \BFy^i \right)  
        = \sum_i p_i^\BFy \tilde{\BFx}(\bar \BFy^i) 
        = \sum_i p_i^\BFy \left( \BFx^*(\bar \BFy^i) + \left( \tilde{\BFx}(\bar \BFy^i) - \BFx^*(\bar \BFy^i) \right) \right) 
        = \sum_i p_i^\BFy  \BFx^*(\bar \BFy^i) + \sum_i p_i^\BFy\left( \tilde{\BFx}(\bar \BFy^i) - \BFx^*(\bar \BFy^i) \right)  
        = \sum_i p_i^\BFy  \BFx^*(\bar \BFy^i) + \sum_i p_i^\BFy z_i   
    $
{where $\norm{z_i} \le \prox[\Qx\mid\Latticex]$} due to \cref{thm:proxProof}.
But this implies that $\norm{ \tilde{\BFx}(\BFy') -\sum_i p_i^\BFy  \BFx^*(\bar \BFy^i) } \leq \prox[\Qx\mid\Latticex] $. 

We are given that the distance between any two points in $S_x$ is at most $L_x$. 
The point $\sum_i p_i^\BFy \BFx^*(\BFy^i)$ (where $\BFy^i \in S_y$ for all $i$) is a convex combination of points in $S_x$. 
This point is also, hence, at most a distance $L_x$ from any other point in $S_x$, due to the convexity of norms.
Formally,  $\norm{\sum_i p_i^\BFy \BFx^*(\BFy^i) - \BFx^i } \le L_x$ for any $\BFx^i \in S_x$.

Combining $\norm{ \sum_i p_i^\BFy \BFx^*(\BFy^i) - \BFx^i } \le L_x$ and $\norm{ \tilde{\BFx}(\BFy') -\sum_i p_i^\BFy  \BFx^*(\bar \BFy^i) } \leq \prox[\Qx\mid\Latticex] $, we get
$ \norm{ \BFx^i - \tilde \BFx(\BFy')} \le \prox[\Qx\mid\Latticex] + L_x $ for any $\BFx^i \in S_x$ through the triangle inequality for norms. 

Now, define $f_x(\BFx) := \frac{1}{2}\BFx^\top \Qx \BFx + (\Cx \BFy' + \dx)^\top \BFx$. 
By definition, $\tilde \BFx(\BFy')$ is the continuous minimizer of $f_x$.
Since $ \norm{ \BFx^i - \tilde \BFx(\BFy')} \le \prox[\Qx\mid\Latticex] + L_x $ for any $\BFx^i \in S_x$, we have that 
$
f_x(\BFx^i) \le 
f_x(\tilde \BFx(\BFy')) + \lambda_1^\BFx(\prox[\Qx\mid\Latticex] + L_x)^2 
            \le 
			f_x(\BFx^*(\BFy')) + \lambda_1^\BFx(\prox[\Qx\mid\Latticex] + L_x)^2 
    $ for any $\BFx^i \in S_x$.
    Here, the first inequality follows from the fact that a point that is at most a distance $g$ away from the minimum of a convex-quadratic, has a function value of at most $ \lambda_1 g^2$ over and above the minimum value of the quadratic, where $\lambda_1$ is the largest eigenvalue of the matrix defining the quadratic.
    The second inequality follows from the fact that the continuous minimizer has an objective value that is not larger than an integer minimizer. 

    Since $f_x(\BFx^i) $ for each $x_i \in S$ is at most suboptimal by $\lambda_1^\BFx(\prox[\Qx\mid\Latticex] + L_x)^2 $, the maximum improvement possible from the mixed strategies which only plays a subset of $S_x$ can have an improvement not more than $\lambda_1^\BFx(\prox[\Qx\mid\Latticex] + L_x)^2 $, giving the value of $\Delta_x$ as needed. 

    Analogous arguments for the $\BFy$-player proves the analogous result for $\Delta_y$.
\hfill
\Halmos
\iftoggle{arxiv}{
    \end{proof}
}{\endproof}
\begin{remark}
    Observe that we cannot directly use \cref{thm:proxProof} (ii) to directly bound $\Delta$ in \cref{thm:intFinImpl}. This is because the best response to the strategy referred to as $\BFy'$ may or may not be a part of the support of MNE.
\end{remark}
While the previous result holds for \emph{any} instance of \cref{eq:intQuad}, we now show that if we have positively adequate objectives, then $L_x$ and $L_y$ themselves can be bounded, providing an {\em ex-ante} guarantee on the error in equilibria. 
\begin{theorem} \label{thm:closeIter} 
    Suppose \cref{alg:BR} terminates finitely and returns iterates $S_x$ and $S_y$ for an instance of  \cref{eq:intQuad} with positively adequate objectives. 
    Let $L_x= \max \left\{ \norm{\BFx^i-\BFx^j} \mid \BFx^i, \BFx^j \in S_x \right\}$ be the maximum of the norm between any two points in $S_x$.
Analgously, let $L_y$ be the maximum of the norm between any two points in $S_y$. Then, 
$L_x
\le 
\frac{ 2\prox[\Qy\mid\Latticey] \sigma_1^\BFx + 2\prox[\Qx\mid\Latticex]}{ 1 - \sigma_1^\BFx  \sigma_1^\BFy }
$ and 
$L_y \le 
\frac{ 2\prox[\Qx\mid\Latticex] \sigma_1^\BFy + 2\prox[\Qy\mid\Latticey]}{ 1 - \sigma_1^\BFx  \sigma_1^\BFy }
$,
where $\sigma_1^\BFx$ and $\sigma_1^\BFy$  are the largest singular values of $\Rx$ and $\Ry$ respectively.       
\end{theorem}
\iftoggle{arxiv}{
    \begin{proof}[Proof of \cref{thm:closeIter}.]}{
    \proof{Proof of \cref{thm:closeIter}.}}
{
Let $\BFx^i, \BFx^j \in S_x$. %
Since $\BFx^i, \BFx^j \in S_x$ there exist $\BFy^i, \BFy^j \in S_y$ such that $\BFx^i \in \Brx[\BFy^i]$ and $\BFx^j \in \Brx[\BFy^j]$.
But observe that for any $\BFy$, $\Brx[\BFy] = -\Qx^{-1} \left( \Cx \BFy + \dx \right) + \tilde \BFz(\BFy)$, where the first term is the continuous minimizer and the second term is the error induced due to minimizing over integers, and $\norm {\tilde{\BFz}(\BFy)} \le \prox[\Qx\mid\Latticex]$.
Thus, for  $\BFx^i, \BFx^j \in S_x$, 
\begin{subequations}
    \begin{align}
        \BFx^i - \BFx^j 
        &=  \left( -\Qx^{-1} \left( \Cx \BFy^i + \dx \right) + \tilde \BFz(\BFy^i) \right) - \left( -\Qx^{-1} \left( \Cx \BFy^j + \dx \right) + \tilde \BFz(\BFy^j) \right) \\
        &=  -\Qx^{-1} \Cx \left( \BFy^i - \BFy^j \right) + \tilde \BFz(\BFy^i) -  \tilde \BFz(\BFy^j) \\
        &=  -\Rx \left( \BFy^i - \BFy^j \right) + \tilde \BFz(\BFy^i) -  \tilde \BFz(\BFy^j) \\
        \implies  \norm{ \BFx^i - \BFx^j }
        &=  \norm{-\Rx \left( \BFy^i - \BFy^j \right) + \tilde \BFz(\BFy^i) -  \tilde \BFz(\BFy^j) }\\
    &\leq \norm{-\Rx \left( \BFy^i - \BFy^j \right)} + \norm {\tilde \BFz(\BFy^i)} + \norm{ \tilde \BFz(\BFy^j) } \\
        &\leq \sigma_1^\BFx \norm {\BFy^i - \BFy^j } + 2\prox[\Qx\mid\Latticex] \\
        &\leq \sigma_1^\BFx  L_y + 2\prox[\Qx\mid\Latticex]  \label{eq:t3}
    \end{align}
    where the first inequality follows from the triangular inequality of norms, the second inequality follows from the fact that the largest singular value of $\Rx$ is less than $1$ and the last inequality follows from the fact $L_y \ge \norm {\BFy^i-\BFy^j}$ for any $\BFy^i, \BFy^j \in S_y$ by definition.
    However, since $\BFx^i, \BFx^j$ were arbitrary vectors in $S_x$, 
    we have $L_x \le \sigma_1^\BFx  L_y + 2\prox[\Qx\mid\Latticex]$.
    Now, following analogous arguments for $\BFy$-player, as we did above for $\BFx$-player, we get $\norm{\BFy^i - \BFy^j} \le L_y \le \sigma_1^\BFy L_x + 2\prox[\Qy\mid\Latticey]$ for any $\BFy^i, \BFy^j \in S_y$. Substituting this in \cref{eq:t3}, we get 
    \begin{align}
        \norm{ \BFx^i - \BFx^j } 
        &\leq \sigma_1^\BFx  \left( \sigma_1^\BFy L_x + 2\prox[\Qy\mid\Latticey] \right) + 2\prox[\Qx\mid\Latticex]   \\
        &= \sigma_1^\BFx  \sigma_1^\BFy L_x + 2\prox[\Qy\mid\Latticey] \sigma_1^\BFx + 2\prox[\Qx\mid\Latticex]   \\
        \implies L_x &\le \sigma_1^\BFx  \sigma_1^\BFy L_x + 2\prox[\Qy\mid\Latticey] \sigma_1^\BFx + 2\prox[\Qx\mid\Latticex]   \\
        \implies L_x - \sigma_1^\BFx  \sigma_1^\BFy L_x &\le   2\prox[\Qy\mid\Latticey] \sigma_1^\BFx + 2\prox[\Qx\mid\Latticex]   \\
        \implies L_x &\leq \frac{ 2\prox[\Qy\mid\Latticey] \sigma_1^\BFx + 2\prox[\Qx\mid\Latticex]}{ 1 - \sigma_1^\BFx  \sigma_1^\BFy }
    \end{align}
\end{subequations}
Notice that the division by $1 - \sigma_1^\BFx  \sigma_1^\BFy$ in the last step is valid since $\sigma_1^\BFx, \sigma_1^\BFy < 1$ and hence $1 - \sigma_1^\BFx \sigma_1^\BFy$ due to positively adequate objectives, proving the bound for $L_{x}$.

Following analogous steps for the $\BFy$-player, the bound for $L_y$ follows.
}
\hfill
\Halmos
    \iftoggle{arxiv}{
        \end{proof}
    }{\endproof}

	Due to \cref{thm:closeIter,thm:intFinImpl,thm:BRintConds}, we now have the following corollary, which captures the complete result in the context of \cref{eq:intQuad} with positively adequate  objectives. 
\begin{corollary}\label{cor:final}
    Given an instance of \cref{eq:intQuad}, \cref{alg:BR} terminates finitely outputting finite sets $S_x$ and $S_y$. Moreover, any MNE of the finite game restricted to $S_x$ and $S_y$ is a $(\Delta_x, \Delta_y)$-MNE to the instance of \cref{eq:intQuad}, where 
    \begin{subequations}
    \begin{align}
    \Delta_x \quad&=\quad \lambda_1^\BFx\left( \prox[\Qx\mid\Latticex] + \frac{ 2\prox[\Qy\mid\Latticey] \sigma_1^\BFx + 2\prox[\Qx\mid\Latticex]}{ 1 - \sigma_1^\BFx  \sigma_1^\BFy } \right)^2 \label{eq:deltaX} \\
    \Delta_y \quad&=\quad \lambda_1^\BFy\left( \prox[\Qy\mid\Latticey] + \frac{ 2\prox[\Qx\mid\Latticex] \sigma_1^\BFy + 2\prox[\Qy\mid\Latticey]}{ 1 - \sigma_1^\BFx  \sigma_1^\BFy } \right)^2 \label{eq:deltaY} 
    \end{align}
    \end{subequations}
\end{corollary}

\section{Computational experiments}\label{sec:comp}
We conduct computational experiments in two families of instances. 
All tests were done in MacBook Air, 2020 with an Apple M1 (3.2 GHz) processor and 16GB RAM.
The primary comparison in both these families of instances is between the best-respose (BR) algorithm in \cref{alg:BR} and the SGM algorithm \citep{carvalho_2020_computing}. 
The initial iterate used for both the algorithms is always the zero vector of appropriate dimension. 
The best-response optimization problems are solved using Gurobi 9.1 \citep{gurobi}.
All finite games, be it the restricted game at the end of the BR algorithm, or the intermediate games solved in SGM algorithm are also solved by posing the problems as mixed-integer programming problems as shown in \citet{sandholm_mixed-integer_2005}.
These mixed-integer programs were also solved using  Gurobi 9.1 \citep{gurobi}.
\subsection{Pricing with substitutes and complements. }
\paragraph{Family description. }
In this family of instances, we consider $n$ retailers who are competitively pricing their products.
Each retailer $i$ has a disjoint set of  products in the set $J_i$. 
The demand for each product depends upon the price of that product, as well as the price of all other products, which could be strategic complements or substitutes. 
In particular, we consider a linear price-response curve given by 
$q_j = a_j - b_j p_j - \sum_{j' \in J\setminus{j}} d_{jj'} p_{j'} $, where $J = \bigcup_i J_i$ is the set of all products, $p_j$ is the price of product $j$ and $q_j$ is the quantity of product $j$ sold. 
The terms $d_{jj'}$ account for the cross elasticities, and $d_{jj'}$ is positive if $j$ and $j'$ are strategic complements and $d_{jj'}$ is negative $j$ and $j'$ are strategic substitutes. 
The player $i$ controls the prices of only the products that they sell. 
Each product could also have a marginal cost $c_j$, and each player maximizes their profit $\sum_{j\in J_i} (p_j - c_j) q_j$. 
Substituting the price-response function for $q_j$ in the above results in a convex-quadratic objective for each player. 
Further, in many realistic situations, prices are required to take discrete values rather than in a continuum. Thus, we enforce that the prices must be integers. 
\begin{figure}[t]
    \centering
    \begin{subfigure}[b]{0.45\textwidth}
        \includegraphics[width=\textwidth]{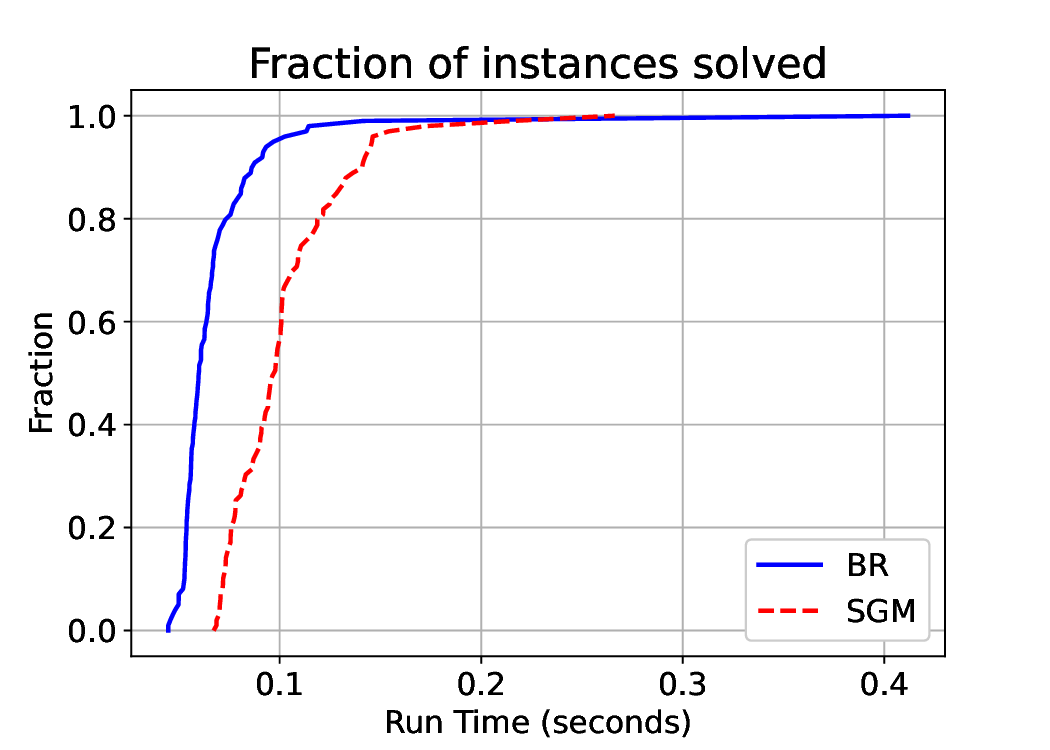}
        \caption{Pricing game with $n=2$ players}\label{fig:pl2}
    \end{subfigure}
    \begin{subfigure}[b]{0.45\textwidth}
        \includegraphics[width=\textwidth]{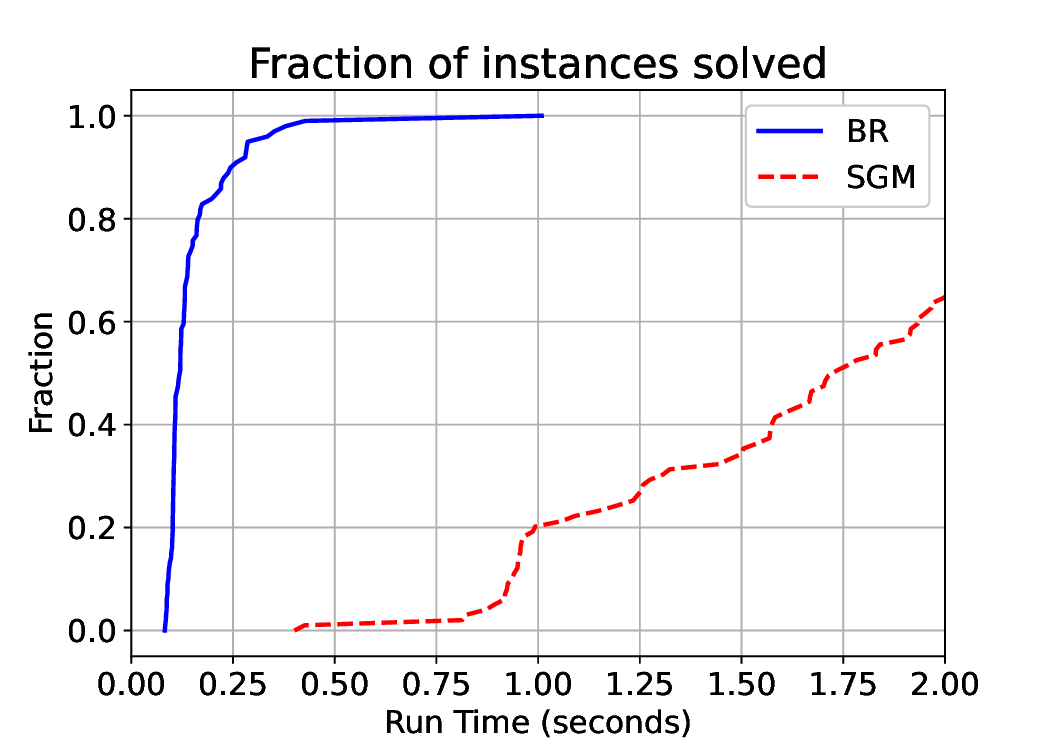}
        \caption{Pricing game with $n=3$ players}\label{fig:pl3}
    \end{subfigure}
    \\
    \begin{subfigure}[b]{0.45\textwidth}
        \includegraphics[width=\textwidth]{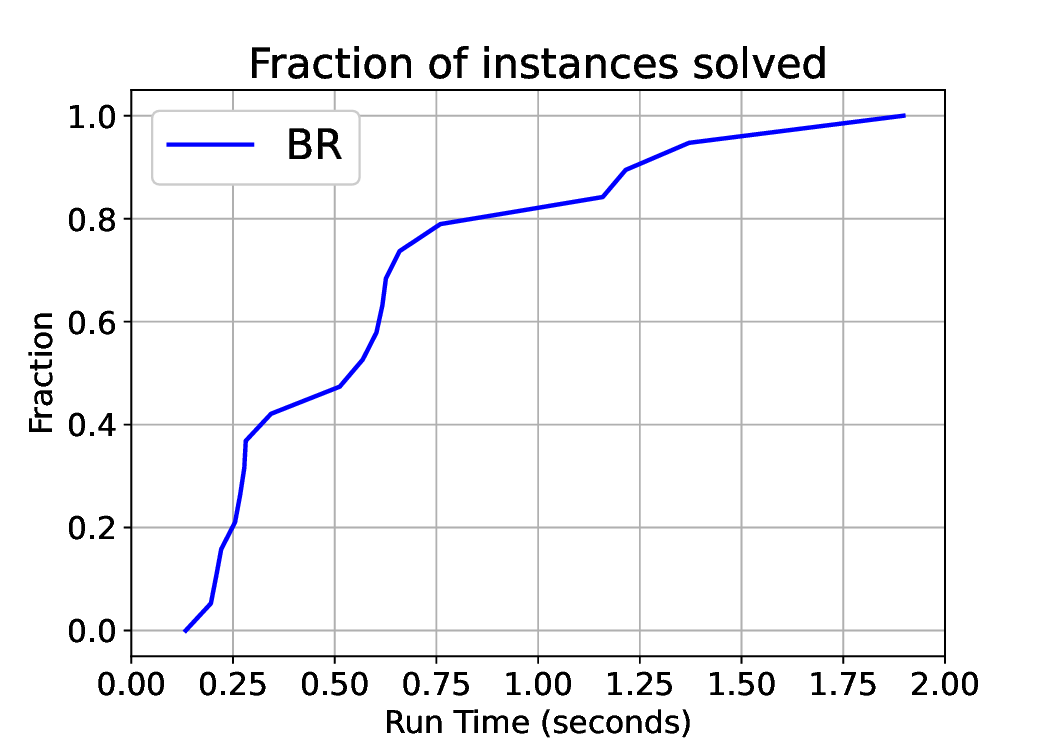}
        \caption{Pricing game with $n=4$ players}\label{fig:pl4}
    \end{subfigure}
    \begin{subfigure}[b]{0.45\textwidth}
        \includegraphics[width=\textwidth]{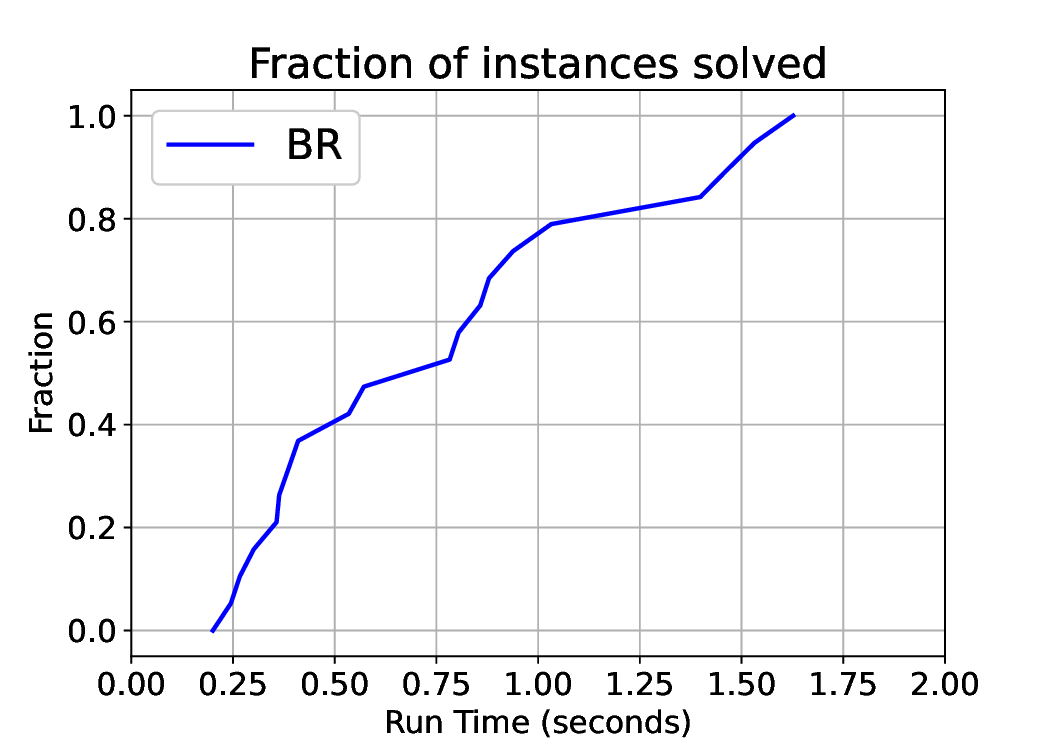}
        \caption{Pricing game with $n=5$ players}\label{fig:pl5}
    \end{subfigure}
    \caption{Pricing game - performance profile. Shows the fraction of instances solved within given time}\label{fig:pl}
\end{figure}
This results in the first family of problems. 
\paragraph{Instance generation. }
We generated 100 instances with two players each, 100 instances with three players each, 20 instances with four players each, and 20 instances with five players each. 
Thus, it adds up to 240 instances over all. 
The number of products each player controls is a random number between three and six. 
All the other parameters $a, b$ for each product as well as $d$ for each pair of products is generated uniformly randomly between appropriate limits. 
As soon as the instances are generated, we check if the instance has positively adequate objectives. 
If not, then the instance is discarded. 
All other instances were retained. 
\paragraph{Results. }
In the two player case, we observe that we are competitive with SGM. 
\cref{fig:pl2} compares the performance profile of our algorithm with SGM. 
The BR algorithm presented in \cref{alg:BR} is slightly faster than SGM. 
The mean run-time for BR is 0.0683 seconds as oppossed to the mean run-time for SGM being 0.1017 seconds. 
The median run-time for BR is 0.0599 seconds, while the median run-time for SGM is 0.0971 seconds. 
This is consistent with the mild speed up discussed before. 
However, the mild improvement is statistically significant that a paired t-test between the run time of the algorithms testing equality of means, the null hypothesis can be rejected with a p-value of $2.683 \times 10^{-11}$. 

However, with three players, there is a considerable speed up when using the BR algorithm. 
The performance profile is depicted in \cref{fig:pl3}. We can see that almost all the instances were solved in less than 0.5 seconds when using BR, while almost no instance is solved within that time in SGM. 
Comparably, the mean run-time for BR and SGM in this set of instances is 0.1506 seconds and 3.3608 seconds. The median run-time for BR and SGM are 0.1192 and 1.7239 seconds, indicating that our algorithm is clearly at least ten times faster than SGM. 

With four and five players, the difference is even more pronounced. 
The mean run-time for BR in the four and five player cases are 0.6087 seconds and 0.7477 seconds. 
The median run-time for BR in the four and five player cases are 0.5407 seconds and 0.6772 seconds. 
The SGM was run on these instances with a maximum allotted time of 120 seconds. 
We observe that not even one of the four-player instance or five-player instance was solved within 120 seconds, hinting at at least that our algorithm provides 100-times speed up when applicable. 

Finally, we also note that in each of the 240 instances, the BR algorithm always terminated after finding a PNE or an MNE, but never a $\Delta$-MNE with $\Delta > 0$ (after allowing for numerical tolerance of $1\times 10^{-6}$). 
We share the instance-by-instance data on run time and the number of iterations in \cref{sec:tables} in the electronic companion. 

\subsection{Random instances. }
\paragraph{Family description. }
In this family of instances, the matrices $\Qq[i], \dq[i], \Cq{i}$ are all randomly generated matrices with integer entries. To enure that $\Qq[i]$s are positive definite, we generate a random integer matrix $P$, and compute $\tilde{\Qq[i]} = PP^\top$, which is now guaranteed to be a positive-definite matrix with integer entries. 
Next, to ensure that the players have positively adequate objectives, we compute $\tilde{\Rq{i}}=\tilde{\Qq[i]}^{-1}\Cq{i}$ and compute the largest singular value of $\tilde{\Rq{i}}$, which we denote as $\tilde{\sigma_1}$. 
Finally, we define $\Qq[i] = \ceil{\sigma_1}\tilde{\Qq[i]} + \BFI$. 
The ceiling ensures that $\Qq[i]$ has integer entries and the addition with the identity matrix ensures that we have each of the singular values of $\Rq{i} = \Qq[i]^{-1}\Cq{i}$ is \emph{strictly} lesser than $1$. 
\begin{figure}[t]
    \centering
    \begin{subfigure}[b]{0.45\textwidth}
        \includegraphics[width=\textwidth]{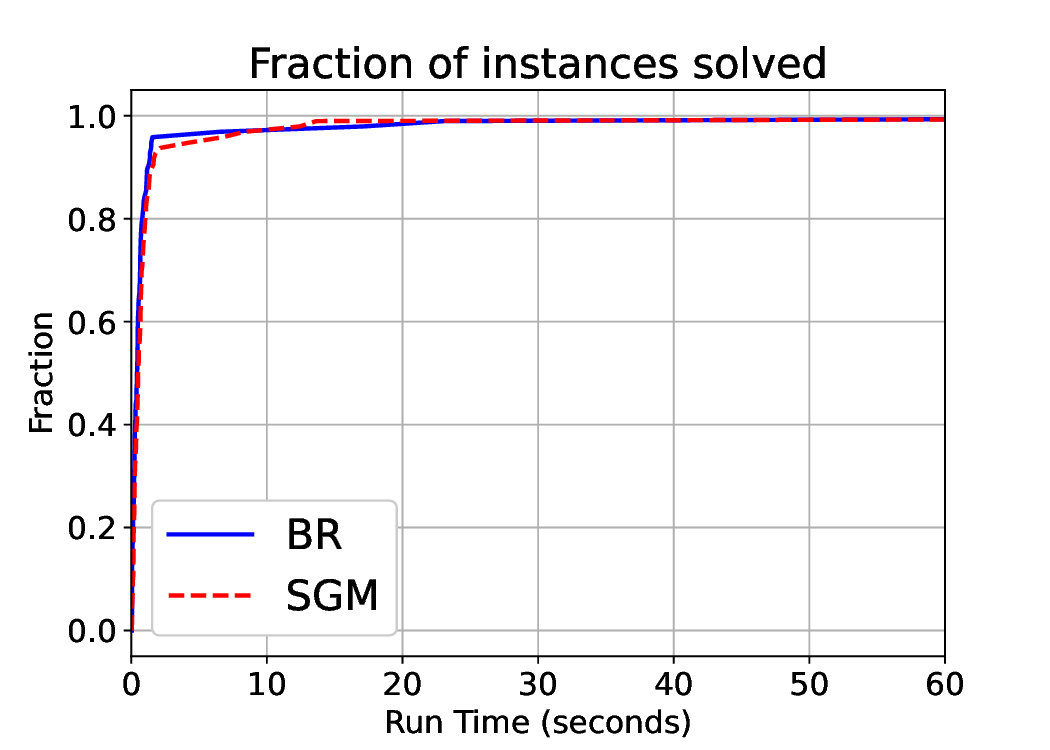}
        \caption{Random game with $n=2$ players}\label{fig:rand2}
    \end{subfigure}
    \begin{subfigure}[b]{0.45\textwidth}
        \includegraphics[width=\textwidth]{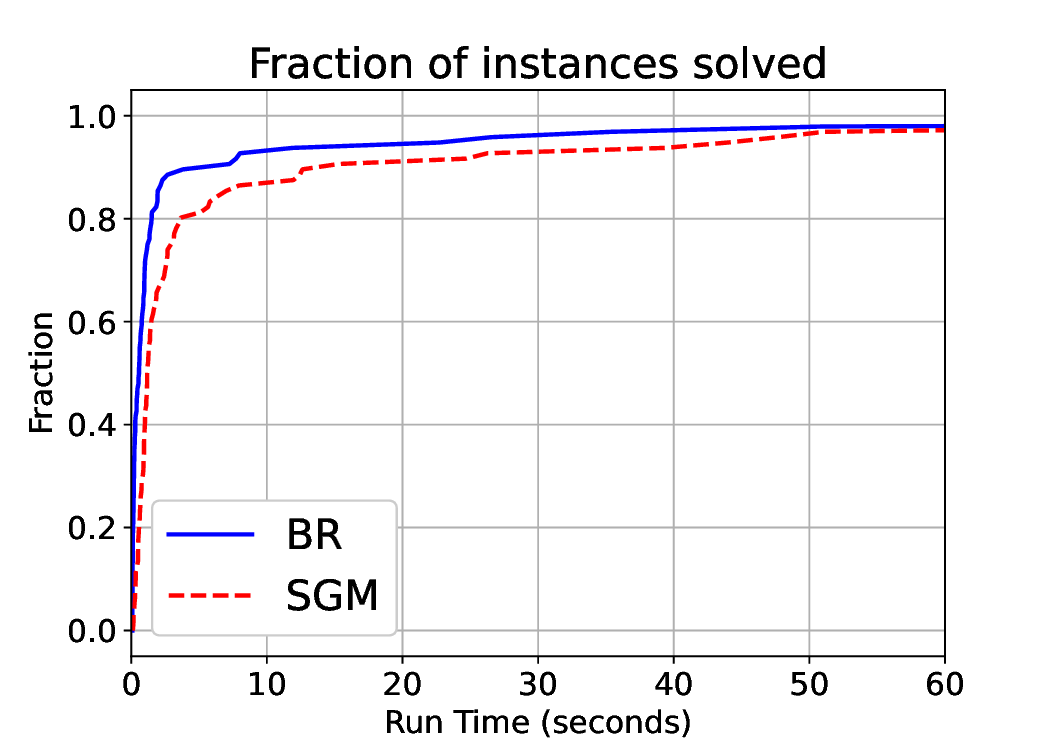}
        \caption{Random game with $n=3$ players}\label{fig:rand3}
    \end{subfigure}
    \\
    \begin{subfigure}[b]{0.45\textwidth}
        \includegraphics[width=\textwidth]{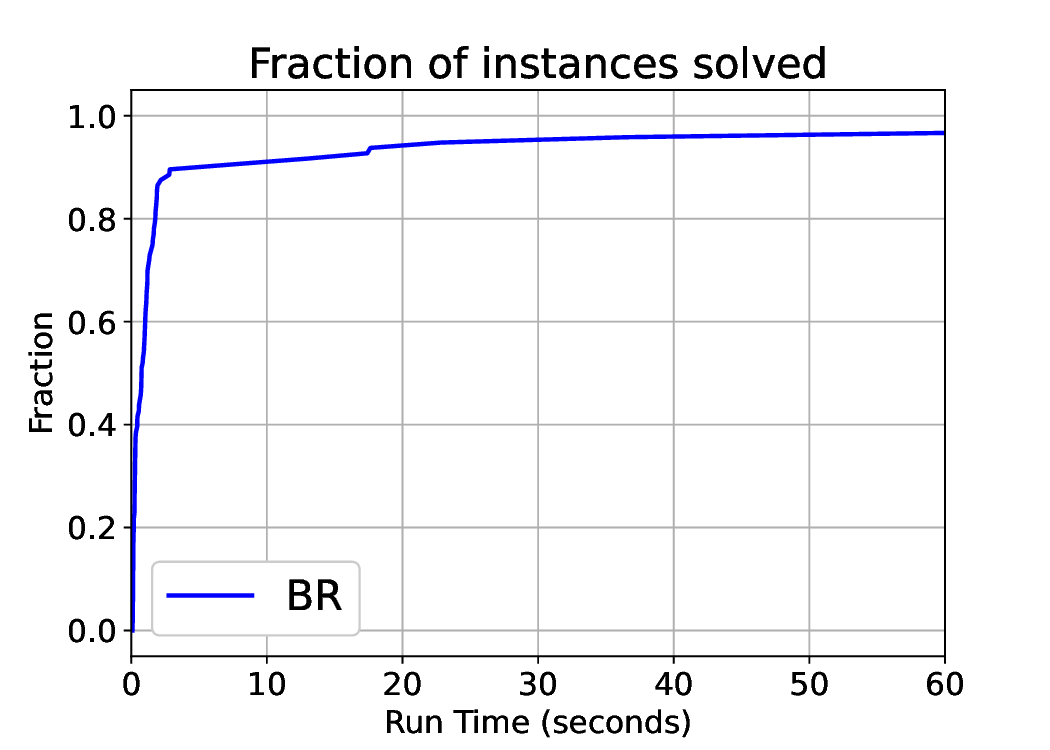}
        \caption{Random game with $n=4$ players}\label{fig:rand4}
    \end{subfigure}
    \begin{subfigure}[b]{0.45\textwidth}
        \includegraphics[width=\textwidth]{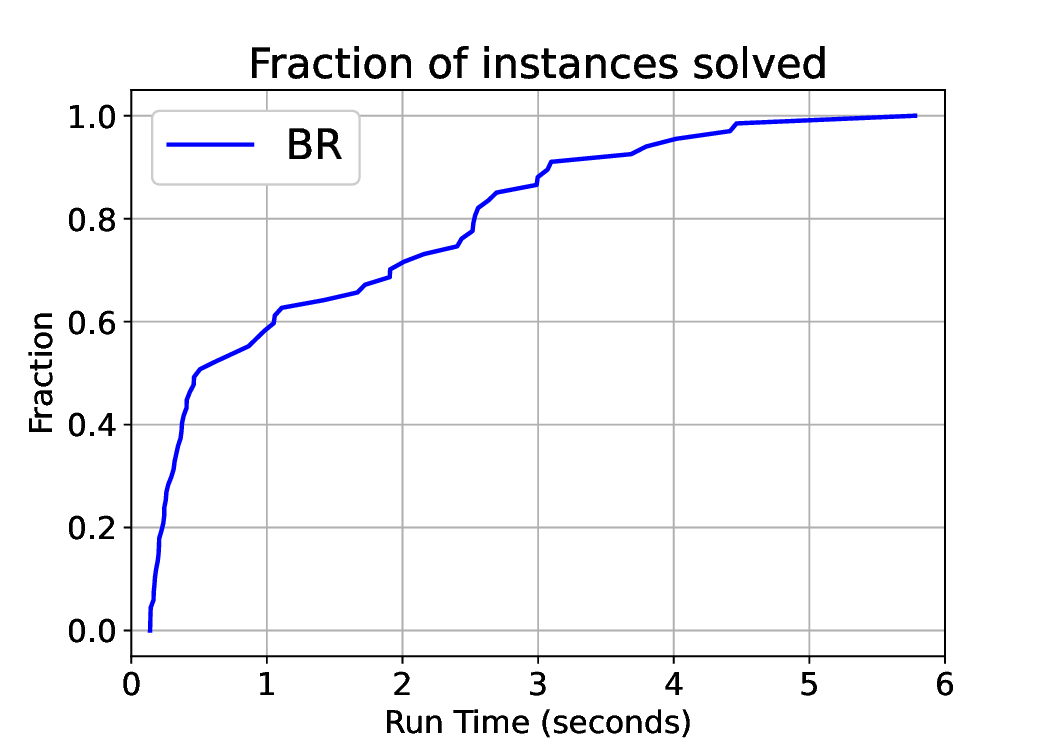}
        \caption{Random game with $n=5$ players}\label{fig:rand5}
    \end{subfigure}
    \caption{Random game - performance profile. Shows the fraction of instances solved within given time}\label{fig:rand}
\end{figure}

\paragraph{Instance generation. }
We vary the number of players between two, three, four, and five. 
Further, for each of the above four situations, we consider the decision vector of each player to be vary from 5, 10, 15, 20 or 25 variables.  
For each of the combinations, for example, three player games with fifteen variables per player or five player games with five variables per player, we generate 20 instances randomly. 
This gives a total of $4\times 5\times 20 = 400$ instances. 
Along with the exact matrices defining these 400 instances, we also share the code used to generate them. 
\paragraph{Results. }
In all subfamilies of instances with two to five players, there were two to four instances in each setting where numerical instabilities called failure of both the BR as well as the SGM algorithm.
These instances were discarded from the below analysis as both the algorithms failed in these instances. 

Out of the remaining instances, in the two player and three player cases, the performance of BR and SGM is comparable. In fact, we do not find any statistically significant difference between the two algorithms. Their performance profiles are plotted in \cref{fig:rand2,fig:rand3}. 
However, when we have four or five players, the BR algorithm is significantly faster than the SGM algorithm. 
The BR always found a solution (except for the cases with numerical instabilities), with a median time of 5.348 seconds for four player-instances and 6.131 seconds for five player-instances, with the maximum time taken for any single instance being 233 seconds approximately. 
However, with the SGM algorithm, only eight of the hundred instances with four players were solved within a time limit of 500 seconds. The quickest one among them took over 290 seconds. 
Moreover, all eight solved instances are the simplest of the four player instances, where the decision vector of each player has five variables. 
Among the five player instances, only two of the hundred instances were solved within a time limit of 500 seconds, both of them taking over 400 seconds. 
Again, both the solved instances correspond to those where each player's decision vector has five variables. 
The complete instance-by-instance details of the computational tests are presented in \cref{sec:tables} in the electronic companion. 

We also note that that, in every single instance that was solved ({\em i.e., } the ones that did not run into numerical error), the MNE of the restricted finite game after running the BR algorithm, had only a profitable deviation with maximum profit in the order of $10^{{-6}}$, which can be considered as errors due to numerical methods used within the solver, thus motivating a conjecture that $\Delta=0$ is provable.

\section{Limitations and Future Work}\label{sec:concl}
We first state some limitations of the work.
When the algorithm terminates in a trap, then the bounds for the MNE suggested in this paper are not very tight. This is because the proximity bound in \cref{thm:proxProof} is inherently not tight. 
Moreover, \cref{thm:intFinImpl,thm:closeIter} also assume the worst case to happen at equilibria, that the farthest strategies in the trap have a support in the MNE, and the lattice minimizers is in a direction from the continuous minimizer so that the error increases more. However, it is unclear, if further arguments can refine these possibilities and can provide a closer error bound. 
However, our contribution's merit derives from our ability to provide a finite error bound. 
In fact, the computational results in \cref{sec:comp} provides preliminary evidence about the weakness of our bound, and motivates us to make the following conjecture, which strengthens \cref{cor:final}.
\begin{conjecture} \label{conj:final}
    Given an instance of \cref{eq:intQuad} with positively adequate objectives, and sets $S_x$ and $S_y$ from \cref{alg:BR}, any MNE of the version of the game restricted to $S_x$ and $S_y$ is an MNE to \cref{eq:intQuad}.
\end{conjecture}
In other words, the conjecture says that \cref{cor:final} holds  with $\Delta_x = \Delta_y = 0$. 
The conjecture is validated by the computational experiments in \cref{sec:comp}. 
This is also consistent with the fact that the family of counterexamples provided in the proof of  \cref{thm:intFinNeg} do not have positively adequate objectives.

Finally, in this paper, we fundamentally use the properties of quadratic functions to prove the results.
It is conceivable then, that the results should hold even if the objective functions are approximated well by quadratic functions. 
For example, $L$-Lipschitz, $\mu$-strongly convex functions are both under-approximated and over-approximated by quadratic functions. 
However, an extension of these results to such functions and identifying the loss in the approximation ratios when considering such functions is non trivial, and is an interesting avenue for future work. 

    \iftoggle{incConj}{
        \input{conjectures}
    }{}

\iftoggle{arxiv}{}{
}
\bibliographystyle{plainnat}
\bibliography{reference}
\iftoggle{arxiv}{\appendix}{
    \ECSwitch
}

\iftoggle{incCont}{
}{ 
}

\section{Auxiliary results}
We state these auxiliary results for ready reference. 
The following results are available (typically in greater generality) in most standard texts on matrix analysis, for example, \citet{horn2012matrix}. 
However, a short proof sketch is provided for the reader's convenience.

\begin{proposition}\label{thm:diagSing}
    Let $\BFA$ be an $m\times n$  matrix and $\BFB$ be an $n\times m$ matrix. Let $\sigma^A_1, \dots, \sigma^A_k$ be the singular values of $\BFA$ and $\sigma^B_1, \dots, \sigma^B_\ell$ be the singular values of $\BFB$. Then, the singular values of the $(m+n)\times (m+n)$ matrix  $ \begin{pmatrix} \BFA & \BFzero \\ \BFzero & \BFB \end{pmatrix} $ are $\sigma^A_1, \dots, \sigma^A_k, \sigma^B_1,\dots,\sigma^B_\ell$. 
\end{proposition}
\iftoggle{arxiv}{
    \begin{proof}[Proof of \cref{thm:diagSing}.]}{
    \proof{Proof of \cref{thm:diagSing}.}}
    Consider the singular value decompositions (SVD) of matrices $\BFA$ and $\BFB$. 
    Let $\BFA = \BFU^A\BFSigma^A \BFV^A$ and $\BFB = \BFU^B\BFSigma^B \BFV^B$. 
    Then, $ \BFC :=
\begin{pmatrix}
    \BFA & \BFzero \\ \BFzero & \BFB
    \end{pmatrix} =  \BFU \BFSigma' \BFV = 
    \begin{pmatrix} \BFU^A & \BFzero \\ \BFzero & \BFU^B \end{pmatrix}
    \begin{pmatrix} \BFSigma^A & \BFzero \\ \BFzero & \BFSigma^B \end{pmatrix}
    \begin{pmatrix} \BFV^A & \BFzero \\ \BFzero & \BFV^B \end{pmatrix}
    $, which is obtained by muiltiplying the block matrices. 
    While 
    $\BFSigma'$
    is not diagonal, its rows and columns can be permuted according to a permutation matrix $\BFP$ to get $\BFSigma'  = \BFP \BFSigma \BFP^\top$ to get $\BFC = (\BFU\BFP) \BFSigma (\BFP^\top \BFV)$ where the only non-zero elements of $\BFSigma$ are along its diagonal. 
    It can also be verified that $\BFU\BFP$ and $\BFP^\top \BFV$ are unitary. 
    The only non-zero elements of $\BFSigma$ are the singular values of $\BFA$ and $\BFB$, making them the singular values of $\BFC$.
    \iftoggle{arxiv}{
        \end{proof}
    }{
\hfill
\Halmos
    \endproof}

\begin{proposition} \label{thm:singularBnd}
    Every singular value of a matrix $\BFM$ is strictly lesser than $1$ if and only if  $\norm{\BFM\BFx}_2 < \norm{\BFx}_2$ for every $\BFx \in \R^n$. 
    Every singular value of a matrix $\BFM$ is strictly greater than $1$ if and only if  $\norm{\BFM\BFx}_2 > \norm{\BFx}_2$ for every $\BFx \in \R^n$. 
\end{proposition}
\iftoggle{arxiv}{
    \begin{proof}[Proof of \cref{thm:singularBnd}.]}{
    \proof{Proof of \cref{thm:singularBnd}.}}
    Let $\BFM$ be a matrix of dimension $m\times n$ with real entries.
    Then, the singular values of $\BFM$ are the square roots of the eigenalues of $\BFM^\top \BFM$.
    Let us call this matrix $\BFM'$. 
    Since $\BFM'$ is symmetric, all its eigenvalues are real. Moreover, from Rayleigh's theorem,  the largest value that  $\BFx^\top \BFM' \BFx$ can take is $\bar \lambda \BFx^\top \BFx$ and the smallest value it can take is $\underline \lambda \BFx^\top \BFx$ where $\bar \lambda$ and $\underline {\lambda}$ are the largest and the smallest eigenvalues of $\BFM'$ respectively.
    Now, observe $\norm{\BFM\BFx}^2_2 = \BFx^\top \BFM^\top \BFM \BFx = \BFx^\top \BFM' \BFx \le \bar \lambda \norm{\BFx}^2_2$. 
    But the largest eigenvalue of $\BFM'$, which is $\bar \lambda$, is the square of the largest singular value of $\BFM$ (denoted as $\sigma_1$). 
    Thus, we have $\norm {\BFM\BFx}^2_2 \le \sigma_1^2 \norm{\BFx}^2_2$, which implies that $\norm{\BFM\BFx}_2 \le \sigma_1 \norm{\BFx}_2$. 
    This proves the first part of the result. The second part can be proved using analogous arguments. 
    \iftoggle{arxiv}{
        \end{proof}
    }{
\hfill
\Halmos
    \endproof}

\section{Extension to multiple players}\label{sec:multiPlayer}
Suppose there are multiple players $1, \dots, k$. We denote the variables of player $i$ as $\BFx^i$ and those of everybody except $i$ as $\BFx^{-i}$.
Now, the objective function if player $i$ is given as
$%
\frac{1}{2} {\BFx^i}^\top \Qq[i] \BFx^i  + \left( \Cq{i} \BFx^{-i} + \dq[i] \right)^\top \BFx^i
$%
. 
For example, in a three player game, suppose the objective function of player $1$ is given as
$
\frac{1}{2} {\BFx^1}^\top \Qq[1] \BFx^1  + \left( \Cq{1,2} \BFx^2 + \Cq{1,3} \BFx^3 + \dq[1] \right)^\top \BFx^1
$, 
we write 
$
\Cq{1} = \begin{pmatrix} \Cq{1,2} & \Cq{1,3} \end{pmatrix}
$. 
Thus,
$
\Cq{1} \BFx^{-1} =
\begin{pmatrix} \Cq{1,2} & \Cq{1,3} \end{pmatrix} 
\begin{pmatrix} \BFx^2 \\ \BFx^3 \end{pmatrix}
= \Cq{1,2} \BFx^2 + \Cq{1,3} \BFx^3
$. 

The successive iterates, as in the proof of \cref{thm:BRintConds}, are given by $\BFx^{i,t+1} \gets -\Qq[i]^{-1} (\Cq{i}\BFx^{-i,t} + \dq[i]) + z_i(\BFx^{-i,t})$. 
The index $i$ refers to the player whose best response is being computed and $t$ refers to the iteration number. 

We provide a proof sketch that even with $k$ players, a theorem analogous to that of \cref{thm:BRintConds} holds. In other words, when the game has positively adequate objectives, then the \BR algorithm terminates. 
Now, analogous to the proof of \cref{thm:BRintConds}, we can write the difference between two strategy profiles as 
$
\norm{ \vecN{ 
    \BFx^{1,t+1}  \\
    \vdots \\
    \BFx^{k,t+1} 
    }}
= \norm { \vecN{
    -\Qq[1]^{-1} (\Cq{1}\BFx^{-1,t} + \dq[1]) + z_1(\BFx^{-1,t})\\
    \vdots\\
    -\Qq[k]^{-k} (\Cq{k}\BFx^{-k,t} + \dq[k]) + z_k(\BFx^{-k,t})
    } }
= \norm { \vecN{
        ( -\Rq{1}\BFx^{-1,t} - \Qq[1]^{-1} \dq[1]) + z_1(\BFx^{-1,t})\\
    \vdots\\
    ( -\Rq{k}\BFx^{-k,t} + - \Qq[k]^{-k}\dq[k]) + z_k(\BFx^{-k,t})
    } }
\le 
\norm{ \vecN{\Rq{1} & \cdots & 0 \\
        \vdots & \ddots & \vdots \\
        0 & \cdots & \Rq{k} } } \norm{ \vecN{\BFx^{1,t} \\ \vdots \\ \BFx^{k,t}} } + \norm{ \vecN{ \Qq[1]^{-1} \dq[1] \\ \vdots \\ \Qq[k]^{-1} \dq[k] } } + \norm{ \vecN{ z_1(\BFx^{-1,t}) \\ \vdots \\ z_k(\BFx^{-k,t}) } }
\le 
(1-\rho) \norm{\vecN{ \vecN{\BFx^{1,t} \\ \vdots \\ \BFx^{k,t}} }}
+ \norm{ \vecN{ \Qq[1]^{-1} \dq[1] \\ \vdots \\ \Qq[k]^{-1} \dq[k] } }
+ \norm {\vecN{ \prox[{\Qq[1]}] \\ \vdots \\ \prox[{\Qq[k]}] }}
$. 
Like before, we observe that if the norm of $\norm{\vecN{ \vecN{\BFx^{1,t} \\ \vdots \\ \BFx^{k,t}} }}
$ is large, then the iterate in the subsequent iteration will necessarily have a smaller norm. 
But this means, the iterates will have to return to a bounded region, if they begin to move towards infinity. But in any bounded region, there will be finitely many feasible integer points, leading to cycling and hence termination. Hence the proof. 

Next, it is straightforward to observe \cref{thm:intFinImpl} translates to multiplayer case naturally.
The proof of \cref{thm:intFinImpl} considers the strategies of player $\BFx$, while using the probabilities and strategies of player $\BFy$. 
In a multi-player setting, the same proof can be adapted, by considering \emph{all the other players}'s strategies $\BFx^{-i}$, when establishing a bound $\Delta_i$. 

Finally, to provide the bounds on $L_x$ (which will now be notated as $L_i$ when there are multiple players), we express \cref{eq:t3} in terms of $L_{-i}$, which is the maximum distance between two valid $\BFx^{-i}$ that appears in the cycle.

\section{Computational experiments data}\label{sec:tables}
For the 240 instances of the pricing game with substitutes and complements and the 500 instances of the random games, we provide the run-time data here below. 
Each instance is recognized by the unique filename that has the data for the instance. 
The second column indicates the number of players in the instance. 
The third and fourth columns titled $t_{BR}$ and $t_{SGM}$ indicate the time taken by the BR and SGM algorithms respectively on the problem. 
An entry saying \emph{TL} here indicates that the maximum time limit is reached but no MNE is found. 
An entry saying \emph{Num Err} indicates that the solver ran into numerical issues as some of the matrices possibly have large entries in them. 
In particular, we state numerical error, if the integer programming solver (Gurobi) declares that the matrix $\Qq$ used is not positive definite. This is not possible from construction, as the instances are generated by choosing $\Qq = \BFA \BFA^{\top} + \BFI$ for some random $\BFA$. 
However, some times, large entries in $\Qq$ makes Gurobi to declare that the matrix is not positive definite, and we report numerical errors in these cases. 
Finally, the last two columns $k_{{BR}}$ and $k_{SGM}$ indicate the number of iterations of the BR and the SGM algorithms that ran before either successful termination or reaching the time limit or running into numerical issues. 

\subsection{Pricing with substitutes and complements}
\begin{tiny}
    \begin{longtable}{|rccccc|}
		\hline
                \textbf{Instance name} & \textbf{nPlay} & \textbf{$t_{BR}$} & \textbf{$t_{SGM}$} & \textbf{$k_{BR}$} & \textbf{$k_{SGM}$} \\  \midrule
			\endhead
        \hline
        asymmMktGame\_N2\_1.json & 2 & 0.2172 & 0.1425 & 4 & 6 \\ \hline
        asymmMktGame\_N2\_2.json & 2 & 0.1812 & 0.1904 & 4 & 5 \\ \hline
        asymmMktGame\_N2\_3.json & 2 & 0.3832 & 0.3613 & 5 & 7 \\ \hline
        asymmMktGame\_N2\_4.json & 2 & 0.1808 & 0.2887 & 5 & 7 \\ \hline
        asymmMktGame\_N2\_5.json & 2 & 0.3706 & 0.5868 & 4 & 5 \\ \hline
        asymmMktGame\_N2\_6.json & 2 & 0.138 & 0.1444 & 4 & 5 \\ \hline
        asymmMktGame\_N2\_7.json & 2 & 0.4024 & 0.9749 & 4 & 6 \\ \hline
        asymmMktGame\_N2\_8.json & 2 & 0.2808 & 0.5777 & 4 & 5 \\ \hline
        asymmMktGame\_N2\_9.json & 2 & 0.9021 & 0.5659 & 4 & 6 \\ \hline
        asymmMktGame\_N2\_10.json & 2 & 0.1653 & 1.3944 & 4 & 6 \\ \hline
        asymmMktGame\_N2\_11.json & 2 & 1.1006 & 0.3641 & 4 & 6 \\ \hline
        asymmMktGame\_N2\_12.json & 2 & 0.1449 & 0.5067 & 4 & 6 \\ \hline
        asymmMktGame\_N2\_13.json & 2 & 0.8202 & 0.9078 & 5 & 7 \\ \hline
        asymmMktGame\_N2\_14.json & 2 & 1.4994 & 0.9507 & 4 & 5 \\ \hline
        asymmMktGame\_N2\_15.json & 2 & 0.2634 & 0.1862 & 6 & 5 \\ \hline
        asymmMktGame\_N2\_16.json & 2 & 0.1037 & 0.1353 & 4 & 5 \\ \hline
        asymmMktGame\_N2\_17.json & 2 & 0.7951 & 2.0591 & 4 & 6 \\ \hline
        asymmMktGame\_N2\_18.json & 2 & 1.4928 & 1.1661 & 4 & 6 \\ \hline
        asymmMktGame\_N2\_19.json & 2 & 0.3455 & 1.1541 & 3 & 5 \\ \hline
        asymmMktGame\_N2\_20.json & 2 & 0.5566 & 0.89 & 4 & 5 \\ \hline
        asymmMktGame\_N2\_21.json & 2 & 0.5051 & 0.5074 & 4 & 5 \\ \hline
        asymmMktGame\_N2\_22.json & 2 & 0.9255 & 0.3385 & 3 & 5 \\ \hline
        asymmMktGame\_N2\_23.json & 2 & 1.8685 & 0.8177 & 4 & 6 \\ \hline
        asymmMktGame\_N2\_24.json & 2 & 1.1305 & 0.9827 & 4 & 7 \\ \hline
        asymmMktGame\_N2\_25.json & 2 & 0.3919 & 1.0244 & 4 & 7 \\ \hline
        asymmMktGame\_N2\_26.json & 2 & 0.6474 & 0.5421 & 4 & 6 \\ \hline
        asymmMktGame\_N2\_27.json & 2 & 0.3688 & 0.5998 & 4 & 6 \\ \hline
        asymmMktGame\_N2\_28.json & 2 & 0.2082 & 0.3656 & 3 & 5 \\ \hline
        asymmMktGame\_N2\_29.json & 2 & 0.2732 & 0.5181 & 4 & 6 \\ \hline
        asymmMktGame\_N2\_30.json & 2 & 0.5284 & 1.5531 & 4 & 5 \\ \hline
        asymmMktGame\_N2\_31.json & 2 & 1.4595 & 0.89 & 4 & 6 \\ \hline
        asymmMktGame\_N2\_32.json & 2 & 0.2795 & 0.3696 & 4 & 5 \\ \hline
        asymmMktGame\_N2\_33.json & 2 & 0.392 & 0.6649 & 4 & 6 \\ \hline
        asymmMktGame\_N2\_34.json & 2 & 1.4271 & 0.5682 & 4 & 5 \\ \hline
        asymmMktGame\_N2\_35.json & 2 & 0.5249 & 1.1689 & 5 & 8 \\ \hline
        asymmMktGame\_N2\_36.json & 2 & 0.2552 & 2.0878 & 4 & 6 \\ \hline
        asymmMktGame\_N2\_37.json & 2 & 2.0422 & 1.1434 & 4 & 6 \\ \hline
        asymmMktGame\_N2\_38.json & 2 & 0.0891 & 0.0963 & 4 & 5 \\ \hline
        asymmMktGame\_N2\_39.json & 2 & 0.0506 & 0.0941 & 3 & 5 \\ \hline
        asymmMktGame\_N2\_40.json & 2 & 0.059 & 0.1413 & 4 & 6 \\ \hline
        asymmMktGame\_N2\_41.json & 2 & 0.0608 & 0.1537 & 4 & 7 \\ \hline
        asymmMktGame\_N2\_42.json & 2 & 0.0668 & 0.1218 & 3 & 5 \\ \hline
        asymmMktGame\_N2\_43.json & 2 & 0.0877 & 0.0879 & 5 & 5 \\ \hline
        asymmMktGame\_N2\_44.json & 2 & 0.0524 & 0.0787 & 4 & 5 \\ \hline
        asymmMktGame\_N2\_45.json & 2 & 0.0574 & 0.0781 & 4 & 5 \\ \hline
        asymmMktGame\_N2\_46.json & 2 & 0.0601 & 0.0781 & 4 & 5 \\ \hline
        asymmMktGame\_N2\_47.json & 2 & 0.049 & 0.0995 & 4 & 6 \\ \hline
        asymmMktGame\_N2\_48.json & 2 & 0.1138 & 0.1121 & 4 & 5 \\ \hline
        asymmMktGame\_N2\_49.json & 2 & 0.075 & 0.0916 & 4 & 5 \\ \hline
        asymmMktGame\_N2\_50.json & 2 & 0.0673 & 0.1295 & 5 & 7 \\ \hline
        asymmMktGame\_N2\_51.json & 2 & 0.0693 & 0.1177 & 4 & 6 \\ \hline
        asymmMktGame\_N2\_52.json & 2 & 0.0941 & 0.167 & 5 & 7 \\ \hline
        asymmMktGame\_N2\_53.json & 2 & 0.0933 & 0.1605 & 5 & 7 \\ \hline
        asymmMktGame\_N2\_54.json & 2 & 0.0594 & 0.0807 & 4 & 5 \\ \hline
        asymmMktGame\_N2\_55.json & 2 & 0.0663 & 0.0752 & 4 & 5 \\ \hline
        asymmMktGame\_N2\_56.json & 2 & 0.1091 & 0.0791 & 3 & 5 \\ \hline
        asymmMktGame\_N2\_57.json & 2 & 0.0952 & 0.3277 & 4 & 6 \\ \hline
        asymmMktGame\_N2\_58.json & 2 & 0.1663 & 0.1887 & 4 & 5 \\ \hline
        asymmMktGame\_N2\_59.json & 2 & 0.4696 & 0.5084 & 5 & 7 \\ \hline
        asymmMktGame\_N2\_60.json & 2 & 0.363 & 0.6538 & 4 & 6 \\ \hline
        asymmMktGame\_N2\_61.json & 2 & 0.1011 & 0.1741 & 4 & 6 \\ \hline
        asymmMktGame\_N2\_62.json & 2 & 0.0979 & 0.1767 & 4 & 7 \\ \hline
        asymmMktGame\_N2\_63.json & 2 & 0.3333 & 0.1599 & 4 & 5 \\ \hline
        asymmMktGame\_N2\_64.json & 2 & 0.0902 & 0.1394 & 4 & 6 \\ \hline
        asymmMktGame\_N2\_65.json & 2 & 0.0752 & 0.3657 & 4 & 6 \\ \hline
        asymmMktGame\_N2\_66.json & 2 & 0.1045 & 0.1041 & 4 & 5 \\ \hline
        asymmMktGame\_N2\_67.json & 2 & 0.077 & 0.3213 & 4 & 5 \\ \hline
        asymmMktGame\_N2\_68.json & 2 & 0.9419 & 0.5485 & 5 & 7 \\ \hline
        asymmMktGame\_N2\_69.json & 2 & 0.7156 & 0.7363 & 5 & 6 \\ \hline
        asymmMktGame\_N2\_70.json & 2 & 0.4242 & 0.7839 & 5 & 6 \\ \hline
        asymmMktGame\_N2\_71.json & 2 & 0.2864 & 0.3569 & 4 & 5 \\ \hline
        asymmMktGame\_N2\_72.json & 2 & 0.4803 & 0.3992 & 3 & 5 \\ \hline
        asymmMktGame\_N2\_73.json & 2 & 0.2708 & 0.587 & 4 & 7 \\ \hline
        asymmMktGame\_N2\_74.json & 2 & 0.1068 & 0.1103 & 4 & 5 \\ \hline
        asymmMktGame\_N2\_75.json & 2 & 0.0553 & 0.0919 & 4 & 6 \\ \hline
        asymmMktGame\_N2\_76.json & 2 & 0.0655 & 0.1096 & 5 & 6 \\ \hline
        asymmMktGame\_N2\_77.json & 2 & 0.0852 & 0.1902 & 4 & 7 \\ \hline
        asymmMktGame\_N2\_78.json & 2 & 0.1082 & 0.1227 & 5 & 6 \\ \hline
        asymmMktGame\_N2\_79.json & 2 & 0.0681 & 0.1019 & 4 & 5 \\ \hline
        asymmMktGame\_N2\_80.json & 2 & 0.0662 & 0.0927 & 4 & 6 \\ \hline
        asymmMktGame\_N2\_81.json & 2 & 0.0559 & 0.0829 & 4 & 5 \\ \hline
        asymmMktGame\_N2\_82.json & 2 & 0.0656 & 0.1007 & 4 & 5 \\ \hline
        asymmMktGame\_N2\_83.json & 2 & 0.0845 & 0.162 & 6 & 9 \\ \hline
        asymmMktGame\_N2\_84.json & 2 & 0.0671 & 0.1092 & 5 & 6 \\ \hline
        asymmMktGame\_N2\_85.json & 2 & 0.0664 & 0.1466 & 5 & 8 \\ \hline
        asymmMktGame\_N2\_86.json & 2 & 0.0925 & 0.1326 & 4 & 6 \\ \hline
        asymmMktGame\_N2\_87.json & 2 & 0.0769 & 0.1228 & 4 & 6 \\ \hline
        asymmMktGame\_N2\_88.json & 2 & 0.0868 & 0.1925 & 5 & 8 \\ \hline
        asymmMktGame\_N2\_89.json & 2 & 0.074 & 0.0801 & 5 & 5 \\ \hline
        asymmMktGame\_N2\_90.json & 2 & 0.065 & 0.1533 & 4 & 6 \\ \hline
        asymmMktGame\_N2\_91.json & 2 & 0.0924 & 0.0929 & 4 & 5 \\ \hline
        asymmMktGame\_N2\_92.json & 2 & 0.0576 & 0.0841 & 4 & 5 \\ \hline
        asymmMktGame\_N2\_93.json & 2 & 0.0537 & 0.1142 & 4 & 7 \\ \hline
        asymmMktGame\_N2\_94.json & 2 & 0.0691 & 0.1286 & 5 & 7 \\ \hline
        asymmMktGame\_N2\_95.json & 2 & 0.0672 & 0.0961 & 4 & 5 \\ \hline
        asymmMktGame\_N2\_96.json & 2 & 0.0915 & 0.1329 & 5 & 6 \\ \hline
        asymmMktGame\_N2\_97.json & 2 & 0.0666 & 0.1194 & 5 & 7 \\ \hline
        asymmMktGame\_N2\_98.json & 2 & 0.0833 & 0.0997 & 6 & 6 \\ \hline
        asymmMktGame\_N2\_99.json & 2 & 0.0578 & 0.1072 & 4 & 6 \\ \hline
        asymmMktGame\_N2\_100.json & 2 & 0.0777 & 0.0895 & 5 & 5 \\ \hline
        asymmMktGame\_N3\_1.json & 3 & 1.0087 & 7.9041 & 5 & 8 \\ \hline
        asymmMktGame\_N3\_2.json & 3 & 0.2208 & 11.3431 & 6 & 9 \\ \hline
        asymmMktGame\_N3\_3.json & 3 & 0.1698 & 7.9446 & 7 & 9 \\ \hline
        asymmMktGame\_N3\_4.json & 3 & 0.0948 & 2.2528 & 5 & 7 \\ \hline
        asymmMktGame\_N3\_5.json & 3 & 0.1604 & 3.1355 & 5 & 7 \\ \hline
        asymmMktGame\_N3\_6.json & 3 & 0.2268 & 2.3477 & 4 & 6 \\ \hline
        asymmMktGame\_N3\_7.json & 3 & 0.2436 & 18.098 & 6 & 8 \\ \hline
        asymmMktGame\_N3\_8.json & 3 & 0.2096 & 0.9865 & 6 & 6 \\ \hline
        asymmMktGame\_N3\_9.json & 3 & 0.1306 & 1.5388 & 5 & 7 \\ \hline
        asymmMktGame\_N3\_10.json & 3 & 0.1033 & 1.9332 & 4 & 7 \\ \hline
        asymmMktGame\_N3\_11.json & 3 & 0.1512 & 1.244 & 6 & 6 \\ \hline
        asymmMktGame\_N3\_12.json & 3 & 0.1609 & 3.2307 & 5 & 8 \\ \hline
        asymmMktGame\_N3\_13.json & 3 & 0.1288 & 28.0043 & 7 & 8 \\ \hline
        asymmMktGame\_N3\_14.json & 3 & 0.1057 & 1.6685 & 5 & 7 \\ \hline
        asymmMktGame\_N3\_15.json & 3 & 0.1312 & 3.6755 & 6 & 8 \\ \hline
        asymmMktGame\_N3\_16.json & 3 & 0.1393 & 3.8776 & 6 & 7 \\ \hline
        asymmMktGame\_N3\_17.json & 3 & 0.2839 & 40.2936 & 6 & 9 \\ \hline
        asymmMktGame\_N3\_18.json & 3 & 0.1229 & 0.9355 & 5 & 6 \\ \hline
        asymmMktGame\_N3\_19.json & 3 & 0.1026 & 6.3381 & 6 & 9 \\ \hline
        asymmMktGame\_N3\_20.json & 3 & 0.0891 & 0.9412 & 5 & 6 \\ \hline
        asymmMktGame\_N3\_21.json & 3 & 0.3346 & 1.9141 & 6 & 7 \\ \hline
        asymmMktGame\_N3\_22.json & 3 & 0.1294 & 1.8399 & 6 & 7 \\ \hline
        asymmMktGame\_N3\_23.json & 3 & 0.0985 & 1.0896 & 5 & 6 \\ \hline
        asymmMktGame\_N3\_24.json & 3 & 0.1317 & 1.7056 & 5 & 7 \\ \hline
        asymmMktGame\_N3\_25.json & 3 & 0.0978 & 0.4013 & 5 & 5 \\ \hline
        asymmMktGame\_N3\_26.json & 3 & 0.0913 & 1.5011 & 5 & 6 \\ \hline
        asymmMktGame\_N3\_27.json & 3 & 0.282 & 1.5047 & 5 & 6 \\ \hline
        asymmMktGame\_N3\_28.json & 3 & 0.1149 & 1.2748 & 6 & 6 \\ \hline
        asymmMktGame\_N3\_29.json & 3 & 0.2207 & 2.1117 & 6 & 7 \\ \hline
        asymmMktGame\_N3\_30.json & 3 & 0.1084 & 1.4443 & 4 & 6 \\ \hline
        asymmMktGame\_N3\_31.json & 3 & 0.1686 & 1.3235 & 5 & 6 \\ \hline
        asymmMktGame\_N3\_32.json & 3 & 0.2579 & 3.4047 & 7 & 7 \\ \hline
        asymmMktGame\_N3\_33.json & 3 & 0.1741 & 2.7879 & 5 & 7 \\ \hline
        asymmMktGame\_N3\_34.json & 3 & 0.0825 & 1.5759 & 4 & 7 \\ \hline
        asymmMktGame\_N3\_35.json & 3 & 0.0834 & 0.8216 & 4 & 6 \\ \hline
        asymmMktGame\_N3\_36.json & 3 & 0.1004 & 1.7345 & 5 & 7 \\ \hline
        asymmMktGame\_N3\_37.json & 3 & 0.3519 & 29.3943 & 6 & 10 \\ \hline
        asymmMktGame\_N3\_38.json & 3 & 0.2801 & 4.3804 & 6 & 7 \\ \hline
        asymmMktGame\_N3\_39.json & 3 & 0.2861 & 4.7138 & 5 & 7 \\ \hline
        asymmMktGame\_N3\_40.json & 3 & 0.2381 & 3.3041 & 6 & 7 \\ \hline
        asymmMktGame\_N3\_41.json & 3 & 0.4278 & 3.7807 & 8 & 7 \\ \hline
        asymmMktGame\_N3\_42.json & 3 & 0.1975 & 1.6091 & 4 & 6 \\ \hline
        asymmMktGame\_N3\_43.json & 3 & 0.0873 & 1.2337 & 4 & 6 \\ \hline
        asymmMktGame\_N3\_44.json & 3 & 0.1617 & 1.1925 & 5 & 6 \\ \hline
        asymmMktGame\_N3\_45.json & 3 & 0.1089 & 2.6424 & 6 & 7 \\ \hline
        asymmMktGame\_N3\_46.json & 3 & 0.3799 & 1.672 & 5 & 6 \\ \hline
        asymmMktGame\_N3\_47.json & 3 & 0.0888 & 0.8927 & 5 & 6 \\ \hline
        asymmMktGame\_N3\_48.json & 3 & 0.1042 & 0.963 & 6 & 6 \\ \hline
        asymmMktGame\_N3\_49.json & 3 & 0.0918 & 0.9928 & 5 & 6 \\ \hline
        asymmMktGame\_N3\_50.json & 3 & 0.0893 & 1.3069 & 5 & 7 \\ \hline
        asymmMktGame\_N3\_51.json & 3 & 0.1509 & 1.6395 & 6 & 7 \\ \hline
        asymmMktGame\_N3\_52.json & 3 & 0.085 & 0.426 & 4 & 5 \\ \hline
        asymmMktGame\_N3\_53.json & 3 & 0.1463 & 2.2021 & 6 & 7 \\ \hline
        asymmMktGame\_N3\_54.json & 3 & 0.1317 & 0.9569 & 5 & 6 \\ \hline
        asymmMktGame\_N3\_55.json & 3 & 0.1208 & 6.0969 & 6 & 9 \\ \hline
        asymmMktGame\_N3\_56.json & 3 & 0.103 & 1.7829 & 5 & 7 \\ \hline
        asymmMktGame\_N3\_57.json & 3 & 0.1041 & 1.5699 & 5 & 7 \\ \hline
        asymmMktGame\_N3\_58.json & 3 & 0.106 & 1.1478 & 5 & 6 \\ \hline
        asymmMktGame\_N3\_59.json & 3 & 0.1018 & 1.7604 & 5 & 7 \\ \hline
        asymmMktGame\_N3\_60.json & 3 & 0.1208 & 0.8129 & 6 & 6 \\ \hline
        asymmMktGame\_N3\_61.json & 3 & 0.1012 & 1.9973 & 5 & 7 \\ \hline
        asymmMktGame\_N3\_62.json & 3 & 0.1019 & 5.97 & 5 & 8 \\ \hline
        asymmMktGame\_N3\_63.json & 3 & 0.112 & 0.949 & 4 & 6 \\ \hline
        asymmMktGame\_N3\_64.json & 3 & 0.1225 & 0.9235 & 6 & 6 \\ \hline
        asymmMktGame\_N3\_65.json & 3 & 0.1228 & 1.9663 & 6 & 7 \\ \hline
        asymmMktGame\_N3\_66.json & 3 & 0.1376 & 1.951 & 7 & 7 \\ \hline
        asymmMktGame\_N3\_67.json & 3 & 0.1179 & 1.9159 & 6 & 7 \\ \hline
        asymmMktGame\_N3\_68.json & 3 & 0.1022 & 2.0993 & 5 & 7 \\ \hline
        asymmMktGame\_N3\_69.json & 3 & 0.087 & 0.9505 & 4 & 6 \\ \hline
        asymmMktGame\_N3\_70.json & 3 & 0.0856 & 1.2578 & 4 & 6 \\ \hline
        asymmMktGame\_N3\_71.json & 3 & 0.107 & 1.0562 & 5 & 6 \\ \hline
        asymmMktGame\_N3\_72.json & 3 & 0.1204 & 3.5317 & 6 & 8 \\ \hline
        asymmMktGame\_N3\_73.json & 3 & 0.116 & 5.5965 & 6 & 8 \\ \hline
        asymmMktGame\_N3\_74.json & 3 & 0.1345 & 0.9562 & 5 & 6 \\ \hline
        asymmMktGame\_N3\_75.json & 3 & 0.1209 & 1.2547 & 6 & 6 \\ \hline
        asymmMktGame\_N3\_76.json & 3 & 0.1075 & 2.2855 & 5 & 7 \\ \hline
        asymmMktGame\_N3\_77.json & 3 & 0.1382 & 2.4462 & 7 & 7 \\ \hline
        asymmMktGame\_N3\_78.json & 3 & 0.1395 & 12.7006 & 7 & 8 \\ \hline
        asymmMktGame\_N3\_79.json & 3 & 0.1297 & 1.935 & 6 & 7 \\ \hline
        asymmMktGame\_N3\_80.json & 3 & 0.1022 & 0.9593 & 5 & 6 \\ \hline
        asymmMktGame\_N3\_81.json & 3 & 0.1037 & 1.5719 & 5 & 7 \\ \hline
        asymmMktGame\_N3\_82.json & 3 & 0.0874 & 0.9517 & 4 & 6 \\ \hline
        asymmMktGame\_N3\_83.json & 3 & 0.1048 & 1.5823 & 5 & 7 \\ \hline
        asymmMktGame\_N3\_84.json & 3 & 0.1058 & 1.9695 & 5 & 7 \\ \hline
        asymmMktGame\_N3\_85.json & 3 & 0.1083 & 0.9189 & 5 & 6 \\ \hline
        asymmMktGame\_N3\_86.json & 3 & 0.1064 & 1.7134 & 5 & 7 \\ \hline
        asymmMktGame\_N3\_87.json & 3 & 0.1405 & 2.083 & 7 & 7 \\ \hline
        asymmMktGame\_N3\_88.json & 3 & 0.093 & 0.872 & 4 & 6 \\ \hline
        asymmMktGame\_N3\_89.json & 3 & 0.1085 & 1.7017 & 5 & 7 \\ \hline
        asymmMktGame\_N3\_90.json & 3 & 0.1055 & 1.8302 & 5 & 7 \\ \hline
        asymmMktGame\_N3\_91.json & 3 & 0.1064 & 1.6668 & 5 & 7 \\ \hline
        asymmMktGame\_N3\_92.json & 3 & 0.1089 & 1.5687 & 5 & 7 \\ \hline
        asymmMktGame\_N3\_93.json & 3 & 0.1205 & 0.9254 & 6 & 6 \\ \hline
        asymmMktGame\_N3\_94.json & 3 & 0.1025 & 0.9178 & 5 & 6 \\ \hline
        asymmMktGame\_N3\_95.json & 3 & 0.1005 & 1.8297 & 5 & 7 \\ \hline
        asymmMktGame\_N3\_96.json & 3 & 0.1029 & 2.0787 & 5 & 7 \\ \hline
        asymmMktGame\_N3\_97.json & 3 & 0.1631 & 2.0128 & 7 & 7 \\ \hline
        asymmMktGame\_N3\_98.json & 3 & 0.1046 & 1.9033 & 5 & 7 \\ \hline
        asymmMktGame\_N3\_99.json & 3 & 0.1221 & 3.2013 & 6 & 8 \\ \hline
        asymmMktGame\_N3\_100.json & 3 & 0.1318 & 1.4718 & 5 & 7 \\ \hline
        asymmMktGame\_N4\_1.json & 4 & 0.2778 & Time Limit & 7 & 4 \\ \hline
        asymmMktGame\_N4\_2.json & 4 & 0.6593 & Time Limit & 7 & 4 \\ \hline
        asymmMktGame\_N4\_3.json & 4 & 1.8977 & Time Limit & 7 & 4 \\ \hline
        asymmMktGame\_N4\_4.json & 4 & 0.7597 & Time Limit & 7 & 4 \\ \hline
        asymmMktGame\_N4\_5.json & 4 & 0.569 & Time Limit & 6 & 4 \\ \hline
        asymmMktGame\_N4\_6.json & 4 & 0.6259 & Time Limit & 6 & 4 \\ \hline
        asymmMktGame\_N4\_7.json & 4 & 0.2087 & Time Limit & 5 & 4 \\ \hline
        asymmMktGame\_N4\_8.json & 4 & 0.2815 & Time Limit & 6 & 4 \\ \hline
        asymmMktGame\_N4\_9.json & 4 & 0.2552 & Time Limit & 7 & 4 \\ \hline
        asymmMktGame\_N4\_10.json & 4 & 1.3711 & Time Limit & 5 & 4 \\ \hline
        asymmMktGame\_N4\_11.json & 4 & 0.3437 & Time Limit & 6 & 4 \\ \hline
        asymmMktGame\_N4\_12.json & 4 & 0.2211 & Time Limit & 8 & 4 \\ \hline
        asymmMktGame\_N4\_13.json & 4 & 0.6027 & Time Limit & 10 & 4 \\ \hline
        asymmMktGame\_N4\_14.json & 4 & 0.2676 & Time Limit & 6 & 4 \\ \hline
        asymmMktGame\_N4\_15.json & 4 & 0.1957 & Time Limit & 7 & 4 \\ \hline
        asymmMktGame\_N4\_16.json & 4 & 0.6172 & Time Limit & 7 & 4 \\ \hline
        asymmMktGame\_N4\_17.json & 4 & 1.2152 & Time Limit & 10 & 4 \\ \hline
        asymmMktGame\_N4\_18.json & 4 & 0.134 & Time Limit & 6 & 4 \\ \hline
        asymmMktGame\_N4\_19.json & 4 & 1.1588 & Time Limit & 5 & 4 \\ \hline
        asymmMktGame\_N4\_20.json & 4 & 0.5124 & Time Limit & 8 & 4 \\ \hline
        asymmMktGame\_N5\_1.json & 5 & 1.3986 & Time Limit & 8 & 3 \\ \hline
        asymmMktGame\_N5\_2.json & 5 & 0.2007 & Time Limit & 7 & 2 \\ \hline
        asymmMktGame\_N5\_3.json & 5 & 0.5717 & Time Limit & 7 & 2 \\ \hline
        asymmMktGame\_N5\_4.json & 5 & 0.4102 & Time Limit & 7 & 2 \\ \hline
        asymmMktGame\_N5\_5.json & 5 & 0.2448 & Time Limit & 7 & 2 \\ \hline
        asymmMktGame\_N5\_6.json & 5 & 0.3572 & Time Limit & 7 & 2 \\ \hline
        asymmMktGame\_N5\_7.json & 5 & 0.2669 & Time Limit & 7 & 2 \\ \hline
        asymmMktGame\_N5\_8.json & 5 & 0.387 & Time Limit & 7 & 2 \\ \hline
        asymmMktGame\_N5\_9.json & 5 & 0.3012 & Time Limit & 7 & 2 \\ \hline
        asymmMktGame\_N5\_10.json & 5 & 0.5347 & Time Limit & 8 & 2 \\ \hline
        asymmMktGame\_N5\_11.json & 5 & 1.5319 & Time Limit & 9 & 2 \\ \hline
        asymmMktGame\_N5\_12.json & 5 & 1.4644 & Time Limit & 8 & 2 \\ \hline
        asymmMktGame\_N5\_13.json & 5 & 0.3636 & Time Limit & 6 & 2 \\ \hline
        asymmMktGame\_N5\_14.json & 5 & 1.0323 & Time Limit & 8 & 2 \\ \hline
        asymmMktGame\_N5\_15.json & 5 & 1.6267 & Time Limit & 7 & 2 \\ \hline
        asymmMktGame\_N5\_16.json & 5 & 0.9383 & Time Limit & 7 & 2 \\ \hline
        asymmMktGame\_N5\_17.json & 5 & 0.8792 & Time Limit & 8 & 2 \\ \hline
        asymmMktGame\_N5\_18.json & 5 & 0.7827 & Time Limit & 8 & 2 \\ \hline
        asymmMktGame\_N5\_19.json & 5 & 0.8046 & Time Limit & 6 & 2 \\ \hline
        asymmMktGame\_N5\_20.json & 5 & 0.8578 & Time Limit & 6 & 2 \\ \hline
    \end{longtable}
\end{tiny}

\subsection{Random instances}
\begin{tiny}
    \begin{longtable}{|rccccc|}
\hline
                \textbf{Instance name} & \textbf{nPlay} & \textbf{$t_{BR}$} & \textbf{$t_{SGM}$}
				& \textbf{$k_{BR}$} & \textbf{$k_{SGM}$} \\ \midrule
			\endhead
        \hline
randGame\_N2\_1.json & 2 & 0.2598 & 0.247 & 7 & 9 \\ \hline
        randGame\_N2\_2.json & 2 & 0.0371 & 0.0618 & 3 & 4 \\ \hline
        randGame\_N2\_3.json & 2 & 0.0612 & 0.153 & 4 & 7 \\ \hline
        randGame\_N2\_4.json & 2 & 0.0537 & 0.1517 & 4 & 6 \\ \hline
        randGame\_N2\_5.json & 2 & 0.0561 & 0.1682 & 5 & 8 \\ \hline
        randGame\_N2\_6.json & 2 & 0.0619 & 0.2017 & 6 & 11 \\ \hline
        randGame\_N2\_7.json & 2 & 0.0357 & 0.0673 & 3 & 4 \\ \hline
        randGame\_N2\_8.json & 2 & 0.0625 & 0.0484 & 4 & 3 \\ \hline
        randGame\_N2\_9.json & 2 & 0.0632 & 0.2244 & 5 & 9 \\ \hline
        randGame\_N2\_10.json & 2 & 0.0519 & 0.0677 & 3 & 3 \\ \hline
        randGame\_N2\_11.json & 2 & 0.3077 & 0.5071 & 7 & 9 \\ \hline
        randGame\_N2\_12.json & 2 & 0.0924 & 0.253 & 5 & 7 \\ \hline
        randGame\_N2\_13.json & 2 & 0.1164 & 0.2388 & 3 & 4 \\ \hline
        randGame\_N2\_14.json & 2 & 0.0988 & 0.0907 & 3 & 4 \\ \hline
        randGame\_N2\_15.json & 2 & 0.0524 & 0.0878 & 4 & 5 \\ \hline
        randGame\_N2\_16.json & 2 & 0.0412 & 0.0613 & 3 & 4 \\ \hline
        randGame\_N2\_17.json & 2 & 0.0627 & 0.141 & 5 & 3 \\ \hline
        randGame\_N2\_18.json & 2 & 0.1867 & 0.2277 & 5 & 8 \\ \hline
        randGame\_N2\_19.json & 2 & 0.1326 & 0.1059 & 3 & 5 \\ \hline
        randGame\_N2\_20.json & 2 & 0.0559 & 0.1085 & 4 & 5 \\ \hline
        randGame\_N2\_21.json & 2 & 0.0958 & 0.312 & 2 & 2 \\ \hline
        randGame\_N2\_22.json & 2 & 0.2786 & 0.3681 & 4 & 6 \\ \hline
        randGame\_N2\_23.json & 2 & 0.1528 & 0.1753 & 3 & 4 \\ \hline
        randGame\_N2\_24.json & 2 & 0.1177 & 0.1633 & 3 & 4 \\ \hline
        randGame\_N2\_25.json & 2 & 0.213 & 0.702 & 2 & 3 \\ \hline
        randGame\_N2\_26.json & 2 & 0.1594 & 0.1681 & 2 & 3 \\ \hline
        randGame\_N2\_27.json & 2 & 0.1173 & 0.3165 & 5 & 8 \\ \hline
        randGame\_N2\_28.json & 2 & 0.1288 & 0.2097 & 3 & 5 \\ \hline
        randGame\_N2\_29.json & 2 & 0.2005 & 0.554 & 4 & 6 \\ \hline
        randGame\_N2\_30.json & 2 & 0.2378 & 0.5475 & 3 & 3 \\ \hline
        randGame\_N2\_31.json & 2 & 0.3077 & 0.4836 & 3 & 3 \\ \hline
        randGame\_N2\_32.json & 2 & 0.4779 & 1.0128 & 4 & 6 \\ \hline
        randGame\_N2\_33.json & 2 & 0.819 & 1.2814 & 3 & 3 \\ \hline
        randGame\_N2\_34.json & 2 & 0.3761 & 0.6844 & 3 & 4 \\ \hline
        randGame\_N2\_35.json & 2 & 0.4412 & 0.8147 & 3 & 4 \\ \hline
        randGame\_N2\_36.json & 2 & 0.4627 & 0.7551 & 4 & 7 \\ \hline
        randGame\_N2\_37.json & 2 & 0.5962 & 0.3668 & 3 & 4 \\ \hline
        randGame\_N2\_38.json & 2 & 0.4831 & 0.2352 & 2 & 3 \\ \hline
        randGame\_N2\_39.json & 2 & 0.2263 & 0.1888 & 3 & 3 \\ \hline
        randGame\_N2\_40.json & 2 & 0.1835 & 0.1618 & 3 & 3 \\ \hline
        randGame\_N2\_41.json & 2 & 0.242 & 0.3951 & 2 & 2 \\ \hline
        randGame\_N2\_42.json & 2 & 0.5582 & 0.8329 & 3 & 4 \\ \hline
        randGame\_N2\_43.json & 2 & 0.2537 & 0.5019 & 2 & 3 \\ \hline
        randGame\_N2\_44.json & 2 & 0.3777 & 0.4589 & 3 & 3 \\ \hline
        randGame\_N2\_45.json & 2 & 0.6063 & 0.9533 & 3 & 4 \\ \hline
        randGame\_N2\_46.json & 2 & 0.3217 & 0.5556 & 2 & 2 \\ \hline
        randGame\_N2\_47.json & 2 & 1.1087 & 1.3295 & 5 & 6 \\ \hline
        randGame\_N2\_48.json & 2 & 0.4534 & 0.7185 & 2 & 3 \\ \hline
        randGame\_N2\_49.json & 2 & 0.531 & 1.0884 & 3 & 5 \\ \hline
        randGame\_N2\_50.json & 2 & Num Err & Num Err & 2 & 2 \\ \hline
        randGame\_N2\_51.json & 2 & Num Err & Num Err & 2 & 2 \\ \hline
        randGame\_N2\_52.json & 2 & 0.4636 & 0.5617 & 3 & 4 \\ \hline
        randGame\_N2\_53.json & 2 & 0.2575 & 0.3534 & 3 & 3 \\ \hline
        randGame\_N2\_54.json & 2 & 0.1825 & 0.2869 & 2 & 3 \\ \hline
        randGame\_N2\_55.json & 2 & 0.0845 & 0.2536 & 2 & 3 \\ \hline
        randGame\_N2\_56.json & 2 & 0.2336 & 0.4225 & 2 & 3 \\ \hline
        randGame\_N2\_57.json & 2 & 0.1975 & 0.2206 & 2 & 2 \\ \hline
        randGame\_N2\_58.json & 2 & 0.5248 & 0.6395 & 3 & 4 \\ \hline
        randGame\_N2\_59.json & 2 & 0.4451 & 0.4754 & 3 & 4 \\ \hline
        randGame\_N2\_60.json & 2 & 0.6563 & 0.2773 & 2 & 2 \\ \hline
        randGame\_N2\_61.json & 2 & 0.4227 & 0.6788 & 2 & 3 \\ \hline
        randGame\_N2\_62.json & 2 & Num Err & Num Err & 2 & 2 \\ \hline
        randGame\_N2\_63.json & 2 & 1.1397 & 1.1299 & 3 & 4 \\ \hline
        randGame\_N2\_64.json & 2 & 0.457 & 0.4256 & 2 & 2 \\ \hline
        randGame\_N2\_65.json & 2 & 0.8943 & 0.8541 & 3 & 3 \\ \hline
        randGame\_N2\_66.json & 2 & 0.7664 & 0.7313 & 2 & 2 \\ \hline
        randGame\_N2\_67.json & 2 & 1.4633 & 8.336 & 3 & 4 \\ \hline
        randGame\_N2\_68.json & 2 & 0.8547 & 1.3668 & 3 & 5 \\ \hline
        randGame\_N2\_69.json & 2 & 0.6389 & 0.6517 & 3 & 2 \\ \hline
        randGame\_N2\_70.json & 2 & 1.3068 & 1.307 & 3 & 4 \\ \hline
        randGame\_N2\_71.json & 2 & 0.6671 & 0.9354 & 3 & 4 \\ \hline
        randGame\_N2\_72.json & 2 & 0.4663 & 0.4703 & 3 & 3 \\ \hline
        randGame\_N2\_73.json & 2 & 0.4123 & 0.697 & 2 & 3 \\ \hline
        randGame\_N2\_74.json & 2 & 0.2197 & 0.047 & 2 & 1 \\ \hline
        randGame\_N2\_75.json & 2 & 0.6426 & 0.7004 & 3 & 4 \\ \hline
        randGame\_N2\_76.json & 2 & 0.4131 & 0.3586 & 3 & 3 \\ \hline
        randGame\_N2\_77.json & 2 & 0.4846 & 0.686 & 2 & 3 \\ \hline
        randGame\_N2\_78.json & 2 & 0.6591 & 0.5017 & 4 & 3 \\ \hline
        randGame\_N2\_79.json & 2 & 0.6708 & 0.6222 & 3 & 3 \\ \hline
        randGame\_N2\_80.json & 2 & 0.7022 & 0.9953 & 3 & 4 \\ \hline
        randGame\_N2\_81.json & 2 & 0.3859 & 0.4403 & 2 & 2 \\ \hline
        randGame\_N2\_82.json & 2 & 0.7183 & 0.8864 & 3 & 3 \\ \hline
        randGame\_N2\_83.json & 2 & 6.5218 & 13.6443 & 3 & 4 \\ \hline
        randGame\_N2\_84.json & 2 & 23.0611 & 12.4179 & 3 & 4 \\ \hline
        randGame\_N2\_85.json & 2 & 1.0975 & 0.8735 & 3 & 3 \\ \hline
        randGame\_N2\_86.json & 2 & 17.0679 & 6.7586 & 3 & 3 \\ \hline
        randGame\_N2\_87.json & 2 & 1.3737 & 1.0909 & 3 & 3 \\ \hline
        randGame\_N2\_88.json & 2 & 0.9864 & 4.3091 & 3 & 3 \\ \hline
        randGame\_N2\_89.json & 2 & 1.1663 & 1.4746 & 3 & 4 \\ \hline
        randGame\_N2\_90.json & 2 & 0.6933 & 0.76 & 2 & 2 \\ \hline
        randGame\_N2\_91.json & 2 & 0.2959 & 0.1681 & 2 & 1 \\ \hline
        randGame\_N2\_92.json & 2 & 0.6501 & 0.5357 & 2 & 2 \\ \hline
        randGame\_N2\_93.json & 2 & 1.3736 & 2.1506 & 3 & 4 \\ \hline
        randGame\_N2\_94.json & 2 & 127.1435 & 163.9926 & 2 & 3 \\ \hline
        randGame\_N2\_95.json & 2 & 1.478 & 1.645 & 3 & 3 \\ \hline
        randGame\_N2\_96.json & 2 & 0.5112 & 0.7518 & 2 & 3 \\ \hline
        randGame\_N2\_97.json & 2 & 1.5585 & 1.7865 & 4 & 5 \\ \hline
        randGame\_N2\_98.json & 2 & 0.7386 & 1.0407 & 3 & 3 \\ \hline
        randGame\_N2\_99.json & 2 & 0.912 & 1.1804 & 3 & 4 \\ \hline
        randGame\_N2\_100.json & 2 & 1.1353 & 1.6631 & 3 & 4 \\ \hline
        randGame\_N3\_1.json & 3 & 0.4414 & 43.9727 & 6 & 10 \\ \hline
        randGame\_N3\_2.json & 3 & 0.0767 & 0.5908 & 4 & 6 \\ \hline
        randGame\_N3\_3.json & 3 & 0.0833 & 0.5658 & 4 & 7 \\ \hline
        randGame\_N3\_4.json & 3 & 0.1178 & 1.8644 & 6 & 8 \\ \hline
        randGame\_N3\_5.json & 3 & 0.1101 & 3.3098 & 7 & 8 \\ \hline
        randGame\_N3\_6.json & 3 & 0.1083 & 1.4666 & 8 & 10 \\ \hline
        randGame\_N3\_7.json & 3 & 0.0859 & 1.1735 & 5 & 7 \\ \hline
        randGame\_N3\_8.json & 3 & 0.1162 & 0.7654 & 4 & 7 \\ \hline
        randGame\_N3\_9.json & 3 & 0.089 & 1.6733 & 5 & 7 \\ \hline
        randGame\_N3\_10.json & 3 & 0.0976 & 1.4009 & 6 & 7 \\ \hline
        randGame\_N3\_11.json & 3 & 0.0941 & 1.6009 & 5 & 7 \\ \hline
        randGame\_N3\_12.json & 3 & 0.1479 & 0.4326 & 5 & 4 \\ \hline
        randGame\_N3\_13.json & 3 & 0.1063 & 0.6982 & 5 & 6 \\ \hline
        randGame\_N3\_14.json & 3 & 0.0775 & 5.6645 & 4 & 9 \\ \hline
        randGame\_N3\_15.json & 3 & 0.0722 & 5.1134 & 5 & 9 \\ \hline
        randGame\_N3\_16.json & 3 & 0.2548 & 6.3728 & 6 & 9 \\ \hline
        randGame\_N3\_17.json & 3 & 0.0723 & 0.2241 & 4 & 5 \\ \hline
        randGame\_N3\_18.json & 3 & 0.0951 & 3.5196 & 5 & 8 \\ \hline
        randGame\_N3\_19.json & 3 & 0.1887 & 2.4177 & 5 & 8 \\ \hline
        randGame\_N3\_20.json & 3 & 0.0885 & 3.1656 & 5 & 9 \\ \hline
        randGame\_N3\_21.json & 3 & 0.1841 & 0.1744 & 4 & 3 \\ \hline
        randGame\_N3\_22.json & 3 & 0.1342 & 12.4326 & 3 & 100 \\ \hline
        randGame\_N3\_23.json & 3 & 0.1406 & 0.2856 & 3 & 4 \\ \hline
        randGame\_N3\_24.json & 3 & 0.1077 & 7.009 & 3 & 100 \\ \hline
        randGame\_N3\_25.json & 3 & 0.2891 & 1.0321 & 6 & 6 \\ \hline
        randGame\_N3\_26.json & 3 & 0.398 & 0.748 & 3 & 5 \\ \hline
        randGame\_N3\_27.json & 3 & 0.1585 & 0.1789 & 2 & 2 \\ \hline
        randGame\_N3\_28.json & 3 & 0.2319 & 0.9005 & 4 & 6 \\ \hline
        randGame\_N3\_29.json & 3 & 0.1818 & 0.2371 & 3 & 3 \\ \hline
        randGame\_N3\_30.json & 3 & 0.1572 & 0.3287 & 3 & 4 \\ \hline
        randGame\_N3\_31.json & 3 & 0.2052 & 0.3114 & 4 & 4 \\ \hline
        randGame\_N3\_32.json & 3 & 0.323 & 0.6602 & 5 & 5 \\ \hline
        randGame\_N3\_33.json & 3 & 0.2034 & 0.6383 & 4 & 5 \\ \hline
        randGame\_N3\_34.json & 3 & 0.0957 & 0.1498 & 2 & 3 \\ \hline
        randGame\_N3\_35.json & 3 & 0.2592 & 0.3239 & 4 & 4 \\ \hline
        randGame\_N3\_36.json & 3 & 0.069 & 0.1114 & 2 & 3 \\ \hline
        randGame\_N3\_37.json & 3 & 0.2911 & 0.3391 & 3 & 3 \\ \hline
        randGame\_N3\_38.json & 3 & 0.2193 & 0.5122 & 3 & 4 \\ \hline
        randGame\_N3\_39.json & 3 & 0.2024 & 0.9445 & 5 & 6 \\ \hline
        randGame\_N3\_40.json & 3 & 0.1593 & 0.9972 & 3 & 6 \\ \hline
        randGame\_N3\_41.json & 3 & 0.292 & 0.5197 & 3 & 3 \\ \hline
        randGame\_N3\_42.json & 3 & 0.9592 & 1.0196 & 4 & 4 \\ \hline
        randGame\_N3\_43.json & 3 & 0.4345 & 0.6816 & 2 & 4 \\ \hline
        randGame\_N3\_44.json & 3 & 0.6823 & 1.1879 & 3 & 5 \\ \hline
        randGame\_N3\_45.json & 3 & 0.2281 & 0.5023 & 2 & 2 \\ \hline
        randGame\_N3\_46.json & 3 & 0.574 & 1.8442 & 3 & 5 \\ \hline
        randGame\_N3\_47.json & 3 & 0.9579 & 0.9095 & 3 & 3 \\ \hline
        randGame\_N3\_48.json & 3 & 1.3496 & 1.2831 & 4 & 4 \\ \hline
        randGame\_N3\_49.json & 3 & 0.1807 & 0.2735 & 2 & 3 \\ \hline
        randGame\_N3\_50.json & 3 & 0.769 & 1.1428 & 4 & 5 \\ \hline
        randGame\_N3\_51.json & 3 & 0.8608 & 1.3787 & 4 & 5 \\ \hline
        randGame\_N3\_52.json & 3 & 0.9488 & 1.1685 & 5 & 5 \\ \hline
        randGame\_N3\_53.json & 3 & 0.89 & 1.2591 & 4 & 5 \\ \hline
        randGame\_N3\_54.json & 3 & 0.395 & 0.5041 & 2 & 3 \\ \hline
        randGame\_N3\_55.json & 3 & 0.7196 & 1.1708 & 3 & 4 \\ \hline
        randGame\_N3\_56.json & 3 & 0.5929 & 1.1109 & 3 & 5 \\ \hline
        randGame\_N3\_57.json & 3 & 0.6146 & 0.7705 & 3 & 4 \\ \hline
        randGame\_N3\_58.json & 3 & 0.5265 & 0.9452 & 3 & 5 \\ \hline
        randGame\_N3\_59.json & 3 & 0.3913 & 0.5479 & 2 & 3 \\ \hline
        randGame\_N3\_60.json & 3 & 0.6109 & 0.9479 & 3 & 4 \\ \hline
        randGame\_N3\_61.json & 3 & 0.889 & 2.5543 & 4 & 7 \\ \hline
        randGame\_N3\_62.json & 3 & 0.8091 & 1.1602 & 3 & 4 \\ \hline
        randGame\_N3\_63.json & 3 & 1.5175 & 2.9208 & 5 & 7 \\ \hline
        randGame\_N3\_64.json & 3 & 0.9724 & 1.2411 & 3 & 4 \\ \hline
        randGame\_N3\_65.json & 3 & 1.0193 & 2.6233 & 3 & 7 \\ \hline
        randGame\_N3\_66.json & 3 & 0.6746 & 0.633 & 3 & 3 \\ \hline
        randGame\_N3\_67.json & 3 & 1.0796 & 1.2864 & 3 & 4 \\ \hline
        randGame\_N3\_68.json & 3 & 0.5718 & 0.8955 & 3 & 4 \\ \hline
        randGame\_N3\_69.json & 3 & 0.7779 & 1.3836 & 3 & 4 \\ \hline
        randGame\_N3\_70.json & 3 & 0.9842 & 0.9761 & 3 & 4 \\ \hline
        randGame\_N3\_71.json & 3 & 1.402 & 1.4921 & 3 & 4 \\ \hline
        randGame\_N3\_72.json & 3 & 0.6315 & 0.9409 & 3 & 3 \\ \hline
        randGame\_N3\_73.json & 3 & 1.3367 & 1.0137 & 3 & 3 \\ \hline
        randGame\_N3\_74.json & 3 & Num Err & Num Err & 2 & 2 \\ \hline
        randGame\_N3\_75.json & 3 & Num Err & Num Err & 2 & 2 \\ \hline
        randGame\_N3\_76.json & 3 & 1.1404 & 0.8586 & 3 & 3 \\ \hline
        randGame\_N3\_77.json & 3 & 1 & 1.8078 & 3 & 4 \\ \hline
        randGame\_N3\_78.json & 3 & Num Err & Num Err & 2 & 2 \\ \hline
        randGame\_N3\_79.json & 3 & 8.0092 & 11.9516 & 3 & 5 \\ \hline
        randGame\_N3\_80.json & 3 & 1.9248 & 2.4797 & 3 & 4 \\ \hline
        randGame\_N3\_81.json & 3 & 2.67 & 2.6686 & 3 & 3 \\ \hline
        randGame\_N3\_82.json & 3 & 11.8882 & 15.2893 & 3 & 3 \\ \hline
        randGame\_N3\_83.json & 3 & 1.9271 & 3.1359 & 3 & 4 \\ \hline
        randGame\_N3\_84.json & 3 & 383.3733 & 470.0893 & 3 & 4 \\ \hline
        randGame\_N3\_85.json & 3 & 1.84 & 12.6218 & 3 & 3 \\ \hline
        randGame\_N3\_86.json & 3 & 1.9534 & 80.4689 & 3 & 4 \\ \hline
        randGame\_N3\_87.json & 3 & 1.505 & 2.0335 & 3 & 4 \\ \hline
        randGame\_N3\_88.json & 3 & 26.5388 & 39.3477 & 2 & 2 \\ \hline
        randGame\_N3\_89.json & 3 & 51.1266 & 51.0833 & 3 & 4 \\ \hline
        randGame\_N3\_90.json & 3 & 7.7242 & 5.7968 & 4 & 4 \\ \hline
        randGame\_N3\_91.json & 3 & 1.472 & 2.2398 & 3 & 4 \\ \hline
        randGame\_N3\_92.json & 3 & 35.3787 & 47.6815 & 4 & 5 \\ \hline
        randGame\_N3\_93.json & 3 & 22.6921 & 24.7038 & 4 & 5 \\ \hline
        randGame\_N3\_94.json & 3 & 1.1923 & 1.11 & 3 & 3 \\ \hline
        randGame\_N3\_95.json & 3 & 174.7423 & 202.2808 & 3 & 3 \\ \hline
        randGame\_N3\_96.json & 3 & 0.5262 & 0.4992 & 2 & 2 \\ \hline
        randGame\_N3\_97.json & 3 & 7.234 & 7.9554 & 3 & 4 \\ \hline
        randGame\_N3\_98.json & 3 & 2.1551 & 3.6649 & 3 & 4 \\ \hline
        randGame\_N3\_99.json & 3 & 3.8018 & 26.256 & 3 & 4 \\ \hline
        randGame\_N3\_100.json & 3 & 2.2863 & 2.6739 & 3 & 4 \\ \hline
		randGame\_N4\_1.json & 4 & 0.2523 & 347.1214 & 5 & 3 \\ \hline
		randGame\_N4\_2.json & 4 & 0.1472 & 421.113 & 7 & 2 \\ \hline
		randGame\_N4\_3.json & 4 & 0.1072 & 322.7234 & 4 & 2 \\ \hline
        randGame\_N4\_4.json & 4 & 0.1323 & TL & 5 & 2 \\ \hline
        randGame\_N4\_5.json & 4 & 0.1137 & TL & 4 & 3 \\ \hline
        randGame\_N4\_6.json & 4 & 0.1164 & TL & 6 & 2 \\ \hline
		randGame\_N4\_7.json & 4 & 0.1332 & 362.011 & 5 & 3 \\ \hline
        randGame\_N4\_8.json & 4 & 0.1375 & TL & 6 & 3 \\ \hline
		randGame\_N4\_9.json & 4 & 0.1621 & 290.113 & 7 & 2 \\ \hline
        randGame\_N4\_10.json & 4 & 0.1313 & TL & 5 & 2 \\ \hline
        randGame\_N4\_11.json & 4 & 0.1369 & TL & 6 & 2 \\ \hline
        randGame\_N4\_12.json & 4 & 0.0853 & TL & 4 & 3 \\ \hline
        randGame\_N4\_13.json & 4 & 0.2833 & TL & 12 & 2 \\ \hline
		randGame\_N4\_14.json & 4 & 0.1111 & 367.1411 & 5 & 2 \\ \hline
        randGame\_N4\_15.json & 4 & 0.0986 & TL & 5 & 2 \\ \hline
        randGame\_N4\_16.json & 4 & 0.1539 & TL & 6 & 2 \\ \hline
        randGame\_N4\_17.json & 4 & 0.19 & TL & 6 & 2 \\ \hline
		randGame\_N4\_18.json & 4 & 0.1439 & 466.9406 & 6 & 3 \\ \hline
		randGame\_N4\_19.json & 4 & 0.1277 & 486.7372 & 6 & 2 \\ \hline
        randGame\_N4\_20.json & 4 & 0.148 & TL & 7 & 2 \\ \hline
        randGame\_N4\_21.json & 4 & 0.4612 & TL & 4 & 2 \\ \hline
        randGame\_N4\_22.json & 4 & 0.1491 & TL & 2 & 2 \\ \hline
        randGame\_N4\_23.json & 4 & 0.3453 & TL & 4 & 2 \\ \hline
        randGame\_N4\_24.json & 4 & 0.1886 & TL & 4 & 3 \\ \hline
        randGame\_N4\_25.json & 4 & 0.2342 & TL & 3 & 3 \\ \hline
        randGame\_N4\_26.json & 4 & 0.2962 & TL & 4 & 2 \\ \hline
        randGame\_N4\_27.json & 4 & 0.4339 & TL & 4 & 3 \\ \hline
        randGame\_N4\_28.json & 4 & 0.2758 & TL & 3 & 3 \\ \hline
        randGame\_N4\_29.json & 4 & 0.1675 & TL & 3 & 2 \\ \hline
        randGame\_N4\_30.json & 4 & 0.2961 & TL & 4 & 2 \\ \hline
        randGame\_N4\_31.json & 4 & 0.5576 & TL & 5 & 2 \\ \hline
        randGame\_N4\_32.json & 4 & 0.2459 & TL & 4 & 2 \\ \hline
        randGame\_N4\_33.json & 4 & 0.2455 & TL & 3 & 3 \\ \hline
        randGame\_N4\_34.json & 4 & 0.3037 & TL & 4 & 3 \\ \hline
        randGame\_N4\_35.json & 4 & 0.1934 & TL & 3 & 3 \\ \hline
        randGame\_N4\_36.json & 4 & 0.2401 & TL & 4 & 2 \\ \hline
        randGame\_N4\_37.json & 4 & 0.2566 & TL & 4 & 2 \\ \hline
        randGame\_N4\_38.json & 4 & 0.2514 & TL & 4 & 2 \\ \hline
        randGame\_N4\_39.json & 4 & 0.271 & TL & 4 & 2 \\ \hline
        randGame\_N4\_40.json & 4 & 0.2672 & TL & 4 & 3 \\ \hline
        randGame\_N4\_41.json & 4 & 1.1902 & TL & 5 & 2 \\ \hline
        randGame\_N4\_42.json & 4 & 0.6315 & TL & 3 & 2 \\ \hline
        randGame\_N4\_43.json & 4 & 0.9992 & TL & 4 & 2 \\ \hline
        randGame\_N4\_44.json & 4 & 0.3146 & TL & 2 & 3 \\ \hline
        randGame\_N4\_45.json & 4 & 0.9278 & TL & 4 & 2 \\ \hline
        randGame\_N4\_46.json & 4 & 0.8623 & TL & 4 & 2 \\ \hline
        randGame\_N4\_47.json & 4 & 1.1013 & TL & 3 & 2 \\ \hline
        randGame\_N4\_48.json & 4 & 0.9648 & TL & 4 & 3 \\ \hline
        randGame\_N4\_49.json & 4 & 0.7384 & TL & 3 & 2 \\ \hline
        randGame\_N4\_50.json & 4 & 0.4339 & TL & 3 & 2 \\ \hline
        randGame\_N4\_51.json & 4 & 1.1604 & TL & 4 & 2 \\ \hline
        randGame\_N4\_52.json & 4 & 1.1829 & TL & 3 & 3 \\ \hline
        randGame\_N4\_53.json & 4 & 0.7451 & TL & 3 & 2 \\ \hline
        randGame\_N4\_54.json & 4 & 0.5663 & TL & 3 & 2 \\ \hline
        randGame\_N4\_55.json & 4 & 1.1166 & TL & 4 & 2 \\ \hline
        randGame\_N4\_56.json & 4 & 0.7567 & TL & 3 & 2 \\ \hline
        randGame\_N4\_57.json & 4 & 0.841 & TL & 3 & 3 \\ \hline
        randGame\_N4\_58.json & 4 & 0.9415 & TL & 3 & 3 \\ \hline
        randGame\_N4\_59.json & 4 & 1.8539 & TL & 4 & 3 \\ \hline
        randGame\_N4\_60.json & 4 & 1.1799 & TL & 4 & 3 \\ \hline
        randGame\_N4\_61.json & 4 & 1.5686 & TL & 3 & 3 \\ \hline
        randGame\_N4\_62.json & 4 & 0.6972 & TL & 2 & 2 \\ \hline
        randGame\_N4\_63.json & 4 & 1.0152 & TL & 3 & 2 \\ \hline
        randGame\_N4\_64.json & 4 & 0.9883 & TL & 3 & 3 \\ \hline
        randGame\_N4\_65.json & 4 & 1.3565 & TL & 4 & 2 \\ \hline
        randGame\_N4\_66.json & 4 & 1.0687 & TL & 3 & 3 \\ \hline
        randGame\_N4\_67.json & 4 & 1.0449 & TL & 3 & 3 \\ \hline
        randGame\_N4\_68.json & 4 & Num Err & Num Err & 2 & 2 \\ \hline
        randGame\_N4\_69.json & 4 & 1.651 & TL & 3 & 2 \\ \hline
        randGame\_N4\_70.json & 4 & 1.8826 & TL & 3 & 3 \\ \hline
        randGame\_N4\_71.json & 4 & 1.7343 & TL & 3 & 3 \\ \hline
        randGame\_N4\_72.json & 4 & 1.8863 & TL & 3 & 2 \\ \hline
        randGame\_N4\_73.json & 4 & 1.9228 & TL & 3 & 3 \\ \hline
        randGame\_N4\_74.json & 4 & 1.6536 & TL & 3 & 2 \\ \hline
        randGame\_N4\_75.json & 4 & 1.5918 & TL & 3 & 2 \\ \hline
        randGame\_N4\_76.json & 4 & 1.0251 & TL & 3 & 2 \\ \hline
        randGame\_N4\_77.json & 4 & 0.7239 & TL & 2 & 3 \\ \hline
        randGame\_N4\_78.json & 4 & 1.7735 & TL & 4 & 2 \\ \hline
        randGame\_N4\_79.json & 4 & 1.2493 & TL & 3 & 3 \\ \hline
        randGame\_N4\_80.json & 4 & 1.3149 & TL & 2 & 2 \\ \hline
        randGame\_N4\_81.json & 4 & 12.9558 & TL & 3 & 2 \\ \hline
        randGame\_N4\_82.json & 4 & 1.7845 & TL & 3 & 2 \\ \hline
        randGame\_N4\_83.json & 4 & 1.4656 & TL & 2 & 2 \\ \hline
        randGame\_N4\_84.json & 4 & 2.8433 & TL & 3 & 2 \\ \hline
        randGame\_N4\_85.json & 4 & 2.7925 & TL & 4 & 2 \\ \hline
        randGame\_N4\_86.json & 4 & 36.5626 & TL & 4 & 2 \\ \hline
        randGame\_N4\_87.json & 4 & 7.8445 & TL & 3 & 2 \\ \hline
        randGame\_N4\_88.json & 4 & 17.3983 & TL & 4 & 2 \\ \hline
        randGame\_N4\_89.json & 4 & 1.8221 & TL & 3 & 2 \\ \hline
        randGame\_N4\_90.json & 4 & 0.7607 & TL & 2 & 2 \\ \hline
        randGame\_N4\_91.json & 4 & 22.8644 & TL & 3 & 3 \\ \hline
        randGame\_N4\_92.json & 4 & 2.1676 & TL & 4 & 3 \\ \hline
        randGame\_N4\_93.json & 4 & Num Err & Num Err & 2 & 2 \\ \hline
        randGame\_N4\_94.json & 4 & Num Err & Num Err & 2 & 2 \\ \hline
        randGame\_N4\_95.json & 4 & 77.0931 & TL & 4 & 3 \\ \hline
        randGame\_N4\_96.json & 4 & 1.1293 & TL & 2 & 2 \\ \hline
        randGame\_N4\_97.json & 4 & 233.0056 & TL & 4 & 2 \\ \hline
        randGame\_N4\_98.json & 4 & 130.3478 & TL & 3 & 2 \\ \hline
        randGame\_N4\_99.json & 4 & 17.6308 & TL & 3 & 2 \\ \hline
        randGame\_N4\_100.json & 4 & 66.078 & TL & 4 & 3 \\ \hline
        randGame\_N5\_1.json & 5 & 0.3292 & TL & 6 & 2 \\ \hline
        randGame\_N5\_2.json & 5 & 0.2012 & TL & 6 & 2 \\ \hline
        randGame\_N5\_3.json & 5 & 0.3148 & TL & 6 & 2 \\ \hline
        randGame\_N5\_4.json & 5 & 0.3527 & TL & 5 & 2 \\ \hline
        randGame\_N5\_5.json & 5 & 0.1963 & TL & 7 & 2 \\ \hline
        randGame\_N5\_6.json & 5 & 0.1408 & TL & 5 & 3 \\ \hline
        randGame\_N5\_7.json & 5 & 0.1615 & TL & 5 & 3 \\ \hline
        randGame\_N5\_8.json & 5 & 0.1628 & TL & 5 & 2 \\ \hline
        randGame\_N5\_9.json & 5 & 0.1682 & TL & 6 & 2 \\ \hline
		randGame\_N5\_10.json & 5 & 0.1372 & 466.228 & 5 & 2 \\ \hline
        randGame\_N5\_11.json & 5 & 0.2199 & TL & 6 & 2 \\ \hline
        randGame\_N5\_12.json & 5 & 0.1807 & TL & 8 & 2 \\ \hline
        randGame\_N5\_13.json & 5 & 0.2404 & TL & 8 & 2 \\ \hline
        randGame\_N5\_14.json & 5 & 0.1378 & TL & 5 & 2 \\ \hline
		randGame\_N5\_15.json & 5 & 0.3488 & 422.1422 & 7 & 3 \\ \hline
        randGame\_N5\_16.json & 5 & 0.1982 & TL & 6 & 2 \\ \hline
        randGame\_N5\_17.json & 5 & 0.2406 & TL & 7 & 2 \\ \hline
        randGame\_N5\_18.json & 5 & 0.1918 & TL & 6 & 2 \\ \hline
        randGame\_N5\_19.json & 5 & 0.1358 & TL & 5 & 2 \\ \hline
        randGame\_N5\_20.json & 5 & 0.1724 & TL & 7 & 2 \\ \hline
        randGame\_N5\_21.json & 5 & 0.2677 & TL & 3 & 2 \\ \hline
        randGame\_N5\_22.json & 5 & 0.2549 & TL & 3 & 3 \\ \hline
        randGame\_N5\_23.json & 5 & 0.3286 & TL & 4 & 3 \\ \hline
        randGame\_N5\_24.json & 5 & 0.2512 & TL & 4 & 2 \\ \hline
        randGame\_N5\_25.json & 5 & 0.4049 & TL & 4 & 2 \\ \hline
        randGame\_N5\_26.json & 5 & 0.3094 & TL & 5 & 2 \\ \hline
        randGame\_N5\_27.json & 5 & 0.4548 & TL & 4 & 2 \\ \hline
        randGame\_N5\_28.json & 5 & 0.4269 & TL & 5 & 3 \\ \hline
        randGame\_N5\_29.json & 5 & 0.2328 & TL & 3 & 2 \\ \hline
        randGame\_N5\_30.json & 5 & 0.4023 & TL & 4 & 2 \\ \hline
        randGame\_N5\_31.json & 5 & 0.3696 & TL & 3 & 3 \\ \hline
        randGame\_N5\_32.json & 5 & 0.2914 & TL & 3 & 2 \\ \hline
        randGame\_N5\_33.json & 5 & 0.6152 & TL & 4 & 2 \\ \hline
        randGame\_N5\_34.json & 5 & 0.4579 & TL & 4 & 2 \\ \hline
        randGame\_N5\_35.json & 5 & 1.034 & TL & 4 & 3 \\ \hline
        randGame\_N5\_36.json & 5 & 0.9702 & TL & 3 & 2 \\ \hline
        randGame\_N5\_37.json & 5 & 0.9179 & TL & 5 & 2 \\ \hline
        randGame\_N5\_38.json & 5 & 0.7389 & TL & 4 & 3 \\ \hline
        randGame\_N5\_39.json & 5 & 0.383 & TL & 5 & 3 \\ \hline
        randGame\_N5\_40.json & 5 & 0.5029 & TL & 4 & 2 \\ \hline
        randGame\_N5\_41.json & 5 & 1.6599 & TL & 4 & 2 \\ \hline
        randGame\_N5\_42.json & 5 & 3.0782 & TL & 3 & 2 \\ \hline
        randGame\_N5\_43.json & 5 & 3.0619 & TL & 4 & 2 \\ \hline
        randGame\_N5\_44.json & 5 & 2.9871 & TL & 3 & 2 \\ \hline
        randGame\_N5\_45.json & 5 & 2.5106 & TL & 4 & 3 \\ \hline
        randGame\_N5\_46.json & 5 & 4.4058 & TL & 4 & 2 \\ \hline
        randGame\_N5\_47.json & 5 & 5.773 & TL & 4 & 2 \\ \hline
        randGame\_N5\_48.json & 5 & 4.0088 & TL & 5 & 2 \\ \hline
        randGame\_N5\_49.json & 5 & 2.5236 & TL & 4 & 2 \\ \hline
        randGame\_N5\_50.json & 5 & 1.9007 & TL & 3 & 2 \\ \hline
        randGame\_N5\_51.json & 5 & 1.1015 & TL & 3 & 3 \\ \hline
        randGame\_N5\_52.json & 5 & 2.3951 & TL & 3 & 3 \\ \hline
        randGame\_N5\_53.json & 5 & 2.0004 & TL & 4 & 2 \\ \hline
        randGame\_N5\_54.json & 5 & 2.152 & TL & 3 & 2 \\ \hline
        randGame\_N5\_55.json & 5 & 2.976 & TL & 4 & 2 \\ \hline
        randGame\_N5\_56.json & 5 & 2.5105 & TL & 4 & 2 \\ \hline
        randGame\_N5\_57.json & 5 & 3.679 & TL & 4 & 2 \\ \hline
        randGame\_N5\_58.json & 5 & 4.2192 & TL & 5 & 2 \\ \hline
        randGame\_N5\_59.json & 5 & 2.429 & TL & 4 & 3 \\ \hline
        randGame\_N5\_60.json & 5 & 1.0492 & TL & 2 & 2 \\ \hline
        randGame\_N5\_61.json & 5 & 3.7919 & TL & 5 & 2 \\ \hline
        randGame\_N5\_62.json & 5 & 2.6304 & TL & 3 & 2 \\ \hline
        randGame\_N5\_63.json & 5 & 0.8615 & TL & 2 & 3 \\ \hline
        randGame\_N5\_64.json & 5 & 2.5485 & TL & 3 & 2 \\ \hline
        randGame\_N5\_65.json & 5 & 1.9033 & TL & 3 & 2 \\ \hline
        randGame\_N5\_66.json & 5 & 2.6877 & TL & 4 & 2 \\ \hline
        randGame\_N5\_67.json & 5 & 1.7178 & TL & 3 & 2 \\ \hline
        randGame\_N5\_68.json & 5 & 1.415 & TL & 3 & 2 \\ \hline
        randGame\_N5\_70.json & 5 & 2.5291 & TL & 3 & 3 \\ \hline
        randGame\_N5\_71.json & 5 & 2.5622 & TL & 3 & 3 \\ \hline
        randGame\_N5\_72.json & 5 & 3.1483 & TL & 4 & 2 \\ \hline
        randGame\_N5\_73.json & 5 & 2.3002 & TL & 3 & 2 \\ \hline
        randGame\_N5\_74.json & 5 & 2.2855 & TL & 3 & 3 \\ \hline
        randGame\_N5\_75.json & 5 & 2.7916 & TL & 4 & 2 \\ \hline
        randGame\_N5\_76.json & 5 & 2.361 & TL & 4 & 3 \\ \hline
        randGame\_N5\_77.json & 5 & 1.4727 & TL & 3 & 2 \\ \hline
        randGame\_N5\_78.json & 5 & 3.5901 & TL & 5 & 3 \\ \hline
        randGame\_N5\_79.json & 5 & 1.9264 & TL & 3 & 2 \\ \hline
        randGame\_N5\_80.json & 5 & 3.2194 & TL & 4 & 2 \\ \hline
        randGame\_N5\_81.json & 5 & 20.7868 & TL & 3 & 2 \\ \hline
        randGame\_N5\_82.json & 5 & 77.052 & TL & 3 & 2 \\ \hline
        randGame\_N5\_83.json & 5 & 2.9141 & TL & 5 & 2 \\ \hline
        randGame\_N5\_84.json & 5 & 13.5376 & TL & 3 & 3 \\ \hline
        randGame\_N5\_85.json & 5 & 51.0934 & TL & 4 & 3 \\ \hline
        randGame\_N5\_86.json & 5 & 4.3119 & TL & 4 & 3 \\ \hline
        randGame\_N5\_87.json & 5 & 15.6461 & TL & 4 & 2 \\ \hline
        randGame\_N5\_88.json & 5 & 1.5185 & TL & 3 & 3 \\ \hline
        randGame\_N5\_89.json & 5 & 4.2633 & TL & 3 & 3 \\ \hline
        randGame\_N5\_90.json & 5 & 48.3602 & TL & 3 & 2 \\ \hline
        randGame\_N5\_91.json & 5 & 60.0801 & TL & 4 & 2 \\ \hline
        randGame\_N5\_92.json & 5 & 3.7866 & TL & 4 & 2 \\ \hline
        randGame\_N5\_93.json & 5 & 117.4824 & TL & 4 & 2 \\ \hline
        randGame\_N5\_94.json & 5 & 1.3942 & TL & 2 & 3 \\ \hline
        randGame\_N5\_95.json & 5 & 2.5022 & TL & 2 & 2 \\ \hline
        randGame\_N5\_96.json & 5 & 14.1286 & TL & 3 & 3 \\ \hline
        randGame\_N5\_97.json & 5 & 1.8665 & TL & 3 & 2 \\ \hline
        randGame\_N5\_98.json & 5 & 7.9392 & TL & 3 & 3 \\ \hline
        randGame\_N5\_99.json & 5 & 2.831 & TL & 3 & 3 \\ \hline
        randGame\_N5\_100.json & 5 & 1.859 & TL & 3 & 2 \\ \hline
    \end{longtable}
\end{tiny}

\iftoggle{arxiv}{}{
}

\end{document}